\documentclass[aps,prd,twocolumn,floatfix,noshowpacs,tightenlines,noshowkeys,superscriptaddress,amsmath,amssymb,nofootinbib, 10pt]{revtex4-2}
\usepackage{amsbsy,epsfig,color,graphicx}
\usepackage{color}
\usepackage{longtable}
\usepackage{array}
\usepackage{dcolumn}   
\usepackage{cellspace}
\usepackage{mathtools}
\usepackage{amstext}
\usepackage{amssymb}
\usepackage{multirow}
\usepackage{mathrsfs}
\usepackage{stmaryrd}
\usepackage{stackrel}
\usepackage{graphicx}
\usepackage{esint}
\usepackage[utf8]{inputenc}
\usepackage{adjustbox}
\usepackage{multirow}
\usepackage{booktabs}
\usepackage{enumitem}
\usepackage{slashed}
\usepackage{hyperref}
\usepackage{etoolbox} 
\usepackage{lipsum} 
\usepackage[capitalize]{cleveref}
\usepackage{lipsum,booktabs}
\usepackage{changepage}
\usepackage{calligra}
\DeclareMathAlphabet{\mathcalligra}{T1}{calligra}{m}{n}
\DeclareFontShape{T1}{calligra}{m}{n}{<->s*[2.2]callig15}{}
\newcommand{\scripty}[1]{\ensuremath{\mathcalligra{#1}}}
\usepackage{multirow}
\usepackage[caption=false]{subfig}
\allowdisplaybreaks
\usepackage{soul}
\usepackage{cancel}
\makeatletter

\appto{\appendix}{%
	\@ifstar{\def\theequation@prefix{A.}}%
	{}%
}
\makeatother

\hypersetup{pdfstartview=FitV,colorlinks=true,linkcolor=midblue,citecolor=midblue,filecolor=midblue,urlcolor=midblue}

\definecolor{midblue}{rgb}{0,0,0.5}
\definecolor{cadmiumorange}{rgb}{0.93, 0.53, 0.18}

\newcommand{\beq}{\begin{equation}\begin{aligned}}
\newcommand{\eeq}{\end{aligned}\end{equation}}
\def\Tmat{\mathsf{T}^{(\textrm{m})}}
\def\Tpp{\mathsf{T}^{(\textrm{p})}}
\def\e{\epsilon}
\def\s{\sigma}
\def\CO{\mathcal{O}(\xi^32,\epsilon^3,\xi \epsilon^2)}
\def\Op{\mathcal{O}(\epsilon)}
\def\Od{\mathcal{O}(\xi\epsilon)}
\def\F{\mathsf{F}_{\Omega}}
\def\tpp{^{(1,0)}}
\def\tdd{^{(0,1)}}
\def\tpd{^{(1,1)}}
\def\todd{_{\textrm{odd}}}
\def\teven{_{\textrm{even}}}
\def\mfa{\mathfrak{a}}
\def\mfb{\mathfrak{b}}
\def\mfc{\mathfrak{c}}

\usepackage{hyperref}
\hypersetup{colorlinks=true}

\def\equationautorefname~#1\null{%
	Eq.~(#1)\null
}
\def\figureautorefname~#1\null{%
	Fig.~#1\null
}
\def\tableautorefname~#1\null{%
	Table.~#1\null
}
\def\sectionautorefname~#1\null{%
	Section #1\null
}
\def\appendixautorefname~#1\null{%
	Appendix #1\null
}

\newcommand{\Mbh}{M_{\textrm{BH}}}
\newcommand{\Mh}{M_{\textrm{Halo}}}
\newcommand{\rhoF}{\rho_{\textrm{gal}}}
\newcommand{\rhoB}{\rho_{\textrm{BH}}}

\begin{document}

	\title{Post-adiabatic waveforms from extreme mass ratio inspirals in the presence of dark matter}

	\author{Mostafizur Rahman}
	\email{rahman@tap.scphys.kyoto-u.ac.jp}
	\affiliation{Department of Physics, Kyoto University, Kyoto 606-8502, Japan}
	
\author{Takuya Takahashi}
	\email{takuya.takahashi@resceu.s.u-tokyo.ac.jp}
	\affiliation{Research Center for the Early Universe (RESCEU), Graduate School of Science, The University of Tokyo, Tokyo 113-0033, Japan}

	\begin{abstract}
   Extreme mass-ratio inspirals (EMRIs), in which a solar mass compact object is whirling around a supermassive black hole, act as precise tracers of the spacetime geometry and astrophysical environment around the supermassive black hole. These systems are highly sensitive to even the smallest deviations from the vacuum general relativity scenario.
However, detecting these signals requires highly accurate waveform modeling up to the first post-adiabatic order, incorporating self-force effects, system parameters, and environmental influences. In this paper, we focus on the impact of dark matter on gravitational waveforms. Cold dark matter in galactic centers can be redistributed by the gravitational pull of a supermassive black hole, forming a dense, spike-like profile. When an EMRI evolves in such an environment, the interaction between the binary and the surrounding dark matter can leave distinctive imprints on the emitted waveform, and thus offers a novel way to probe the nature and distribution of dark matter. We specifically examine how dark matter modifies the background spacetime. By treating these modifications perturbatively, we present a framework to incorporate dark matter environmental effects into gravitational waveform modeling at the first post-adiabatic order.

	\end{abstract}
	
	\maketitle
\section{Introduction}

The detection of gravitational waves has marked a monumental shift in observational astrophysics by granting direct access to the highly dynamical and strong gravity regimes \cite{LIGOScientific:2016aoc, LIGOScientific:2017vox}. Ground-based detectors such as LIGO, Virgo, and KAGRA have already detected signals from around a hundred binary black hole and neutron star coalescence events \cite{LIGOScientific:2018mvr, LIGOScientific:2020ibl, KAGRA:2021vkt}. These detectors are sensitive to high-frequency gravitational wave signals emitted by black hole and/or neutron star binaries. Deep buried in the detector noise lies a treasure trove of information, not only about the binary systems themselves but also about their surrounding environments and the underlying physical theory that governs their dynamics. When this information is extracted using matched filtering techniques, enabled by precise modeling of gravitational waveforms, it offers unprecedented insights into the nature of compact objects and the spacetime geometry \cite{Carson:2020rea, Berti:2015itd, Perkins:2020tra, Barack:2018yly, LIGOScientific:2021sio,Cardoso:2019rvt, Liebling:2012fv, Cardoso:2016oxy, Chakravarti:2024ncc, DeFalco:2024ojf, Battista:2023znv, Battista:2021rlh}.
\par
The breakthrough in numerical relativity simulation was a significant milestone in waveform modeling of comparable mass black hole/neutron star binaries \cite{Pretorius:2005gq, Campanelli:2005dd, Buonanno:2006ui, Centrella:2010mx}. However, despite its overwhelming success for waveform modeling such binaries, the technique becomes less practical for highly asymmetric binaries. 
Extreme mass-ratio inspirals (EMRIs) are one such class of systems \cite{Amaro-Seoane:2007osp, Gair:2017ynp, Babak:2017tow,Gair:2010yu, PhysRevD.78.064028}. These are binary systems consisting of a compact stellar-mass object of mass $\mu$ (hereafter referred to as the secondary) slowly inspiraling into a supermassive black hole of mass $\Mbh$ (hereafter referred to as the primary), typically with a mass ratio $q = \mu : \Mbh \sim 10^{-4} - 10^{-7}$. The extreme mass disparity in such systems makes full numerical relativity simulations computationally extremely expensive, as they must resolve length scales that differ by a factor of $\sim \Mbh / \mu = 1/q$. As a result, evolving a single EMRI over the inspiral timescale ($\sim \Mbh/q$) would require an unrealistically large number of timesteps, far beyond what current computational resources allow (in practice, a single EMRI would take billions of years to simulate with present numerical relativity codes).
\par
However, EMRIs are one of the most important sources for space-based gravitational wave detectors like the \textit{Laser Interferometer Space Antenna} (LISA), \textit{TianQin}, \textit{Taiji} and \textit{DECIGO} \cite{PhysRevD.95.103012,  Amaro-Seoane:2007osp, Babak:2017tow, Amaro_Seoane_2012, Luo_2016, TianQin:2015yph, Ruan:2018tsw, Kawamura:2011zz}. The secondary completes millions of orbital cycles around the primary before merger, and in this process, it effectively maps the spacetime geometry of the primary with extreme precision \cite{PhysRevD.78.064028}. That makes such systems highly sensitive even to the tiniest deviation from vacuum general relativity \cite{Babak:2017tow, Apostolatos:2009vu, Amaro-Seoane:2007osp, Barack:2009ux, Hinderer:2008dm, Drasco:2005kz, Gair:2004iv, Sopuerta:2009iy,Lukes-Gerakopoulos:2010ipp, Rahman:2021eay, Rahman:2022fay,Fan:2022wio,Liang:2022gdk,Destounis:2020kss,Maselli:2021men,Destounis:2021mqv,Destounis:2021rko,Drummond:2023loz, PhysRevD.102.024041,Destounis:2023gpw,Destounis:2023khj}. During this inspiral process, the systems emit low-frequency gravitational waves in the milli-Hertz band, which makes them primary targets for space-based gravitational wave detectors\cite{PhysRevD.95.103012,  Amaro-Seoane:2007osp, Babak:2017tow, Amaro_Seoane_2012, Luo_2016, TianQin:2015yph, Ruan:2018tsw, Kawamura:2011zz}. Thus, it is extremely important to accurately model gravitational waves from such system. 
\par
Currently, the self-force formalism provides the only practical and accurate framework for modeling gravitational waveforms from EMRIs \cite{Barack:2018yly}. The underlying idea is that the background spacetime $\bar g_{\mu\nu}$ of the primary gets perturbed in the presence of the secondary. The perturbed spacetime can be expressed as $g_{\mu\nu}=\bar g_{\mu\nu}+\e h^{(1,0)}_{\mu\nu}+\e^2 h^{(2,0)}_{\mu\nu}+\mathcal{O}(\e^3)$, where the parameter $\e$ counts the power of the mass ratio and $ h^{(n,0)}_{\mu\nu}$ is the $n$th-order perturbation term in vacuum background spacetime. Because of the interaction of the secondary object with its own perturbations, the object experiences a \textit{force} that drives it away from the geodesic trajectory of the background spacetime. This force term is referred to as the self-force. Considering a vacuum background spacetime, the acceleration can be expressed as follows \cite{ Mino:1996nk, Quinn:1996am, PhysRevD.103.064048,PhysRevLett.109.051101, PhysRevD.95.104056}
\beq\label{sf_vacc}
\frac{dz^\mu}{d\tau}=\e F^{\mu}_{(1,0)}+\e^2 F^{\mu}_{(2,0)}+\mathcal{O}(\e^3)
\eeq
where $z^{\mu}$ represents the worldline of the secondary object and $\tau$ represents the proper time in the background spacetime. The term $F^{\mu}_{(1,0)}$ captures the linear order gravitational self-force and the spin effect of the secondary, whereas the term $F^{\mu}_{(2,0)}$ contains quadratic order self-force terms and also the deformation effects of the secondary. Using the two-timescale analysis, Hinderer and Flanagan showed that over the long inspiral timescale $T_i \sim \mathcal{O}(\Mbh/\epsilon)$, the total accumulated gravitational wave phase can be expanded as $\varphi(T_i)=\left[\varphi_0(T_i)+\e \varphi_1(T_i)+\mathcal{O}(\e^2)\right]/\e$ \cite{PhysRevD.78.064028}. Here, the the leading (adiabatic, or 0PA) order term $\varphi_0/\e$ is determined by the orbit-averaged dissipative part of the linear force term $F^{\mu}_{(1,0)}$. The computation of the subleading (first post-adiabatic, or 1PA) term $\varphi_1(T_i)$ requires knowledge of both the conservative and dissipative components of linear-order force term $F^{\mu}_{(1,0)}$ and the orbit-averaged dissipative component of the quadratic-order force term $F^{\mu}_{(2,0)}$. Higher-order corrections to the gravitational wave phase can be ignored as these corrections appear at $\mathcal{O}(\e)$.

While substantial progress has been made in modeling 1PA waveforms in vacuum general relativity, incorporating gravitational self-force, and  spin effects \cite{Barack:2018yvs, Sago:2005fn, Isoyama:2018sib, vandeMeent:2017bcc, PhysRevD.96.084057, Witzany:2019dii, Rahman:2021eay, PhysRevD.102.064013, PhysRevLett.109.051101, PhysRevD.89.104020, PhysRevD.92.104047,DeLuca:2023laa,Berezhiani:2023vlo, Gliorio:2025cbh, Datta:2025ruh}, much less attention has been paid to EMRIs evolving in astrophysical environments. Recently, however, there has been growing interest in this direction \cite{PhysRevD.105.L061501,Jaramillo:2020tuu, Cardoso:2022whc,Barausse:2014tra,Cheung:2021bol,Courty:2023rxk,Sarkar:2023rhp, Vicente:2025gsg, Tomaselli:2023ysb, Coogan:2021uqv, Speeney:2024mas, Mitra:2025tag, Rahman:2023sof, Chakraborty:2024gcr}. This is due to the fact that the supermassive black holes typically reside at galactic centers, where environmental effects, such as the presence of accretion disks, surrounding stars, and dark matter, can significantly alter the orbital dynamics and, consequently, the emitted gravitational waveforms. Furthermore, it has been noted that the cold dark matter in galactic centers redistributes due to the gravitational influence of the supermassive black hole to form a dark matter spike profile around the black hole \cite{Gondolo:1999ef, Sadeghian:2013laa}. In this work, we aim to investigate how such a dark matter environment modifies the covariant acceleration equation, given in \autoref{sf_vacc}, and, in turn, affects the gravitational waveform.

We adopt a perturbative approach to incorporate the influence of dark matter. This is motivated by the observation that, in a self-gravitating system where a static black hole adiabatically grows within a dark matter environment, the leading-order modification to the Schwarzschild spacetime can be characterized by the parameter
\beq
\xi \approx \left(\frac{\Mh}{\Mbh}\right)\left(\frac{4\Mbh}{r_c}\right)^{3-\mfb} \ll 1,
\eeq
where $\Mh$ is the total mass of the dark matter spike profile, $r_c$ is its characteristic length scale, and $\mfb$ is a parameter determined by the dark matter density profile. We refer to $\xi$ as the dark matter parameter, which typically takes small values ($\xi \sim \epsilon$) in realistic astrophysical settings. Hence, we treat $\xi$ as a perturbation parameter which captures the perturbative corrections to the background geometry induced by dark matter. Using the two-timescale analysis \cite{PhysRevD.78.064028} within the fixed-frequency formalism \cite{ Mathews:2021rod, Mathews:2025nyb}, we develop a framework to model 1PA waveforms in a dark matter environment, thereby extending current waveform models built in vacuum spacetimes. We show that the presence of dark matter introduces additional force terms in \autoref{sf_vacc}. The leading-order contribution to the force term $F_{(0,1)}^\mu$ appears at order $\mathcal{O}(\xi)$ and is purely conservative; thus, it causes a shift in the conserved quantities (such as energy and angular momentum) of the secondary's orbit. The next-order correction, $F_{(1,1)}^\mu$, which arises at $\mathcal{O}(\epsilon\xi)$, includes both conservative and dissipative components. However, for constructing 1PA waveforms, only the orbit-averaged dissipative component is required.

We also examine the modification of gravitational wave fluxes due to dark matter and quantify the resulting changes in orbital evolution. Finally, we estimate the gravitational wave dephasing induced by the presence of dark matter, and highlight the potential observability of such effects through the space-based detectors.

The paper is organized as follows. In \autoref{sec: Adaibatic_growth}, we describe the adiabatic growth of a black hole in a dark matter environment. \autoref{sec: BH sol} presents a static, spherically symmetric black hole solution in the presence of dark matter. In \autoref{sec: emri_in_dm}, we outline the framework for modeling EMRI systems embedded in dark matter. \autoref{sec: two_timescale} introduces the two-timescale framework and the fixed-frequency formalism. The perturbation equations in the Regge–Wheeler–Zerilli formalism are discussed in \autoref{Sec:RWZ_formalism}, while the hyperboloidal method used to solve these equations is detailed in \autoref{hyperboloidal}. The main results of this work are presented in \autoref{sec: results}, followed by concluding remarks in \autoref{conclusion}. Additional technical details are provided in the appendices. \ref{app:1} contains the derivation of the perturbation equations, \ref{app: Dirac delta} summarizes the distributional properties of the Dirac delta function.

In this paper, we have adopted the positive sign convention for the metric $(-,+,+,+)$ and use geometrized units $G=c=1$. Greek letters are used to represent four-dimensional indices. Additionally, quantities with bracketed superscripts or subscripts, such as $Q^{(n,k)}$, represent coefficients of the perturbative expansion in powers of $\epsilon^n \xi^k$.
\section{Dark Matter density profile in the presence of black hole}\label{sec: Adaibatic_growth}
\subsection{Dark matter density profile}
In this paper, we consider a dark matter distribution around a black hole, known as a dark matter spike, in which the density of the dark matter halo is compressed due to the adiabatic growth of the black hole at the center.
Following Ref.~\cite{Sadeghian:2013laa}, we first briefly review how the dark matter spike distribution is constructed.
We start by assuming a spherically symmetric halo of collisionless dark matter supported solely by self-gravity, with the density profile given by\cite{Speeney:2022ryg}
\begin{equation}\label{density_galaxy}
    \begin{aligned}     \rhoF(r)=\rho_0\left(\frac{r}{a_0}\right)^{-\gamma}\left[1+\left(\frac{r}{a_0}\right)^\alpha\right]^\frac{\gamma-\beta}{\alpha}
    \end{aligned}
\end{equation}
where, $a_0$ is the typical length scale of the halo, $\rho_0$ is the scale density which is related to $a_0$ by the following relation $\rho_0=2^{(\beta-\gamma)/\alpha}\rhoF(a_0)$. The parameters $\alpha$, $\beta$ and $\gamma$ controls the slope of the distribution. \autoref{tab:dark_matter_profile} shows dark matter density profile corresponding to different models. We consider the Hernquist model \cite{1990ApJ...356..359H} and the Navarro–Frenk–White (NFW) profile \cite{Navarro:1996gj} in this paper. The total mass of the NFW distribution is logarithmically divergent. Thus, we need to set a cut-off radius $r_c$, such that the total mass of the halo $\Mh(r>r_c)=0$ \cite{Speeney:2024mas}.\par
\begin{table}[htb!]
\centering
 \def\arraystretch{1.1}      	
	\setlength{\tabcolsep}{1.2em}
\begin{tabular}{|l|l|l|l|}
\hline\hline
Profile          & $\alpha$ & $\beta$     & $\gamma$                                                                        \\\hline\hline
Hernquist        & 1                     & 4                        & 1                                                                                            \\
NFW              & 1                     & 3                        & 1                                                                                           
\\\hline\hline
\end{tabular}
\caption{Parameter values for different dark matter distribution profiles.}
\label{tab:dark_matter_profile}
\end{table}
Once the initial dark matter density is specified as in \autoref{density_galaxy}, the corresponding phase-space distribution function can be determined.
At this stage, the system can be described within Newtonian gravity, and the distribution function can be computed using the Eddington's inversion formula~\cite{Lacroix:2018qqh}
\begin{equation}
    f(E)=\frac{1}{\sqrt{8}\pi^2}\frac{d}{dE}\int_E^0 d\Phi\frac{1}{\sqrt{\Phi-E}}\frac{d\rho_{\rm gal}}{d\Phi}
\end{equation}
where $E$ is the energy and $\Phi$ is the Newtonian gravitational potential.
For the Hernquist profile, an analytical expression for $f(E)$ can be obtained~\cite{Sadeghian:2013laa}, whereas for the NFW profile, it must be computed numerically.

Next, we consider the dark matter profile after a seed black hole has grown adiabatically into a supermassive black hole at the center of the halo.
The distribution function (defined apart from a Jacobian factor) remains the same functional form before and after the evolution.
Therefore, to obtain the distribution function in terms of the constants of motion in the final state, it is sufficient to know the relation between the constants of motion before and after the evolution. 
This can be achieved by comparing the action variables, which are adiabatic invariants.
In the initial state of the dark matter halo described by the Newtonian gravity, the radial, azimuthal and polar action variables are given by
\begin{align}
    I_{r}^{\rm ini}&=\oint dr\sqrt{2(E-\Phi(r))-\frac{L}{r^2}}~, \\
    I_{\theta}^{\rm ini}&=\oint d\theta \sqrt{L^2-\frac{L_z^2}{\sin^2\theta}}=2\pi(L-|{L_z|})~, \\
    I_{\phi}^{\rm ini}&=\oint d\phi L_z=2\pi L_z~,
\end{align}
respectively.
Here, $E,\ L$ and $L_z$ denotes the energy, total angular momentum and angular momentum along the $z$-direction of the dark matter particle, respectively.
In the final state, assuming the gravitational field is dominated by the central black hole and can be described by the Schwarzschild spacetime, action variables are given by
\begin{align}
    I_{r}^{\rm fin}&=\oint dr \left(1-\frac{2\Mbh}{r}\right)^{-1} \notag \\
    &\qquad\times\sqrt{{\cal E}^2-\left(1-\frac{2\Mbh}{r}\right)\left(1+\frac{L^2}{r^2}\right)}~, \\
    I_{\theta}^{\rm fin}&=\oint d\theta \sqrt{L^2-\frac{L_z^2}{\sin^2\theta}}=2\pi(L-|{L_z|})~, \\
    I_{\phi}^{\rm fin}&=\oint d\phi L_z=2\pi L_z~,
\end{align}
where ${\cal E},\ L$ and $L_z$ are defined in the same way above.

\begin{figure*}[t!]
	\centering
	\minipage{1.0\textwidth}
	\includegraphics[width=\linewidth]{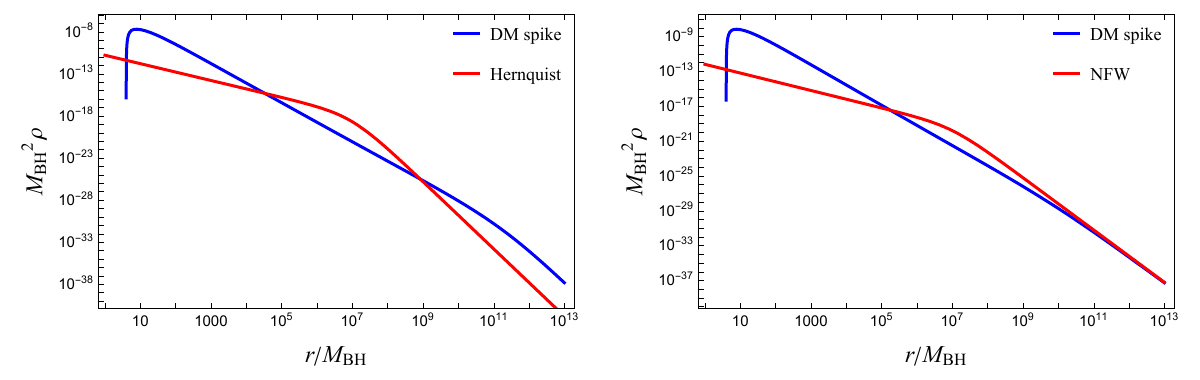}
	\endminipage
	\caption{Dark matter spike density profile (blue line) for the cases where the initial profile (red line) is Hernquist (left) and NFW (right). In this analysis, we assume a total dark matter mass of $M_{\rm Halo} = 10^4 M_{\rm BH}$, with $M_{\rm Halo}/a_0 = 10^{-3}$, and a cutoff radius given by $r_{\rm c} = 100 M_{\rm Halo} a_0 / M_{\rm BH}$. }\label{fig_spike}
\end{figure*}

By equating $\theta$ and $\phi$ components of action variables, it follows that $L$ and $L_z$ remain unchanged between the initial and final states, which reflects the spherical symmetry of the system.
Meanwhile, the energy varies, and from the equation for the radial action variables, the initial energy can be expressed as a function of the final-state energy and the total angular momentum, $E=E({\cal E}, L)$.
As a result, the dark matter distribution after the black hole growth is expressed in terms of the final-state constants of motion.
With this distribution function, the final dark matter spike density profile is given by~\cite{Sadeghian:2013laa}
\begin{align}
    \rho_{\rm BH}(r)=&\frac{4\pi}{r^2}\left(1-\frac{2\Mbh}{r}\right)^{-1/2} \notag \\
    &\times\int d{\cal E}dL\frac{{\cal E}Lf(E({\cal E},L))}{\sqrt{{\cal E}^2-(1-2\Mbh/r)(1+L^2/r^2)}}~.
\end{align}
In \autoref{fig_spike}, we present the numerical results for the dark matter spike profile for the cases where the initial profiles are assumed to be Hernquist and NFW, respectively.
In this analysis, we adopt a total dark matter mass of $M_{\rm Halo} = 10^4 M_{\rm BH}$, with $M_{\rm Halo}/a_0 = 10^{-3}$, and a cutoff radius defined as $r_{\rm c} = 100 M_{\rm Halo} a_0 / M_{\rm BH}$.

\subsection{Fitting formula}
Here, we present a fitting formula for the dark matter density profile constructed numerically.
After the growth of a black hole inside the dark matter distribution in an adiabatic process, the dark matter density takes the following form\cite{Speeney:2024mas}
\begin{equation}\label{density_Cardoso}
    \begin{aligned}
   \rhoB(r)
   =\Mh\frac{\bar{\rho}_s(r)}{I(r_c)}
    \end{aligned}
\end{equation}
where 
\begin{equation}\label{density_Cardoso1}
    \begin{aligned}
   \bar{\rho}_s(r)
   &=\left(1-\frac{4 \Mbh}{r}\right)^{ \mfa}\left(\frac{R_S}{r}\right)^{\mfb}\left(1+\frac{r}{R_S}\right)^{-\mfc}\\
    I(r)
   &=\int_{4\Mbh}^{r}4\pi r^2 \bar{\rho}_s(r) dr
    \end{aligned}
\end{equation}
and $R_S=(\Mh a_0/\Mbh)\gg \Mbh$ for realistic scenario, $\mfa$, $\mfb$ and $\mfc$ are numerical parameters obtained by fitting the numerical curve obtained from the adiabatic growth of the black hole inside the dark matter profile with different values of parameter space $(\Mh,~ a_0, \Mbh,~r_c)$. In \autoref{tab:Fitting_Parameters}, we present the values of the fitting parameters considering  total dark matter mass of $M_{\rm Halo} = 10^4 M_{\rm BH}$, with $M_{\rm Halo}/a_0 = 10^{-3}$, and a cutoff radius given by $r_{\rm c} = 100 M_{\rm Halo} a_0 / M_{\rm BH}$. Note that the dark matter density vanishes for $r < 4\Mbh$. This reflects the fact that, in our analysis, we consider dark matter particles on bound orbits around the black hole, which exist only for $r \geq 4\Mbh$, where $4\Mbh$ corresponds to the innermost bound circular orbit (IBCO) \cite{Sadeghian:2013laa}. \par
It is more convenient to represent the dark matter profile in terms of the variable $y=1-4\Mbh/r$, which leads to the following expression for the density profile
\begin{equation}
    \begin{aligned}
      \rhoB(y)&=\frac{\rho_{H}}{ L(y_c)}\bar{\rho}_s(y)
    \end{aligned}
\end{equation}
where $\rho_{H}=\Mh/4\pi (4\Mbh)^3$ and 
\begin{equation} \label{density}
    \begin{aligned}
       \bar{\rho}_s(y)
   &=\frac{y^\mfa(1-y)^{\mfb+\mfc}}{(1-\zeta y)^\mfc}\\
    L(y)&=\int_{0}^{y}\frac{\bar y^\mfa(1-\bar y)^{\mfb+\mfc-4}}{(1-\zeta \bar y)^\mfc}d\bar y\\
        &=\frac{y^{\mfa+1}}{\mfa+1}F_{1}(\mfa+1,-(\mfb+\mfc-4),\mfc,\mfa+2,y,\zeta y)
    \end{aligned}
\end{equation}
 with $\zeta=R_S/(R_S+4\Mbh)$, $y_c=1-4\Mbh/r_c$ and   $F_1(\alpha,\beta,\gamma,\delta,y,\zeta y)$ is the Appell hypergeometric function \cite{nakagawa2023appelllauricella}. Note that, for astrophysically realistic situations, $R_S\gg \Mbh$. As a result, we can make further simplication, noting $\zeta\sim 1$. Thus, the density profile can be approximated as 
 \begin{equation} 
    \begin{aligned}
       \bar{\rho}_s^{\textrm{approx}}(y)
   &=\left[y^\mfa(1-y)^{\mfb}\right]\\
    L^{\textrm{approx}}(y)&=\int_{0}^{y}\left[\bar y^\mfa(1-\bar y)^{\mfb-4}\right]d\bar y\\
        &=B(1-y; \mfa+1, \mfb-3)
    \end{aligned}
\end{equation}
where $B(1-y; \mfa+1, \mfb-3)$ is the incomplete beta function. In this paper, however, we consider density distribution given by \autoref{density}, without making this approximation.
\begin{table}[htb!]
\centering
 \def\arraystretch{1.1}      	
	\setlength{\tabcolsep}{1.2em}
\begin{tabular}{|l|l|l|l|}
\hline\hline
Fitting Parameters          & $\mfa$ & $\mfb$     & $\mfc$                                                                        \\\hline\hline

Hernquist        & 2.637                     & 2.330                        & 1.344                                                                                            \\
NFW              & 2.640                    & 2.332                        & 0.5446                                                                                          
\\\hline\hline
\end{tabular}
\caption{The fitting parameters $\mfa$, $\mfb$, and $\mfc$ for the dark matter spike profile corresponding to the Hernquist and NFW cases. In this analysis, we assume a total dark matter mass of $M_{\rm Halo} = 10^4 M_{\rm BH}$, with $M_{\rm Halo}/a_0 = 10^{-3}$, and a cutoff radius given by $r_{\rm c} = 100 M_{\rm Halo} a_0 / M_{\rm BH}$. }
\label{tab:Fitting_Parameters}
\end{table}
\section{Black hole solution in the presence of dark matter}\label{sec: BH sol}

In the previous section, following the prescription of \cite{Sadeghian:2013laa}, we have shown that if we adiabatically grow a black hole within a cold dark matter environment, the dark matter particles are redistributed under the gravitational influence of the black hole, leading to the formation of a dark matter spike profile. In this section, we aim to construct a static, spherically symmetric spacetime that describes a black hole of mass $\Mbh$ inside a dark matter distribution.
For this, we consider a static and spherically symmetric black hole spacetime described by the line element 
\beq\label{BH_sol}
    ds^{2}&=  g_{\mu\nu}dx^{\mu}dx^{\nu}\\&=-a(r) dt^{2} + \frac{1}{b(r)} dr^{2} + r^{2} \left[d\theta^{2}+\sin^{2}\theta d\phi^{2}\right],\\
    b(r)&=\left(1-\frac{2m(r)}{r}\right)\,,\quad
    a(r)=b(r)~e^{q(r)}~,
\eeq
 where $m(r)$ denotes the mass function and $q(r)$ denotes the red shift function \cite{PhysRevD.105.L061501}. These functions are determined by solving the Einstein field equations, $G_{\mu\nu} = 8\pi \Tmat_{\mu\nu}$, where $\Tmat_{\mu\nu}$ represents the energy-momentum tensor associated with the dark matter spike profile. Following \cite{PhysRevD.105.L061501, PhysRevD.67.104017}, we model the dark matter surrounding the black hole as an anisotropic fluid, such that the energy-momentum tensor of the dark matter takes the form
\begin{align}\label{stt1}
\Tmat_{\mu\nu} = \rho_{\textrm{BH}} \mathfrak{u}_{\mu}\mathfrak{u}_{\nu}+p_{r}k_{\mu}k_{\nu}+p_{t}\Pi_{\mu\nu}. 
\end{align}
Here, $\rho_{\textrm{BH}}$ denotes the dark matter density, the expression for which is given in \autoref{density_Cardoso1}. The quantities $p_r$ and $p_t$ represent the radial and tangential pressures, respectively. The four-velocity of the fluid is given by $\mathfrak{u}^{\mu} = \delta^{\mu}_{t} / \sqrt{a}$, and $k_{\mu}= \sqrt{b}~\delta^{r}_{\mu}$ denotes a unit spacelike vector which is orthogonal to $\mathfrak{u}^{\mu}$. These quantities satisfy the relations $\mathfrak{u}^{\mu} \mathfrak{u}_{\mu} = -1$, $\mathfrak{u}^{\mu} k_{\mu} = 0$ and $k^{\mu} k_{\mu} = 1$. Additionally, the projection operator $\Pi_{\mu\nu} = g_{\mu\nu} + \mathfrak{u}_{\mu} \mathfrak{u}_{\nu} - k_{\mu} k_{\nu}$ is orthogonal to both $\mathfrak{u}^{\mu}$ and $k^{\mu}$. Following the Einstein cluster prescription presented in \cite{PhysRevD.105.L061501, PhysRevD.67.104017}, we consider that  the dark matter profile has vanishing radial pressure, i.e., $p_r=0$. From the Einstein field equations, we find that the mass function, the redshift factor and the tangential pressure satisfies the following relations 
\begin{align}
      m'(r)&=4\pi r^2\rhoB(r)\label{G11}~,\\
      q'(r)&=\frac{2m'(r)}{r-2m(r)}\label{Gtt1}~,\\
      2p_t&=\frac{m(r)\rhoB(r)}{r-2m(r)}~,\label{G33}
    \end{align}
   where ``prime'' denotes derivative of a function with respect to its argument.  Let us first consider the \autoref{G11}. 
    For the density profile  given by \autoref{density}, the mass function takes a simple form which can be written as follows
\begin{equation}\label{mass_func}
    \begin{aligned}
        m(r)&=\Mbh+\Mh \frac{L(y)}{L(y_c)}
        &=\Mbh+\xi\delta m(r)
    \end{aligned}
\end{equation}
where $y=1-4\Mbh/r$ and
\beq\label{delta_m}
\delta m(r)=\Mbh L(y)~,\\
\xi=\left(\frac{\Mh}{\Mbh}\right)\frac{1}{L(y_c)}~.
\eeq
Note that at $r = 4\Mbh$, we find $m(r) = \Mbh$, since $L(0) = 0$. This is expected, as there is no dark matter enclosed within $r = 4\Mbh$. In what follows, we refer to $\xi$ as the dark matter parameter, as it characterizes the modification to the black hole spacetime induced by the presence of dark matter. To further illustrate its physical interpretation, we note that in astrophysical situation, the cut off radius $r_c$ is much larger than the typical size of a black hole $~\Mbh$, which implies that $y_c\approx1$. Thus, if we expand $L(y_c)$ around $y_c=1$, $L(y_c)$ assumes the value $\sim( r_c/4\Mbh)^{3-\mfb}$ ($\mfb<3$). This allows us to write the dark matter parameter as 
\begin{equation}
    \begin{aligned}
\xi=\left(\frac{\Mh}{\Mbh}\right)\frac{1}{L(y_c)}\sim\left(\frac{\Mh}{\Mbh}\right)\left(\frac{4\Mbh}{r_c}\right)^{3-\mfb}\ll 1
    \end{aligned}
\end{equation}
Recall that the parameter $\mfb$ is determined by the initial dark matter distribution. Thus, given a specific spike profile, the dark matter parameter $\xi$ is controlled by both the spike-to-black hole mass ratio and the spike-to-black hole length scale ratio. In realistic astrophysical scenarios, this parameter is typically very small.\par
To find the expression for the red shift factor $q(r)$, we consider
the equation \autoref{Gtt1} with the mass function given by \autoref{mass_func}.
\begin{figure*}[t!]
	\centering
	\minipage{0.450\textwidth}
	\includegraphics[width=1.1\linewidth]{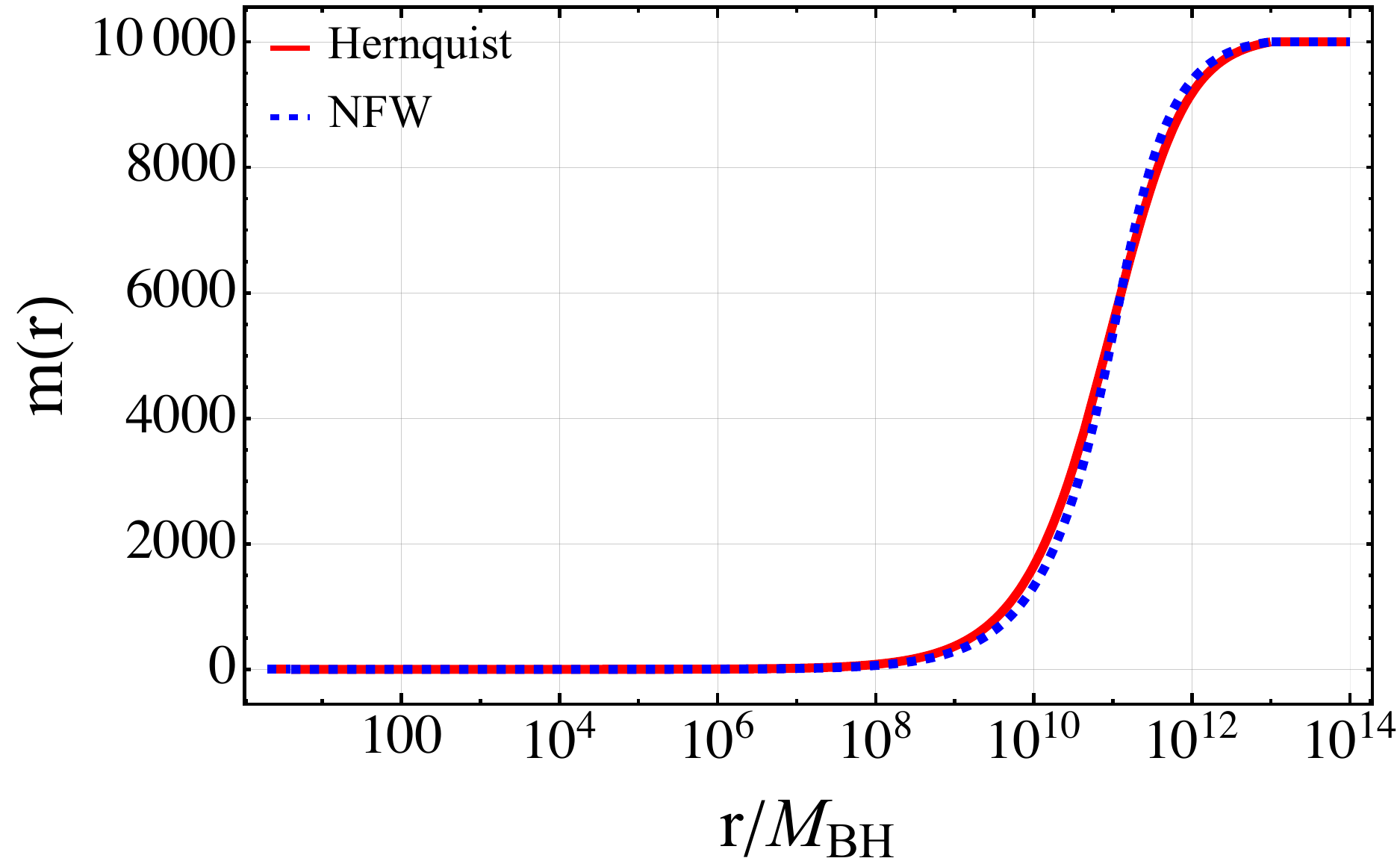}
	\endminipage\hfill
    \minipage{0.450\textwidth}
	\includegraphics[width=1.1\linewidth]{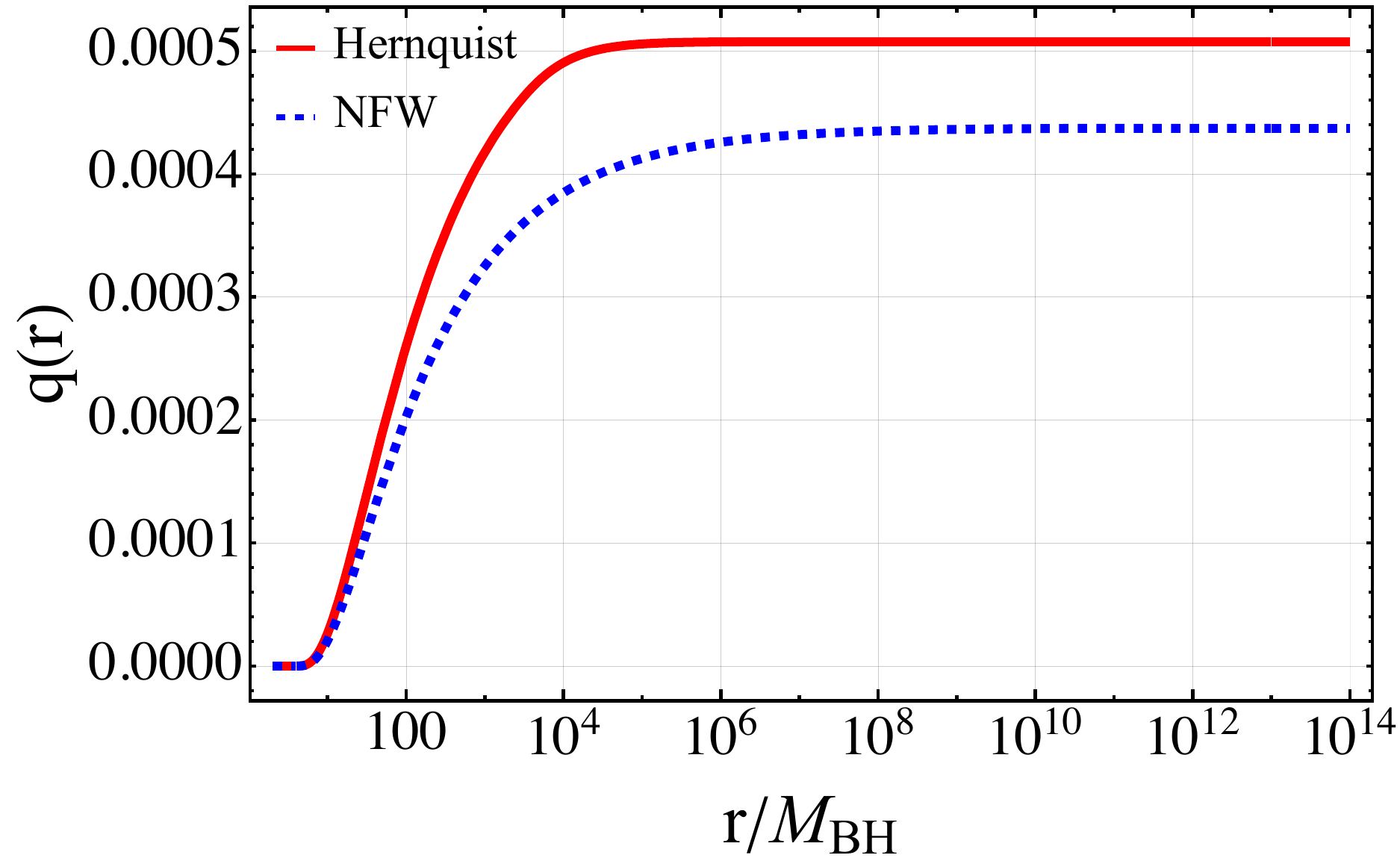}
	\endminipage
	\caption{The plots of the mass function $m(r)$ (left panel) and the redshift function $q(r)$ (right panel) as functions of radius are shown for the Hernquist (solid red line) and NFW (blue dashed line) dark matter density profiles, using the fitting parameters listed in \autoref{tab:Fitting_Parameters}. Since the dark matter density vanishes for $r \leq 4\Mbh$, the mass function $m(r)=\Mbh$ in this region. Beyond $4\Mbh$, $m(r)$ gradually increases and asymptotically approaches to $\Mbh+\Mh$ near the cutoff radius. The redshift function $q(r)$ remains small for both the Hernquist and NFW profiles and approaches a constant value near the cutoff radius.}\label{fig_mass_function}
\end{figure*}	
Thus, we can rewrite this equation in the following manner
\begin{equation}\label{gval1}
    \begin{aligned}
   q(r) &=\xi \delta f(r)~,\\
\delta f'(r) &=   \frac{2\delta m'(r)}{r-2m(r)}~.
    \end{aligned}
\end{equation}
 We obtain the expression of $q(r)$ by numerically integrating the above equation. In \autoref{fig_mass_function}, we plot the mass function $m(r)$  and the redshift function $q(r)$ as functions of $r$ for the Hernquist and NFW dark matter density profiles, using the fitting parameters presented in \autoref{tab:Fitting_Parameters}. As shown in the plot, the mass function $m(r)=4\Mbh$ when $r\leq 4\Mbh$ and gradually increases to reach the value $\Mbh+\Mh$ near the cutoff radius. The redshift function $q(r)$ remains small for both profiles and asymptotes to a constant value beyond the cutoff radius.\par
 As discussed above, the dark matter parameter is typically very small for real astrophysical situations. Given that our motivation is to construct 1PA EMRI waveform model in the presence of dark matter, it is sufficient to consider the leading order corrections to the Schwarzschild spacetime due the presence of dark matter. Thus, we consider $\xi$ as a perturbation parameter  which modifies the Schwarzschild spacetime as $g_{\mu\nu}=\bar{g}_{\mu\nu}+\xi h^{(0,1)}_{\mu\nu}+\mathcal{O}(\xi^2)$, where $\bar{g}_{\mu\nu}$ is the spacetime metric of Schwarzschild black hole. The non-vanishing components of $h^{(0,1)}_{\mu\nu}$ is given by
\begin{equation}\label{Gtt1p}
    \begin{aligned}
    h^{(0,1)}_{rr}&=f^{-2}(r)\frac{2\delta m(r)}{r}\\
     h^{(0,1)}_{tt}&=f(r)\delta f(r)-\frac{2\delta m(r)}{r}\\
    \end{aligned}
\end{equation}
where 
\begin{equation}\label{Gtt1pd}
    \begin{aligned}
    f(r)&=\left(1-\frac{2\Mbh}{r}\right)~,\\
    \xi\delta m'(r)&=4\pi r^2\rhoB(r)~,\\
   \delta f'(r)&= \frac{2\delta m'(r)}{\left(r-2\Mbh\right)}
    \end{aligned}
\end{equation}
 It is important to note that, beyond order $\mathcal{O}(\xi)$, the presence of dark matter modifies only the redshift function.  Higher-order corrections can be systematically incorporated by expanding the expression in \autoref{gval1} as a power series in $\xi$.
\section{Extreme mass ratio inspirals in dark matter environment}\label{sec: emri_in_dm}
\subsection{Equation of motion of the secondary}
In the previous section, we demonstrated that the presence of a matter distribution modifies the background vacuum spacetime as $g_{\mu\nu}=\bar{g}_{\mu\nu}+\xi h^{(0,1)}_{\mu\nu}+\mathcal{O}(\xi^2)$, where the perturbation is dictated by a small parameter $\xi$. In this section, we analyze the motion of a secondary object of mass $\mu$ inspiraling into a supermassive black hole in the presence of this matter distribution. 
However, we do not assume a priori that the matter distribution corresponds to dark matter. The only assumption we make is that the presence of some matter distribution characterized by its energy-momentum tensor $\Tmat_{\mu\nu}$ perturbs the background spacetime, with the perturbation being characterized by the small parameter $\xi$, and the asymptotic structure of spacetime remains unchanged due to the presence of the matter distribution.\par
The secondary object perturbs both the physical spacetime and the matter distribution.
The resulting spacetime in the presence of the secondary can be expressed as  \cite{Dyson:2025dlj}
\beq\label{metric_emri}
\mathbf{g}_{\mu\nu}=&\bar{g}_{\mu\nu}+\xi h^{(0,1)}_{\mu\nu}+\epsilon h^{(1,0)}_{\mu\nu}+\epsilon \xi h^{(1,1)}_{\mu\nu} \\&+\epsilon^2 h^{(2,0)}_{\mu\nu}+\mathcal{O}(\xi^2,\epsilon^3,\xi \epsilon^2)
\eeq
where the order counting parameter $\epsilon$ counts the power of $\mu$. We assume that $\xi \sim \epsilon$, which is a reasonable approximation for realistic astrophysical scenarios. For example, for the configuration presented in \autoref{tab:Fitting_Parameters}, the dark matter parameter takes the value $\xi = 5.65 \times 10^{-4}$ ($\xi = 1.53 \times 10^{-4}$) for the Hernquist (NFW) dark matter density profile. Hereafter, we use bracketed superscripts or subscripts, for example, $h^{(n,k)}$, to denote coefficients in the perturbative expansion in powers of $\epsilon^n \xi^k$.
 Note that the term $h^{(0,2)}_{\mu\nu}$ does not contribute at the first post-adiabatic order; thus, we omit its contribution.\par
The energy momentum tensor of the secondary object $\Tpp_{\mu\nu}$ corresponds to that of point particle which can be written as  \cite{Poisson:2011nh}
\beq\label{pp_em}
\Tpp_{\mu\nu} = \mu \int d\hat\tau \frac{\delta^{(4)}(x-z(\hat\tau))}{\sqrt{-\hat g}} \hat{u}_{\mu} \hat{u}_{\nu},
\eeq
where $z^{\mu}$ denotes the world line,  $\hat\tau$ is the proper time along the worldline with respective to an effective metric $\hat{g}_{\mu\nu}$, and $\hat{u}_{\mu} = dz^{\mu}/d\hat\tau$  denotes the four-velocity.
 The effective metric can be written as $\hat{g}_{\mu\nu}=\bar{g}_{\mu\nu}+\xi h^{(0,1)}_{\mu\nu}+h^{R}_{\mu\nu}$ \cite{ PhysRevD.103.064048,PhysRevLett.109.051101, PhysRevD.95.104056}. Here,   $h^{R}_{\mu\nu}$ denotes the regular piece of metric perturbation which can be obtained by subtracting the singular fields $h^{S(n,k)}_{\mu\nu}$ from the the physical metric perturbation \cite{ PhysRevD.103.064048,PhysRevLett.109.051101, PhysRevD.95.104056}
\beq
h^{R}_{\mu\nu}&=\epsilon\left( h^{(1,0)}_{\mu\nu}- h^{S(1,0)}_{\mu\nu}\right)+\epsilon \xi \left(h^{(1,1)}_{\mu\nu}-h^{S(1,1)}_{\mu\nu}\right) \\&+\epsilon^2 \left(h^{(2,0)}_{\mu\nu}-h^{S(2,0)}_{\mu\nu}\right)+\mathcal{O}(\xi^2,\epsilon^3,\xi \epsilon^2)~.
\eeq
%
Substituting \autoref{metric_emri} into the Einstein field equation $G_{\mu\nu}[\mathbf{g}]=8 \pi (\Tmat_{\mu\nu}+\Tpp_{\mu\nu})$ and equating the coefficients of power of $\epsilon$ and $\xi$, we obtain the following sets of equations
\begin{align}
G_{\mu\nu}[\bar{g}]&=0,\label{vac}\\
\delta G_{\mu\nu}[h^{(0,1)}]&=8\pi \Tmat_{\mu\nu}[\bar g]~,\label{mat}\\
\delta G_{\mu\nu}[h^{(1,0)}]&=8\pi \Tpp_{\mu\nu}[\bar g]~,\label{pp}\\
\delta G_{\mu\nu}[h^{(2,0)}]&=8\pi \delta\Tpp_{\mu\nu}[ h^{(1,0)}]-\delta^2 G_{\mu\nu}[h^{(1,0)},h^{(1,0)}]~,\label{mat2}\\
\delta G_{\mu\nu}[h^{(1,1)}]&=8\pi \delta\Tpp_{\mu\nu}[ h^{(0,1)}]+8\pi \delta\Tmat_{\mu\nu}[ h^{(1,0)}] \notag \\&-\delta^2 G_{\mu\nu}[h^{(0,1)},h^{(1,0)}]~,\label{matpp}
\end{align}
where $\delta G_{\mu\nu}$ is the linearized Einstein tensor, $\delta^2 G_{\mu\nu}[h^{(A)},h^{(B)}]$ is the piece of Einstein tensor that contains terms like, $\partial h^{(A)}\partial h^{(B)}$, $h^{(A)}\partial^2 h^{(B)}$ etc (see Eq.~(29) of \cite{PhysRevD.103.064048}).
In \autoref{mat2}, and \autoref{matpp}$, \delta \mathsf{T}^{\textrm{(m),(p)}}[h^{(A)}]$ denotes the piece of $\mathsf{T}^{\textrm{(m),(p)}}_{\mu\nu}[g+h^{(A)}]$ that appears at order $\Od$. In addition to \autoref{vac}-\autoref{matpp}, we have the covariant conservation law for the matter energy momentum tensor
\beq\label{cov_cons_law}
&\nabla_{\nu}\mathsf{T}^{\mu\nu}_{\textrm{(m)}}[\bar g]=0\,,\\
&\nabla_{\nu}\delta\mathsf{T}^{\mu\nu}_{\textrm{(m)}}[ h^{(1,0)}]+C^{\mu}_{\nu\alpha}[h^{(1,0)}]\mathsf{T}_{\textrm{(m)}}^{\alpha\nu}[\bar g]\\&+C^{\nu}_{\nu\alpha}[h^{(1,0)}]\mathsf{T}_{\textrm{(m)}}^{\mu\alpha}[\bar g]=0\,,
\eeq
where $C^{\mu}_{\nu\alpha}[h]=g^{\mu\lambda}(2h_{\lambda(\nu;\alpha)}-h_{\nu\alpha;\lambda})/2$, and $\nabla_{\nu}$ and the symbol  ``$;$'' denotes covariant derivative with respect to the background metric.\par
The secondary object moves as a test particle in the effective spacetime described by the metric $\hat{g}_{\mu\nu}$ and, consequently, follows a geodesic equation $\hat{u}^\mu \hat{\nabla}_\mu \hat{u}^\nu = \mathcal{O}(\xi^2,\epsilon^3,\xi \epsilon^2)$, where the covariant derivative is taken with respect to $\hat{g}_{\mu\nu}$ \cite{PhysRevD.103.064048, PhysRevLett.109.051101, PhysRevD.95.104056}. 
It is customary to write the equation of motion in terms of the background metric $\bar  g_{\mu\nu}$. Using the relation between the proper time of the effective and background metric $d\hat{\tau}=d\tau \sqrt{1- \mathfrak{h}_{\mu\nu} u^{\mu}u^{\nu}}$, we obtain the equation of motion with respect to the background spacetime \cite{ Mino:1996nk, Quinn:1996am, PhysRevD.103.064048,PhysRevLett.109.051101, PhysRevD.95.104056}
\beq\label{sf_eqn}
\frac{D u^{\nu}}{d\tau}&=u^{\alpha}\nabla_{\alpha}u^{\nu}=F^{\nu}[\mathfrak{h}]\\
F^{\nu}[\mathfrak{h}]&\equiv -\frac{1}{2}P^{\mu\nu}(g_{\nu}^{\lambda}-\mathfrak{h}_{\nu}^{\lambda})\left(2 \mathfrak{h}_{\lambda \rho ; \sigma}-\mathfrak{h}_{\rho \sigma ; \lambda}\right) u^{\rho} u^{\sigma}
\eeq
where $\mathfrak{h}_{\mu\nu}=\xi h^{(0,1)}_{\mu\nu}+h^{R}_{\mu\nu}$, and $u^{\mu}=d z^{\mu}/d\tau$ is the four-velocity with respect to the background metric and $P^{\mu}_{\nu}=\bar{g}^{\mu}_{\nu}+u^{\mu}u_{\nu}$ is the projection tensor.
The coefficient of the force term in the perturbative expansion in powers of $\epsilon^n \xi^k$, namely $F_{(n,k)}^{\mu}$, is given by
\beq\label{force_term_coef}
F_{(n,k)}^{\mu}=\frac{1}{n!k!}\dfrac{\partial^{n+k}F^{\mu}[\mathfrak{h}]}{\partial \e^n\partial \xi^k}~.
\eeq
In the absence of dark matter ($\xi=0$), \autoref{sf_eqn} becomes the self-force equation or the Mino-Sasaki-Tanaka-Quinn-Wald (MiSaTaQuWa ) equation  for a point particle in vacuum spacetime \cite{Mino:1996nk, Quinn:1996am}. However, in the presence of dark matter,  $F^{\mu}$ includes a purely conservative component,  $F^{\mu}_{(0,1)}$ , along with additional contributions that contain both conservative and dissipative parts \cite{PhysRevD.103.064048}.   \par
The advantage of expressing the field equation in this form is that one can first evaluate the solution for a point particle in vacuum independently of the dark matter contribution. The effects of dark matter, as encoded in \autoref{mat}, \autoref{mat2}, \autoref{matpp}, and \autoref{sf_eqn}, can then be added to determine its overall influence. However, in order to single out the contribution due to dark matter, we need to adopt the \textit{fixed-frequency} formalism \cite{PhysRevD.103.064048, Mathews:2021rod, Mathews:2025nyb}, which we discuss in the next section.\par
\section{Two-time scale analysis and fixed frequency formalism}\label{sec: two_timescale}
In what follows, we consider the matter energy-momentum tensor given in the previous section, which is expressed by \autoref{stt1}. Due to the interaction of the point particle with its own gravitational field, the object slowly inspirals towards the supermassive black hole, with the inspiral time scale being $T_i\sim\Mbh/\epsilon$. As a result, both the orbital frequency $\Omega$ and the dark matter parameter $\xi$ evolve slowly on this timescale \cite{Hinderer:2008dm}.
In contrast, the orbital phase of the particle evolves on the fast timescale $T_o\sim \Mbh$. We assume that time dependence  in the particle's motion and the spacetime metric arises solely  through their dependence on $(\phi_p,\Omega,\xi)$. For the sake of simplicity, we further assume that the absorption of radiation or dark matter induces negligible changes in the mass and spin of the supermassive black hole  (see \cite{PhysRevD.103.064048, Mathews:2021rod} for a discussion of how changes in black hole mass and spin impact EMRI dynamics). This assumption will later be used to determine the appropriate boundary conditions for the dark matter density perturbation. 
\subsection{Parameterization of the orbit in the fixed-frequency formalism}
In this section, we consider a particle inspiraling to the supermassive black hole in dark matter environment in equatorial, circular orbit. 
In the absence of dark matter and self-force, the point-particle follows geodesic trajectory of the background Schwarzschild spacetime $z^{\mu}=\left(t, z^{i}\right)$, where
\beq\label{traj_geo}
z^i=z_0^i&=\left(r_p,\pi/2,\phi_p\right)\,,
 \eeq
with 
\beq
r_p=r_{\Omega}\equiv\frac{\Mbh}{\left(\Mbh\Omega_0\right)^{2/3}}\,,\quad \frac{d\phi_p}{dt}=\Omega_0~.
\eeq
However, in the presence of dark matter and self-force, the trajectory gets shifted. The conservative pieces of the force term $F^{\mu}$ introduces shift in the orbital parameters, energy  $E_{\textrm{orb}}=-u^{\mu}\Xi^{(t)}_\mu$, angular momentum $L^z_{\textrm{orb}}=u^{\mu}\Xi^{(\phi)}_\mu$ and the orbital frequency $\Omega_0$, where $\Xi^{(t)}_\mu$ and $\Xi^{(\phi)}_\mu$ are the time-like and azimuthal Killing vectors associated with Schwarzschild spacetime. When written in \textit{fixed-radius} formalism, the shift in orbital frequency can be written as \cite{ Mathews:2021rod}
\beq
r &=r_p\,,\\ \Omega &=\Omega_0+\xi \Omega_{(0,1)}(r_p)+\e \Omega_{(1,0)}(r_p)\\&+\e \xi \Omega_{(1,1)}(r_p)+\CO
\eeq
An alternative way to paramerise the shifted trajectory is the \textit{fixed-frequency} formalism, where we instead write \cite{Mathews:2021rod}
\beq\label{r_expand}
\Omega &=\Omega_0\\
r &=r_\Omega+\xi r_{(0,1)}(\phi_p, \Omega)+\e r_{(1,0)}(\phi_p, \Omega)\\&+\e\xi r_{(1,1)}(\phi_p, \Omega)+\CO\,,\quad 
\eeq
 In what follows, we adopt the fixed-frequency formalism to study the orbital evolution of the point particle. For the sake of brevity, we drop the subscript in $\Omega_0$, and denote the frequency as $\Omega$ instead. Thus, in the fixed-frequency formalism, the trajectory of the point particle can be written as $z^{\mu}=\left(t, z^{i}(\phi_p,\Omega,\xi)\right)$, where
 \beq\label{trajectory}
z^i(\phi_p, \Omega,\xi)&=z_0^i(\phi_p, \Omega)+\xi  z_{(0,1)}^i(\phi_p, \Omega)+\e  z_{(1,0)}^i(\phi_p, \Omega)\\&+\e~ \xi z_{(1,1)}^i(\phi_p, \Omega)+\CO
 \eeq
where the terms $ z_{(0,1)}^i=r_{(0,1)}\delta^i_r$, $ z_{(1,0)}^i=r_{(1,0)}\delta^i_r$, and $ z_{(1,1)}^i=r_{(1,1)}\delta^i_r$ account for the corrections to the trajectory due to presence of dark matter and self-force. The frequency is still governed by the equation
 \beq\label{frequency}
\dot\phi_p=\frac{d\phi_p}{dt}\equiv \Omega~,
 \eeq
where the ``overdot'' represents derivative with respect to $t$.
The rate of change of orbital frequency is given by \cite{Mathews:2021rod}
\beq\label{freq_change}
\dot\Omega=\frac{d\Omega}{dt}=\e \F^{(1,0)}(\Omega)+\e\xi \F^{(1,1)}(\Omega)+\CO
\eeq

To obtain the expression for $\F^{(n,k)}$, we first note that using \autoref{frequency} and \autoref{freq_change} and applying chain rule we can write 
\beq\label{dt}
\frac{d}{dt}=\Omega \partial_{\phi_p}+\dot\Omega \partial_{\Omega}
\eeq
Furthermore, using \autoref{dt} and the expression $u^\mu u_{\mu}=(u^0)^2\dot z^\mu\dot z_{\mu}=-1$, we can obtain 
\beq\label{u_expand}
u^0=\frac{1}{\sqrt{1-3w}}+\e\xi\frac{3 r_{(1,0)}r_{(0,1)}}{(1-3w)^{3/2}}\,,\quad u^3=\Omega u^0
\eeq
where, $u^0=dt/d\tau$ and $w=(\Mbh\Omega)^{2/3}$. Furthermore, in terms of non-affine parameter $t$, the equation of motion \autoref{sf_eqn} becomes \cite{PhysRevD.103.064048,Mathews:2025nyb}
 \beq\label{force_eqn_na}
 \ddot{z}^\mu+\frac{\dot{u}^0}{u^0}\dot{z}^{\mu}+\Gamma^{\mu}_{\nu\alpha}\dot{z}^{\nu}\dot{z}^{\alpha}=\frac{1}{(u^0)^2}F^{\mu}~.
 \eeq
From the $r$ component of the equation, we obtain that 
\beq\label{r_shift}
r_{(0,1)}&=-\frac{F^{r}_{(0,1)}}{3(u^0)^2\Omega^2 f_\Omega}\,,\quad
r_{(1,0)}=-\frac{F^{r}_{(1,0)}}{3(u^0)^2\Omega^2 f_\Omega}\\
r_{(1,1)}&=-\frac{F^{r}_{(1,1)}}{3(u^0)^2\Omega^2 f_\Omega}+\frac{2 r_{(0,1)}r_{(1,0)}w(1-4w)}{\Mbh f_\Omega}
\eeq
where $F^{r}_{(0,1)}$ , $F^{r}_{(1,0)}$, $F^{r}_{(1,1)}$ are the coefficient of $\xi$, $\e$ and $\e\xi$ in the force term $F^{r}$ and $f_\Omega=f(r_\Omega)$. 
\begin{figure}[t!]
	\centering
	\minipage{0.50\textwidth}
	\includegraphics[width=\linewidth]{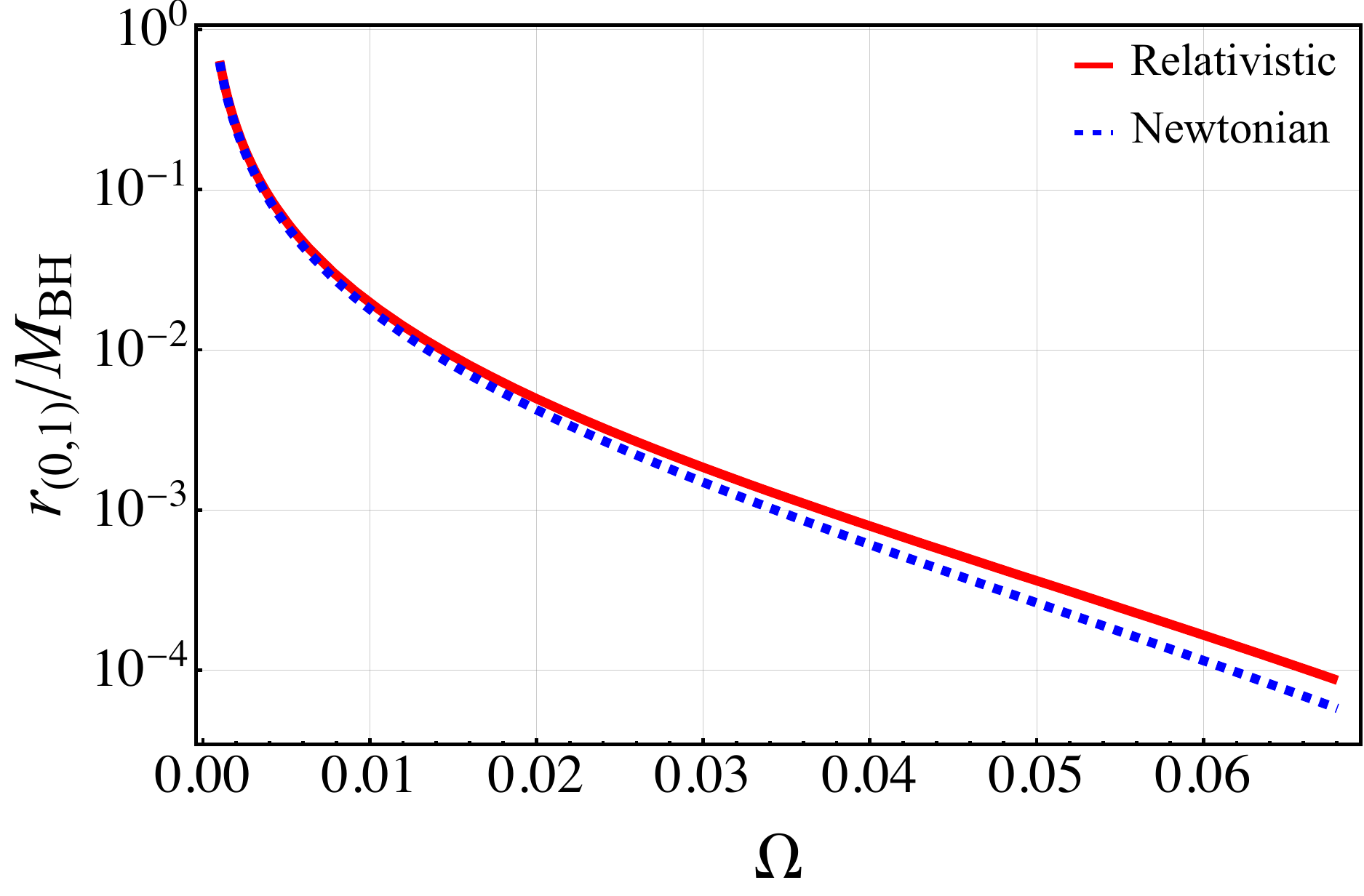}
	\endminipage
	\caption{The plot of radial shift in fixed frequency formalism as a function of orbital frequency. Here, the red line represents the relativistic result whereas the dashed line represents its Newtonian counterpart. For low orbital frequency values, there is a significant radial shift. At higher frequencies, however, the enclosed dark matter mass becomes negligible, resulting in a reduced radial shift.}\label{fig: radial_shift}
\end{figure}	
At this point, we calculate the shift in orbital radius $r_{(0,1)}$ from vacuum configuration due to the presence dark matter. With the configuration presented in \autoref{sec: BH sol}, we can calculate the force term $F_{(0,1)}^r$ using \autoref{sf_eqn}, the expression of which is given by 
\beq
F_{(0,1)}^r=u_0^2\left[-\frac{1}{2} f_{\Omega }^2\delta f'(r_\Omega)+\frac{f_{\Omega } \delta m'(r_\Omega)}{r_\Omega}-\frac{\delta m(r_\Omega)}{r_\Omega^2}\right]~.
\eeq
Replacing the above equation in \autoref{r_shift}, we find the radial shift due the presence of dark matter is given by the following expression 
\beq
r_{(0,1)}=\frac{r_\Omega \delta m(r_\Omega)}{3 \Mbh f_{\Omega }}
\eeq
In \cite{Speeney:2022ryg}, the authors presented an Newtonian approximation for the radial shift $r_{(0,1)}$ for binaries in circular orbit due to the presence of dark matter. Starting with the Newtonian equation of motion
\beq
\ddot{r}=-\frac{1}{r^2}\left[\Mbh+\xi \delta m(r)\right]+\Omega^2 r
\eeq
 and solving this equation perturbatively in the order $\xi$, they find the radial shift as \cite{Speeney:2022ryg}
 \beq
 r_{(0,1)}^{\textrm{N}}=\frac{r_\Omega \delta m(r_\Omega)}{3 \Mbh}
 \eeq
In \autoref{fig: radial_shift}, we present a plot comparing the relativistic results with the Newtonian ones. At small values of orbital frequency (i.e., at large orbital separations), the radial shift is significant. Moreover, in this regime, the Newtonian results agree quite well with the relativistic predictions. However, at larger orbital frequencies, the radial shift decreases. This behavior is expected because, at smaller separations (or higher frequencies), the dark matter mass enclosed within the radius $r_\Omega = \Omega^{-2/3}$ becomes negligible, resulting in a small radial shift. In this regime, the Newtonian approximation no longer agrees well with the relativistic results. \par
With the expression of radial shift in hand, we can calculate the orbital energy and angular momentum in fixed frequency formalism; the expression of which is given by 
\beq\label{orbital_energy}
E_{\textrm{orb}}&=-u^{\mu}\Xi^{(t)}_\mu=u^0 f_\Omega-\xi\frac{2  u^0 \delta m(r_{\Omega})}{3 f_{\Omega } r_{\Omega }}\\
L_{\textrm{orb}}^z &=u^{\mu}\Xi^{(\phi)}_\mu=r_\Omega^2 u^3 +\xi \frac{  r_\Omega^3 u^3  2 \delta m(r_{\Omega})}{3 \Mbh f_{\Omega } r_{\Omega }}~,
\eeq
where we have used \autoref{Gtt1pd} to obtain the above expression.
Finally, from the $t$ component of the forced geodesic equation, we obtain that expressions for $\F^{(n,k)}$ as
\beq\label{Omegadot0}
\F^{(1,0)} &= -\frac{3 f_\Omega\Omega F^t_{(1,0)}(\Omega)}{w(u^0)^4(1-6w)},\\
\F^{(1,1)} &= -\frac{3 f_\Omega\Omega F^t_{(1,1)}}{w(u^0)^4(1-6w)} 
 - \frac{2 \partial_{\Omega}F_{(0,1)}^r(\Omega)}{\sqrt{w}(u^0)^4 f_\Omega(1-6w)}\\
&\quad - \frac{4(1-6w+12w^2)F_{(0,1)}^r F_{(1,0)}^t(\Omega)}{w^{3/2} (u^0)^6f_\Omega(1-6w)^2}
\eeq
An alternative approach to computing these quantities is to invoke a flux balance law, under the assumption that higher-order corrections to the black hole spacetime due to the presence of dark matter can be neglected. Additionally, we ignore any accretion of matter onto the secondary object, so its mass and spin remain constant throughout the inspiral. Under these assumptions, the rate of change of the orbital energy $\dot{E}_{\textrm{orb}}$ and angular momentum $\dot{L}^z_{\textrm{orb}}$ due to the local force is assumed to be balanced by the combined losses from gravitational wave emission and dynamical friction \cite{Speeney:2022ryg}
\begin{equation}\label{flux_balance_law_1}
\dot{E}_{\textrm{orb}} = -\dot{E}_{\textrm{GW}} - \dot{E}_{\textrm{DF}}.
\end{equation}
In the subsequent sections, we detail the calculation of energy loss $\dot{E}_{\textrm{GW}}$ due to gravitational wave emission. The additional energy loss from dynamical friction arises from the interaction between the secondary object and the surrounding dark matter \cite{Chandrasekhar:1943ys}. As the secondary moves through the dark matter environment, it generates a slight overdensity, so-called the \textit{gravitational wake}, trailing behind it. The gravitational pull of this wake exerts a drag force on the secondary, resulting in a dissipative loss of energy. For a secondary in a circular orbit at large orbital separation, the energy loss due to dynamical friction can be approximated as \cite{Barausse:2007ph,Speeney:2022ryg}
\begin{equation}\label{dynamic_friction}
\dot{E}_{\textrm{DF}} = 4\pi \gamma^2(1 + \Omega^2 r_\Omega^2) \frac{m^2 \rho_{\textrm{BH}}}{\Omega r_\Omega} \ln{\Lambda},
\end{equation}
where $\ln{\Lambda} \approx 3$ is the Coulomb logarithm and $\gamma = 1/\sqrt{1 - \Omega^2 r_\Omega^2}$ is the Lorentz factor associated with circular orbital motion.
In this paper, we do not attempt to generalize this expression to the strong gravity regime; we leave such an extension to future work. However, we do compute the dark matter overdensity and the gravitational perturbation it induces.\par
Note that, the orbital energy can be written as $E_{\textrm{orb}}=E^{(0)}(\Omega)+\xi E\tdd(\Omega)+\e E^{(1,0)}(\Omega)$, where $E^{(0)}=f_\Omega u^0$ and $E\tdd(\Omega)$ is the coefficient of $\xi$ in \autoref{orbital_energy}. Additional contribution to the orbital energy comes from the conservative part of the gravitational self-force.
We can calculate the rate of change of orbital energy by substituting the above expression onto the balance law \autoref{flux_balance_law_1} and then applying the chain rule \autoref{dt}. At $\Op$, the flux balance law gives the expression for adiabatic evolution of frequency in a vacuum Schwarzschild spacetime 
\beq\label{flux_bal_adia}
\F^{(1,0)}=-\frac{\mathcal{F}\tpp}{\partial_\Omega E^{(0)}}
\eeq
 where $\mathcal{F}\tpp\equiv\dot{E}_{\textrm{GW}}\tpp$ represents the gravitational wave flux for a point particle inspirational onto a vacuum Schwarzschild black hole.  However, at $\Od$, we have to consider the effect of both dynamic friction and gravitational wave emission.
 The flux balance law at $\Od$ give rise to the 

 \beq\label{flux_bal_DM}
\F^{(1,1)}&=\mathsf{F}_{\Omega,\textrm{GW}}\tpd+\mathsf{F}_{\Omega,\textrm{DF}}\tpd\\
&=-\frac{\mathcal{F}\tpd+\F^{(1,0)} \partial_\Omega E^{(0,1)}}{\partial_\Omega E^{(0)}}-\frac{\dot{E}_{\textrm{DF}}\tpd}{\partial_\Omega E^{(0)}}
 \eeq
$\mathcal{F}\tpd\equiv\dot{E}_{\textrm{GW}}\tpd$ represents the modification of gravitational wave flux due to the presence of dark matter. In the above equation, $\mathsf{F}_{\Omega,\textrm{GW}}\tpd$ and $\mathsf{F}_{\Omega,\textrm{DF}}\tpd$ represents correction to $\dot\Omega$ due to gravitational wave emission and dynamic friction, respectively. 
\section{Perturbation equation: Regge-Wheeler-Zerilli Formalism}\label{Sec:RWZ_formalism}
We adopt the Regge–Wheeler–Zerilli formalism to study the perturbation of spacetime due to the presence of the secondary object \cite{PhysRev.108.1063, PhysRevLett.24.737, PhysRevD.69.044025}. In this approach, the metric perturbation is decomposed into axial (odd-parity) and polar (even-parity) sectors, i.e., $h_{\mu\nu} = h_{\mu\nu}^{ \textrm{odd}} + h_{\mu\nu}^{ \textrm{even}}$, based on their transformation properties under $(\theta, \phi) \to (\pi - \theta, \pi + \phi)$. The even-parity components remain invariant under this transformation, whereas the odd-parity components acquire a factor of $-1$ \cite{PhysRev.108.1063, PhysRevLett.24.737, PhysRevD.2.2141}. In \ref{app:1}, we provide the detailed derivation of the perturbation equations. Here, we highlight only the main results and the governing equations relevant to the perturbation analysis.
\par
We can expand the metric perturbations $h_{\mu\nu}$ and the point-particle energy-momentum tensor $\Tpp_{\mu\nu}$ in tensor basis (see \autoref{prt1} and \autoref{harmonicexp}) \cite{PhysRevD.67.104017, Cardoso:2022whc}. In the Regge-Wheeler gauge, the axial perturbations are described by the functions $h\todd^{lm} = (h_0^{lm}, h_1^{lm})$, while the polar perturbations are described by $h\teven^{lm} = (K^{lm}, H_0^{lm}, H_1^{lm}, H_2^{lm})$, all of which are functions of $r$. For notational simplicity, we drop the superscript $lm$ when referring to the axial and polar modes $h\todd^{lm}$ and $h\teven^{lm}$. Furthermore, to highlight the effect of dark matter, we express these perturbations as
\beq
h\todd &= h\todd\tpp + \xi h\todd\tpd, \\
h\teven &= h\teven\tpp + \xi h\teven\tpd,
\eeq
where $h\todd\tpp$ and $h\teven\tpp$ denote the perturbations at order $\mathcal{O}(\epsilon)$, and $h\todd\tpd$ and $h\teven\tpd$ represent the corrections at order $\mathcal{O}(\xi\epsilon)$. We can also express the perturbation in the matter energy momentum tensor $\Tmat_{\mu\nu}$ in angular basis. The secondary object perturbs the density and pressure of the fluid as 
\beq
\rho\tpd &=\sum_{lm}\delta \rho^{lm}Y_{lm}\,,\\\delta p_r\tpd &=\sum_{lm}\delta p_{r}^{lm}Y_{lm}\,,\\\delta p_t\tpd  &=\sum_{lm}\delta p_{t}^{lm}Y_{lm}
\eeq
Again, for the notational simplicity, we drop the superscript $lm$ and represent the density, radial and tangential pressure fluctuations as $\delta \rho$ , $\delta p_r$ and  $\delta p_t$, respectively. Further relation between the perturbed density and pressure is obtained through barotropic equation of state which is given as follows \cite{Cardoso:2022whc}
\beq
\delta p_{r}=c_{r}^2 (r) \delta \rho\,,\quad \delta p_{t}=c_{t}^2 (r) \delta \rho~,
\eeq
where $c_{r}$ and $c_t$ are radial and tangential sound speeds, respectively . Following \cite{Cardoso:2022whc}, we consider these as constants.\par
The metric perturbation equations for both the axial and polar sectors, as well as the fluid perturbation equations can be described in terms of five master functions: two related metric perturbation in the axial sector $(\Psi_{R}^{(1,0)}, \Psi_{R}^{(1,1)})$, two related to metric perturbation in the polar sector $(\Psi_{Z}^{(1,0)},\Psi_{Z}^{(1,1)})$ and one corresponds to the fluid perturbation in the polar sector 
$\Psi_{F}^{(1,1)}$. At order $\mathcal{O}(\epsilon)$, the perturbation equations correspond to those of a vacuum Schwarzschild spacetime sourced by a point particle. The governing equation at this order is given by \cite{PhysRevD.67.104017}
\beq\label{eq_o_e}
\mathcal{L}_R\Psi_R\tpp &\equiv \left[\partial^2_r{*}+\omega^2-V_{R}(r)\right]\Psi_R\tpp=S_R\tpp~\,,
\\
\mathcal{L}_Z\Psi_Z\tpp &\equiv \left[\partial^2_r{*}+\omega^2-V_{Z}(r)\right]\Psi_Z\tpp=S_Z\tpp~\,,
\eeq
where $r_{*}$ is the tortoise coordinate which satisfies the relation $dr_*=dr/f$, and $V_{R}(r)$ and $V_{Z}(r)$ are the Regge-Wheeler and Zerilli potential, respectively. We presented the explicit expressions for the potential and source term in \autoref{odd_pot} and \autoref{even_pot}.  It is important to note that, the source terms in both polar and axial sector are distribution source, the form of which can be written in the following manner (see \ref{app:1} for detailed discussion)
\beq\label{st10}
S^{(1,0)}_{R,Z}=G^{(1,0)}_{R,Z}\delta_r+F^{(1,0)}_{R,Z}\delta'_r
\eeq
where $\delta_r\equiv\delta(r-r_\Omega)$ is the Dirac delta function and $\delta'_r\equiv\delta'(r-r_\Omega)$ denotes the derivative of the Dirac delta function with respect to $r$ . In the above equation, the functions $G_{R,Z}, F_{R,Z}$  depend only on the $r_\Omega$. We obtain the expressions for these terms using the distributional properties of Dirac delta function, outlying in \autoref{Dirac_delta_1}. The relation between the master functions $(\Psi_{R}^{(1,0)},\Psi_{Z}^{(1,0)})$ and $(h\todd\tpp, h\teven\tpp)$ is also presented in \ref{app:1}.
\par
The effects of dark matter enter at order $\mathcal{O}(\xi\epsilon)$. We begin by considering the fluid perturbation, which is characterized by the master function $\Psi_{F}^{(1,1)}$. This function satisfies the following differential equation
\beq\label{eq_o_ez_f}
\mathcal{L}_{F}\Psi_{F}^{(1,1)}&\equiv \left[\frac{d^2}{dr_*^2}+\left(\frac{\omega^2}{c_r^2}-V_{F}(r)\right)\right]\Psi_{F}^{(1,1)}=S^{(1,1)}_{F}\,,
\eeq
where $V_F(r)$ is potential corresponding to fluid perturbation,  the expression of which is given by 
\beq
V_{F}(r)&=\frac{\left(c_r^2-1\right){}^2 f'^2}{16 c_r^4}+f(r) \left(\frac{\left(3 c_r^2-1\right) c_t^2 f'}{2 r c_r^4}\right)\\&+f(r) \left(\frac{r^2 \left(c_r^2-1\right) f''(r)+8 (n+1) c_t^2}{4 r^2 c_r^2}\right)\\&+f(r) \left(\frac{f(r) \left(c_t^4-c_r^2 c_t^2\right)}{r^2 c_r^4}\right)
\eeq
It is important to note that $V_F(r)$ vanishes at infinity, similar to the Regge–Wheeler and Zerilli potentials. However, unlike the Regge–Wheeler and Zerilli cases, $V_F(r)$ approaches a non-zero value near the horizon, given by $V_F=\left(c_r^2-1\right)^2 /(64 c_r^4)$. Thus, near to the horizon, the master function $\Psi_F\tpd$ effectively behaves  like a massive scalar field \cite{Cardoso:2022whc}.
It is interesting to note that the source term $S^{(1,1)}_{F}$ for the fluid perturbation equation has a distributional and a unbounded part $S^{(1,1)}_{F}=S^{D(1,1)}_{F}+S^{U(1,1)}_{F}$, with the distributional part similar to \autoref{st10}
\beq\label{stD10}
S^{D(1,1)}_{F}=G^{(1,1)}_{F}\delta_r+F^{(1,1)}_{F}\delta'_r+H^{(1,1)}_{F}\delta''_r
\eeq
and the  unbounded part $S^{U(1,1)}_{F}$ is a function of the master function for polar perturbation $\Psi_{Z}^{(1,0)}$ and its derivative, i.e.,
\beq\label{stUZ11}
S^{U(1,1)}_{F}=S^{U(1,1)}_{F}\left(\Psi_{Z}^{(1,0)},\partial_r\Psi_{Z}^{(1,0)}\right)\,.
\eeq
Note that, unlike the distributional sources in \autoref{st10} and \autoref{stD10}, the unbounded sources have support over the entire spatial domain. The presence of such sources poses significant computational challenges. In the next section, we discuss how this issue can be addressed using hyperboloidal methods. The relationship between the master function and the fluid perturbation variables are presented in \ref{app:1}. In particular, the density perturbation $\delta \rho$ is related to  
the master function $\Psi_F^{(1,1)}$ through the expression
\beq
\delta \rho=\Psi^{(1,1)}_F r^{-2+\frac{c_t^2}{c_r^2}}f^{-\frac{3}{4}-\frac{1}{4c_r^2}}
\eeq
At $\Od$, the homogeneous part of the metric perturbation master functions $\Psi_R\tpd$ and $\Psi_Z\tpd$ still satisfies the homogeneous  Regge-Wheeler and Zerilli equations
\beq\label{eq_o_ez}
\mathcal{L}_{R}\Psi_{R}^{(1,1)}&=S^{(1,1)}_{R}\,,
\\
\mathcal{L}_{Z}\Psi_{Z}^{(1,1)}&=S^{(1,1)}_{Z}\,.
\eeq
 where the source terms $S^{(1,1)}_{R,Z}$ at this order has a distributional and unbounded part $S^{(1,1)}_{R,Z}=S^{D(1,1)}_{R,Z}+S^{U(1,1)}_{R,Z}$ similar to fluid perturbation case. We can write the  distributional source term as 
\beq\label{stD11}
S^{D(1,1)}_{R,Z}=G^{(1,1)}_{R,Z}\delta_r+F^{(1,1)}_{R,Z}\delta'_r+H^{(1,1)}_{R,Z}\delta''_r
\eeq
In the axial perturbation case, the unbounded source term is a function of $\Psi_R\tpp$ and its derivative
\beq\label{stUR11}
S^{U(1,1)}_{R}=S^{U(1,1)}_{R}\left(\Psi_{R}^{(1,0)},\partial_r\Psi_{R}^{(1,0)}\right)\,.
\eeq
However, polar perturbation equation is sourced by both metric perturbation $\Psi_Z\tpp$ at $\Op$ and the fluid perturbation $\Psi_F\tpd$ at $\Od$. Thus, we can write the source terms for the polar perturbations as 
\beq\label{stUF11}
S^{U(1,1)}_{Z} &=S^{U(1,1)}_{Z,F}\left(\Psi_F\tpd,\partial_r \Psi_F\tpd\right)\\&+S^{U(1,1)}_{Z,G}\left(\Psi_{Z}^{(1,0)},\partial_r\Psi_{Z}^{(1,0)}\right)\,.
\eeq
 where $S^{U(1,1)}_{Z,F}$ and $S^{U(1,1)}_{Z,G}$ represent the contributions to the source term from the fluid perturbation and the metric perturbation, respectively. Given that the Zerilli operator is linear, we can write the solution of the polar equation \autoref{eq_o_ez} in a similar manner, i.e., $\Psi^{(1,1)}_{Z}=\Psi^{(1,1)}_{Z,F}+\Psi^{(1,1)}_{Z,G}$. Here, $\Psi^{(1,1)}_{Z,F}$ represents the contribution to the metric perturbation induced by the fluid (i.e., density) fluctuation as the secondary object moves through it, while $\Psi^{(1,1)}_{Z,G}$ corresponds to the perturbation arising from the modification of the background spacetime due to the presence of dark matter.
The contribution $\Psi^{(1,1)}_{Z,F}$ gives rise to phenomena like dynamical friction \cite{Barausse:2007ph}.\par
The boundary conditions for the \autoref{eq_o_e}, and \autoref{eq_o_ez} is given by the ``in'' and ``up'' solutions i.e
., the solution are purely incoming at the horizon and purely outgoing at the infinity \cite{Martel:2005ir}
\beq\label{bc_grav}
\Psi_{R,Z}^{\pm}=e^{\pm i \omega r_{*}}\,,\quad r_{*}\to\pm \infty
\eeq
To set the boundary condition for the fluid equation \autoref{eq_o_ez_f}, we make use of the assumption that the infall of dark matter and radiation onto the supermassive black hole changes its mass and angular momentum negligibly. As a result, the location of the innermost bound circular orbit (IBCO) remains at $4\Mbh$, where the radial pressure vanishes.
Now consider the Lagrangian variation of the pressure, $\Delta p_r = \delta p_r + \mathcal{L}_{\chi} p_r$ (see \ref{app:1})\cite{1975ApJ...200..204F}. Given that $p_r = 0$ at this surface, the Lagrangian variation reduces to the Eulerian variation, i.e., $\Delta p_r = \delta p_r$. If $\delta p_r$ were nonzero, it would imply that the surface of zero pressure has shifted. However, this contradicts our assumption that the surface remains fixed at $4\Mbh$. Therefore, $\delta p_r$ must vanish at $4\Mbh$, which in turn implies that $\delta \rho$ also vanishes at $4\Mbh$. This provides the boundary condition for the fluid master function: $\Psi_F\tpd = 0$ at $4\Mbh$. We impose the boundary condition $\Psi_F\tpd = 0$ at $r_c$ based on similar arguments.

\subsection{Gravitational wave flux}

In this section, we calculate the energy flux emitted in the form of gravitational wave to the infinity and down to the horizon. Let us assume $\Sigma_\infty$ ($\Sigma_\mathcal{H}$) is a timelike hypersurface with $r_{*}\gg \Mbh$ ($r_{*}\ll - \Mbh$). If $d\Sigma_{\infty,\mathcal{H}}$ denotes the portion of $\Sigma_{\infty,\mathcal{H}}$ in a small time span, 
the amount of energy flux through $\Sigma_{\infty,\mathcal{H}}$ over time $dt$ is given by the following expression \cite{PhysRevD.81.084039}
\begin{equation}\label{energy_flux_formula}
\begin{aligned}
dE_{\infty,\mathcal{H}}=\mp\oint T^{\alpha}_{(\textrm{gw})\beta}\Xi^{\beta}_{t}d\Sigma_{\alpha}^{\infty,\mathcal{H}}
\end{aligned}  
\end{equation}
where $\Xi^{\beta}_t$ is the time-like Killing vector and $T^{\alpha}_{(\textrm{gw})\beta}$ denotes the Isaacson energy-momentum tensor for gravitational wave, the expression of which is given by 
\begin{equation}\label{energy_flux_gw_tensor}
\begin{aligned}
T_{\mu\nu}^{(\textrm{gw})}=\frac{1}{64\pi}\left\langle h^{\alpha\beta}_{(\textrm{rad});\mu}h_{\alpha\beta;\nu}^{(\textrm{rad})}\right\rangle
\end{aligned}  
\end{equation}
where $h_{\alpha\beta}^{(\textrm{rad})}$ denotes the perturbation tensor in the radiation gauge.
In \autoref{energy_flux_formula}, the sign is so chosen to represent the outflow (inflow) of energy through $\Sigma_{\infty}$ ($\Sigma_{\mathcal{H}}$). Note that,
\begin{equation}\label{dSigma}
\begin{aligned}
d\Sigma_{\alpha}^{\infty,\mathcal{H}}=\sqrt{-g^{(3)}} \scripty{r}_{\alpha}~d\theta d\varphi dt
\end{aligned}  
\end{equation}
where $g^{(3)}=-a r^4\sin^{2}\theta$ is the determinant of the induced metric on $\Sigma_{\infty,\mathcal{H}}$ and $\scripty{r}_{\alpha}=\delta^r_{\alpha}/\sqrt{b}$ is the (outward pointing) unit normal vector to the hypersurface. Here, using \autoref{dSigma} in \autoref{energy_flux_formula}, we obtain the expression of energy flux as \cite{PhysRevD.81.084039}
\begin{equation}\label{final_energy_flux}
\begin{aligned}
\frac{dE_{\infty,\mathcal{H}}}{dt}=\mp \lim_{r_*\to \pm \infty}r^2 \sqrt{a b}\oint T_{rt}^{\textrm{(rad)}}~\sin\theta~d\theta d\varphi
\end{aligned}  
\end{equation}
Note that the spacetime outside $r_c$ is a Schwarzschild spacetime with a redefined time coordinate $t_1 = (1 + \xi \delta f_c) t$ and Arnowitt-Deser-Misner (ADM) mass $\Mbh + \Mh$, by virtue of Birkhoff's theorem, where $\delta f_c = \delta f(r_c)$ is constant \cite{PanossoMacedo:2023qzp}. Thus, we can follow the calculation presented in \cite{Martel:2005ir}, originally formulated for the vacuum Schwarzschild black hole case, to compute the gravitational wave flux. By considering only the leading-order contribution from the dark matter environment, i.e., retaining terms up to $\mathcal{O}(\xi)$, we find that the energy flux carried out to infinity is given by
 \begin{widetext}
\beq\label{flux_formula}
\frac{dE_{\infty}}{dt}=&\frac{1}{64\pi}\sum_{lm}\frac{(l+2)!}{(l-2)!}\left[\omega^2\left|\Psi_Z\tpp\right|^2+\xi\left\{\left(\frac{\delta f_c}{2}\right)\omega^2\left|\Psi_Z\tpp\right|^2+\omega^2\textrm{Re}\left(2\Psi_Z\tpp \overline{\Psi_Z\tpd}\right)\right\}\right]\\
&+\frac{1}{64\pi}\sum_{lm}\frac{(l+2)!}{(l-2)!}\left[4\left|\Psi_R\tpp\right|^2+\xi\left\{\left(\frac{\delta f_c}{2}\right)4\left|\Psi_R\tpp\right|^2+4\textrm{Re}\left(2\Psi_R\tpp \overline{\Psi_R\tpd}\right)\right\}\right]\\
\frac{dE_{\mathcal{H}}}{dt}=&\frac{1}{64\pi}\sum_{lm}\frac{(l+2)!}{(l-2)!}\left[\omega^2\left|\Psi_Z\tpp\right|^2+\xi\left\{\omega^2\textrm{Re}\left(2\Psi_Z\tpp \overline{\Psi_Z\tpd}\right)\right\}\right]\\
&+\frac{1}{64\pi}\sum_{lm}\frac{(l+2)!}{(l-2)!}\left[4\left|\Psi_R\tpp\right|^2+\xi\left\{4\textrm{Re}\left(2\Psi_R\tpp \overline{\Psi_R\tpd}\right)\right\}\right]
\eeq
where ``overline'' represents the complex conjugate of the quantity.
Note that, when we ignore the effect of dark matter by setting $\xi=0$, the flux formula exactly correspond to the vacuum Schwarzschild case \cite{Martel:2005ir}.
 \end{widetext}
\section{Hyperboloid method to solve the perturbation equations}\label{hyperboloidal}

In \autoref{Sec:RWZ_formalism}, we introduce the perturbation equations for an EMRI system in the presence of dark matter. At the order $\mathcal{O}(\epsilon)$, the perturbation equations describe the effects due to the presence of a point particle in a vacuum Schwarzschild spacetime. The contribution from dark matter appears at the order $\mathcal{O}(\xi\epsilon)$. The form of these equations remains relatively simple: the gravitational perturbations are governed by the Regge-Wheeler/Zerilli equations, while the perturbations in matter sector is described in terms of a single master function $\Psi_F\tpd$, which satisfies a Regge-Wheeler-Zerilli type equation \autoref{eq_o_ez_f}. However, a trade-off of this formulation is the emergence of unbounded source terms $S^{U,(1,1)}_{R,Z,F}$ at $\mathcal{O}(\xi\epsilon)$ in addition to the distributional parts $S^{D,(1,1)}_{R,Z,F}$, as shown in  \autoref{stUZ11}, \autoref{stUR11}, and \autoref{stUF11}. The solution of inhomogeneous Regge-Wheeler-Zerilli type equations with a distributional source can be obtained through standard variation of parameter method. If $\hat\Psi^+(r)$ and $\hat\Psi^-(r)$ represents the two homogeneous solutions of the Regge-Wheeler-Zerilli type equations, then the solution of the inhomogeneous equation can be written as $\Psi=C^+ \hat\Psi^+(r)\Theta(r-r_\Omega)+C^- \hat\Psi^-(r)\Theta(r_\Omega-r)$, where $\Theta(r)$ denotes the Heaviside function and the constants $C^{\pm}$ can be obtained through appropriate jump conditions at the particle's location $r_\Omega$ \cite{Hopper:2012ty, Warburton:2013lea}. 
However, unlike the distributional source terms, the unbounded source terms are defined over the entire radial domain. As a result, solving the Regge-Wheeler-Zerilli type equations with unbounded sources using the variation of parameters method poses significant computational challenges. One approach to circumvent this issue, introduced in \cite{Hopper:2012ty}, is the \textit{partial annihilator method}, in which the unbounded source terms are made compact by applying higher-order differential operators \cite{Durkan:2022fvm}. However, for more complicated systems, this method also becomes computationally expensive. An alternative approach is to adopt the \textit{multi-domain spectral method} \cite{Ansorg:2003br, Ansorg:2006gd, Ansorg:2016ztf, PanossoMacedo:2022fdi, Leather:2024mls, PanossoMacedo:2024pox}, which we employ in this paper.
\subsection{Hyperboloidal coordinates}
The starting point of the multi-domain spectral method for solving the frequency-domain equations described in \autoref{Sec:RWZ_formalism} is the introduction of a hyperboloidal coordinate system $(\tau, \sigma, \theta, \phi)$, where the constant $\tau$ hypersurfaces, referred to as hyperboloidal slices, foliate the spacetime, and $\sigma$ is a compactified radial coordinate defined over the interval $[0,1]$ \cite{Zenginoglu:2007jw, Zenginoglu:2011jz, PanossoMacedo:2018hab, Jaramillo:2020tuu, Sarkar:2023rhp, PanossoMacedo:2023qzp}. For an asympotically flat spacetime, these hyperboloidal slices  penetrate  both the future null infinity $\mathscr{I}^+$ and the future event horizon $\mathscr{H}^+$. In this compactified coordinate system, the event horizon is located at $\sigma=1$, whereas the future null infinity corresponds to $\sigma=0$. These hyperboloidal coordinates are related to the original Schwarzschild coordinates $(t,r,\theta,\phi)$ through the relation \cite{Jaramillo:2020tuu, Sarkar:2023rhp, PanossoMacedo:2023qzp}
\beq\label{Hyp_Coordi}
t=\lambda\left(\tau-H(\sigma)\right)\,,\qquad{r}=\lambda \frac{\varrho(\sigma)}{\sigma}
 \eeq
where $\lambda$ is a length scale associated with the problem, which we set $4\Mbh$ following\cite{Jaramillo:2020tuu}. 
The height function $H(\sigma)$ is introduced to ensure that constant $\tau$ hypersurfaces penetrate  both $\mathscr{H}^+$ and $\mathscr{I}^+$ and $\varrho(\sigma)$ is  the areal radius in  the conformal spacetime $d\bar s^2=\sigma^2 ds^2/\lambda^2$. The expression of the height function can be obtained by studying the behavior of the incoming/ outgoing null geodesics around $\mathscr{H}^+$ and $\mathscr{I}^+$. We start with rewriting \autoref{BH_sol} in the ingoing Eddington-Finkelstein (IEF) coordinate \cite{Sarkar:2023rhp, PanossoMacedo:2023qzp} 
\beq\label{IEF}
ds^2 &=-a(r)\lambda^2 dv^2-2\lambda^2 \sqrt{\frac{a(\sigma)}{b(\sigma)}}\frac{\beta(\sigma)}{\sigma^2}dv d\sigma\\&+\lambda^2 \frac{\varrho^2}{\sigma^2}\left[d\theta^{2}+\sin^{2}\theta d\phi^{2}\right]
\eeq
 where $v=(t+r_*)/\lambda$ is the dimensionless advance time, $r_*$ is the tortoise coordinate, defined by $dr_*=dr/\sqrt{ab}$, $a(\sigma)=a(r(\sigma))$ and $\beta=\varrho(\sigma)-\sigma\varrho'(\sigma)$. Here, we have written the line element in terms of the compact radial coordinate $\sigma$. The null vectors associated with the IEF coordinate system are
\beq
k_\mu=\mathfrak{N} \left\{-1,0,0,0\right\}\,,\quad l_\mu=\frac{\lambda^2}{\mathfrak{N}}\left\{-\frac{a}{2},-\frac{\beta_m}{\sigma^2},0,0\right\}
\eeq
where $\beta_m=\beta \sqrt{a/b}$ and $\mathfrak{N}$ is an arbitrary normalization factor. Note that,  the null vectors satisfies the relation $l^\mu l_\mu=k^\mu k_\mu=0$, and  $l^\mu k_\mu=-1$. At this stage, we define $\tau$ as
\beq\label{hyp_time}
\tau\equiv v+H_0(\sigma)
\eeq
 where $H_0(\sigma)$ is an arbitrary function of $\sigma$. We further make a conformal transformation $d\bar s^2=\sigma^2 ds^2/\lambda^2$.  In this new coordinate system $(\tau,\sigma,\theta,\phi)$, the conformal line element takes the form 
 \beq\label{Hyp_LE}
d\bar s^2 &=-a\sigma^2 d\tau^2+2\sigma^2 \left(a H_0'-\frac{\beta_m}{\sigma^2}\right)d\tau d\sigma\\&+\sigma^2 \left(2\frac{\beta_m}{\sigma^2}H_0'-a H_0'^2\right) d\sigma^2+\varrho^2 \left[d\theta^{2}+\sin^{2}\theta d\phi^{2}\right]
\eeq
The associated null vectors are given by the following expression \cite{Sarkar:2023rhp, PanossoMacedo:2023qzp},
\begin{align}\label{nullvectors}
\bar{k}^{\alpha} &=\left(1,\dfrac{1}{H_{0}'},0,0\right)=\delta^{\alpha}_{\tau}+\dfrac{1}{H_{0}'}\delta^{\alpha}_{\sigma}~;
\\
\bar{\ell}^{\alpha} &=\dfrac{H_{0}'}{\beta_m}\left(1-\dfrac{\sigma^{2}a H_{0}'}{2\beta_m}\right)\delta^{\mu}_{\tau}
-\dfrac{\sigma^{2}H_{0}'a}{2\beta_m^{2}}\delta^{\mu}_{\sigma}~.
\end{align}
In the above expression, we fix the normalization factor $\mathfrak{N}$ by imposing the condition $\lim_{\sigma\to0} k^{\alpha} \partial_{\alpha} \tau = 1$, which yields $\mathfrak{N} = \lambda^{2} \beta_m / (\sigma^{2} h_{0}')$. The above condition guarantees that the constant $\tau$ hypersurfaces penetrates the null infinity. Furthermore, to ensure that $\sigma=0$ is indeed a null hypersurface, we set the condition $\lim_{\sigma\to0}\bar{k}^\mu=\delta^\mu_\tau$, which leads to the following condition $\lim_{\sigma\to0}1/H_0'(\sigma)=0$. We have to further ensure that this condition does not jeopardize the regularity of the outgoing null vector $\bar\ell^\mu$. To this end, 
we demand that $\bar\ell^\tau$ remains finite near the null infinity. This leads to the following condition for $H_{0}'(\sigma)$ near $\sigma=0$ \cite{Sarkar:2023rhp, PanossoMacedo:2023qzp}
\begin{align}\label{h0diff}
H_{0}'(\sigma)=\dfrac{2\beta_m(\sigma)}{\sigma^{2}a(\sigma)}+\mathcal{O}(\sigma^{2})~=\dfrac{2\beta(\sigma)}{\sigma^{2}\sqrt{a(\sigma)b(\sigma)}}+\mathcal{O}(\sigma^{2})~.
\end{align}
We expand the function $\beta$ as $\beta=\beta_0+\beta_1 \sigma+\mathcal{O}(\sigma^2)$ around $\sigma=0$, which gives the following expression of the areal radius $\varrho=\beta_0+\rho_1\sigma-\beta_1\sigma\ln{\sigma}$. To avoid logarithmic singularity at $\sigma=0$, we set $\beta_1=0$. Furthermore, expanding the term $\sqrt{ab}$ up to $\mathcal{O}(\xi)$ and integrating the above equation we find \cite{Sarkar:2023rhp, PanossoMacedo:2023qzp}
\beq\label{Hf_0}
H_0(\sigma)&=-2\beta_0\left[\frac{1}{\sigma}-\frac{2\Mbh}{\lambda \beta_0}\ln{\sigma}\right]+\xi A_1(\sigma)+A_2(\sigma)\\
A_1(\s)&=\frac{\delta f \beta_0}{\sigma }+\frac{\ln{\s}}{\lambda}\left[-\lambda \beta_0 \delta f' +2 \delta m-2 \delta f \Mbh \right]
\eeq
where $A_1(\sigma)$ captures the dark matter-induced modification to the height function, and $A_2(\sigma)$ represents the residual gauge freedom in choosing the hyperboloidal slices. Considering the so-called \textit{minimal gauge}, we take $\beta$ to be constant and  set $\rho_1=0$ and $A_2=0$. Furthermore, demanding the event horizon is located at $\s=1$, we fix $\beta_0=\rho_0=2\Mbh/\lambda$. Note that, the term $\delta f'(\s)$ does not contribute to the height function as both $\delta f$ and $\delta m$ constant beyond the cut-off radius $r_c$.\par
From \autoref{Hyp_Coordi} and \autoref{hyp_time}, we find that the full height function can be written as $H(\sigma)=H_0(\sigma)+r_*(\sigma)/\lambda$. Note that, we can write the the tortoise coordinate in terms of its singular $r_{*,\textrm{sing}}$ and regular $r_{*,\textrm{reg}}$ pieces,.i.e.,\cite{PanossoMacedo:2023qzp}
\beq\label{tortoise}
r_{*,\textrm{sing}}(\s)&= 2\Mbh\left[\ln{(1-\sigma)}-\ln{\sigma}+\frac{1}{\s}\right]\\&+\frac{\xi \left(-\delta f-\delta m ~ \sigma  \ln{\s}+\delta f \Mbh \sigma  \ln{\s}\right)}{\sigma }\\
r_{*,\textrm{reg}}'(\s) &=-\frac{2\Mbh}{\s^2\sqrt{ab}}-r_{*,\textrm{sing}}'(\s)
\eeq
where we have considered terms up to $\mathcal{O}(\xi)$. Denoting the correction in the tortoise coordinate due the presence of dark matter by $A_3(\s)$, we can write the full expression of the height function as 
\beq
H(\sigma)&=\frac{\Mbh}{\lambda}\left[\ln{\sigma}+\ln{(1-\sigma)}-\frac{1}{\sigma}\right]+\xi A_1(\sigma)+\xi A_3(\sigma)\\
&=H_S(\sigma)+\xi A(\sigma)
\eeq
where $A(\sigma) = A_1(\sigma) + A_3(\sigma)$. Note that the inclusion of the term $A(\sigma)$ is necessary when solving the frequency-domain equations using hyperboloidal methods, as it helps suppress unphysical divergent contributions to the master function at $\mathscr{I}^+$. The reason behind this can be understood through the following argument: outside the cutoff radius $r_c$, the spacetime is still described by the Schwarzschild geometry, with the ADM mass given by $\Mbh + \Mh$. Moreover, the presence of the redshift factor introduces a redefinition of time as $t_1 = {(1+\delta f / 2)} t$. Therefore, to ensure that the $\tau = \text{const.}$ hypersurfaces continue to intersect future null infinity $\mathscr{I}^+$ at $\Od$, the inclusion of the term $A(\sigma)$ is essential.
\begin{figure*}[t!]
	\centering
	\minipage{0.48\textwidth}
	\includegraphics[width=\linewidth]{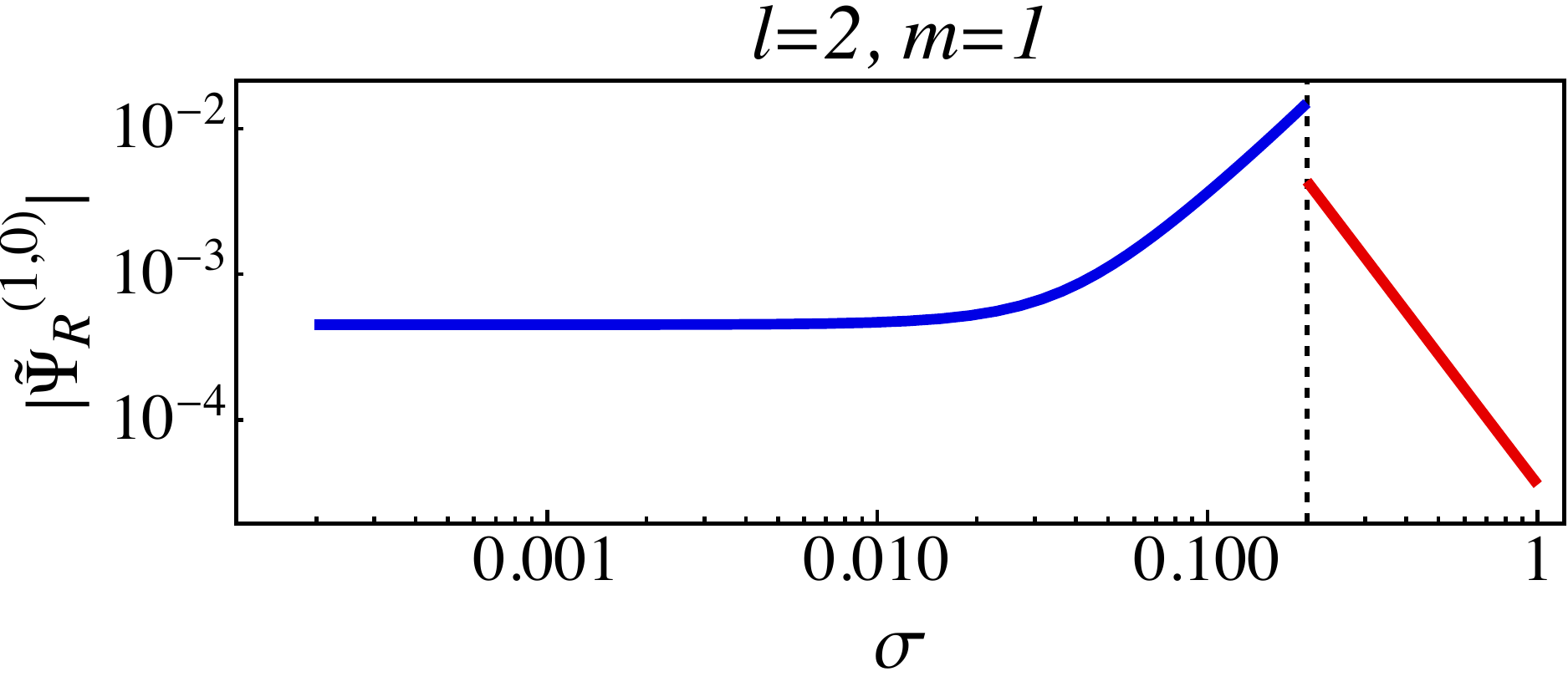}
	\endminipage\hfill
	\minipage{0.48\textwidth}
	\includegraphics[width=\linewidth]{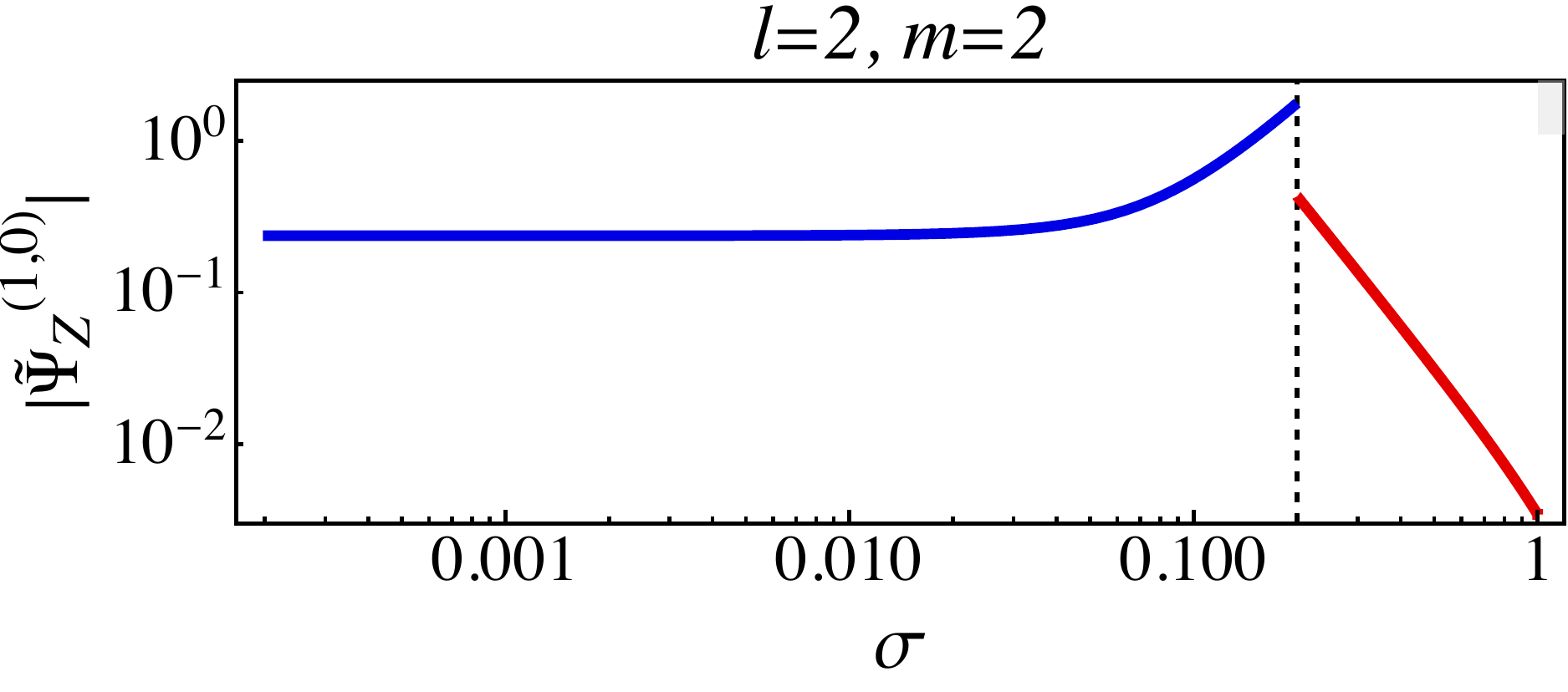}
	\endminipage
	\caption{The plots for the rescaled axial $\tilde{\Psi}_R\tpp$ (for $l=2,~m=1$ mode) and polar $\tilde{\Psi}_Z\tpp$ (for $l=2,~m=2$ mode) master functions as functions of the compactified radial coordinate. The particle is assumed to be located at $\sigma = 0.2$ (corresponding to $r_\Omega = 10\Mbh$), indicated by the black dashed line. Due to the distributional nature of the source term, both master functions exhibit a discontinuity at the particle’s location.
 }\label{fig:master_G_Oe}
\end{figure*}
\subsection{Frequency domain equations in hyperboloidal formalism}

In this section, we rewrite the frequency domain equations introduced in \autoref{Sec:RWZ_formalism} in the hyperboloidal formalism. We adopt the minimal gauge formalism discussed in the previous section to construct the hyperboloidal coordinates. Let us first consider the perturbation at $\Op$. At this order,  we can write the perturbation equations in the following generic form
\beq\label{CS_10}
\mathcal{L}\Psi\tpp &\equiv \left[\partial^2_r{*}+\omega^2-V(r)\right]\Psi\tpp=S\tpp~,
\eeq
where the potential $V=(V_R,V_Z)$ vanishes at both infinity and at horizon $r_*\to\pm\infty$.
As discussed above, the source term $S\tpp$ is distributional.
Given that, the time domain waveform has the following form $\Psi(t,r)=\exp{(-i\omega t)}\Psi(r)$, we can make the following rescaling for the master function using \autoref{Hyp_Coordi} \cite{PanossoMacedo:2023qzp,PanossoMacedo:2022fdi, Leather:2024mls, PanossoMacedo:2024pox}
\beq\label{rescale}
\Psi\tpp(r)=\mathscr{Z}(\sigma) \tilde \Psi\tpp(\sigma)\,,\quad \mathscr{Z}=e^{\zeta H}\,,\quad \zeta=-i\omega \lambda
\eeq
Hereafter, we use ``\textit{tilde}'' to represent the rescaled quantities.  Under this transformation, the rescaled master function $\tilde \Psi\tpp$ satisfies the following equation 
\beq\label{HS_10}
\tilde{\mathcal{L}} \tilde{\Psi}\tpp=\left[\alpha_2\frac{d^2}{d\sigma^2}+\alpha_1\frac{d}{d\sigma}+\alpha_0\right]\tilde{\Psi}\tpp=\tilde{ S}\tpp
\eeq
where the rescaled operator $\tilde{\mathcal{L}}$ and the 
source term are related to the original frequency domain operators and the source terms through a rescaling factor in the following relation
\beq \label{HyED10}
\tilde{\mathcal{L}} \tilde{\Psi}\tpp &=\frac{1}{\mathscr{F}}{\mathcal{L}} {\Psi}\tpp\,,\\ \tilde{ S}\tpp &=\frac{1}{\mathscr{F}}{ S}\tpp=\tilde G^{(1,0)}\delta_\s+\tilde F^{(1,0)}\delta'_\s
\eeq
where $\delta_\s\equiv\delta(\s-\s_\Omega)$ is the Dirac delta function, $\delta'_\s$ denotes derivative of the derivative of the Dirac delta function with respect to $\s$ and  $\s_\Omega=2M/r_\Omega$. The rescaling factor is given by 
\beq
\mathscr{F}=\frac{\mathscr{Z}(1-\s)}{r(\s)^2}~.
\eeq
In \autoref{HyED10}, we use the distributional properties of Dirac delta function to obtain the expressions for $\tilde G\tpp$, and  $\tilde F\tpp$ (see \ref{app: Dirac delta}).
The coefficients $\alpha_0(\sigma)$, $\alpha_1(\sigma)$, and $\alpha_2(\sigma)$ of the rescaled operator $\tilde{\mathcal{L}}$ are functions of the compactified coordinate $\sigma$. The explicit forms of these coefficients are as follows \cite{PanossoMacedo:2023qzp,PanossoMacedo:2022fdi, Leather:2024mls, PanossoMacedo:2024pox} \footnote{Note that there are typos in the expressions for $\alpha_1$ and $\alpha_0$ in \cite{PanossoMacedo:2022fdi, Leather:2024mls}.}
\beq
\alpha_2&=\sigma ^2-\sigma ^3\,,\\
\alpha_1 & = \zeta  \left(1-2 \sigma ^2\right)+\sigma  (2-3 \sigma ),\\
\alpha_0&=-\zeta ^2 (\sigma+1)-2 \zeta  \sigma -V_l(\sigma )
\eeq
where $V=(1-\s) V_l/r(\sigma)^2$.
 Furthermore, the rescaling \autoref{rescale}, give rise to the following equation at $\Od$ for the metric perturbation master functions introduced in \autoref{eq_o_ez}
\beq\label{hypFDEq}
\mathcal{\tilde L}\tilde\Psi\tpd &=\tilde S\tpd\\
\tilde S\tpd &=\frac{1}{\mathcal{F}_f}\bigg[S\tpd-\zeta A S\tpp \\&-\zeta A_{,\s} \left(\alpha_1 \Psi\tpp+2\alpha_0 \Psi\tpp_{,\s}\right)-\zeta \alpha_0 A_{,\s\s} \Psi\tpp\bigg]
\eeq
As discussed before, $S\tpd$ contains both distributional and bounded contributions. Thus, in the hyperboloidal framework, the distributional part of the source term can be written as 
\beq \label{HyED11}
\tilde S^{D(1,1)} &=\frac{1}{\mathcal{F}_f}\left[S^{D(1,1)}-\zeta A S\tpp \right]
\\
&=\tilde G^{(1,1)}\delta_\s+\tilde F^{(1,1)}\delta'_\s+\tilde H^{(1,1)}\delta''_\s
\eeq
 Again, we use the distributional properties of Dirac delta function to obtain the expressions for $\tilde G\tpd$, $\tilde F\tpd$ and $\tilde H\tpd$.
The unbounded part of the source term is given by 
\beq
\tilde S^{U,(1,1)}& =\frac{1}{\mathcal{F}_f}\bigg[S^{U,(1,1)}-\zeta A_{,\s} \left(\alpha_1 \Psi\tpp+2\alpha_0 \Psi\tpp_{,\s}\right)\\&-\zeta \alpha_0 A_{,\s\s} \Psi\tpp\bigg]
\eeq
The distributional source term in \autoref{HyED10} and \autoref{HyED11} leads to discontinuity (jumps) in the master functions $\tilde \Psi\tpp$ and $\tilde \Psi\tpd$ and their derivatives at $\s_{\Omega}$. These jumps uniquely determines the master function. The solution to \autoref{HyED10} and \autoref{hypFDEq} can be written in a generic fashion as $\tilde\Psi(\s)=\tilde\Psi^+(\s)\Theta(\s-\s_\Omega)+\tilde\Psi^-(\s)\Theta(\s_\Omega-\s)+\tilde\Psi^\delta(\s)\delta(\s-\s_\Omega)$, where $\tilde\Psi^+(\s)=C^+ \hat{\tilde\Psi}^+(\s)$ and $\tilde\Psi^-(\s)=C^- \hat{\tilde\Psi}^-(\s)$, with $\hat{\tilde\Psi}^+(\s)$ ($\hat{\tilde\Psi}^-(\s)$) are homogeneous solution of the equation  \autoref{HyED10} and \autoref{HyED11} in the domain $\s\in [\s_\Omega,1]$ ($\s\in [0,\s_\Omega]$) and $C^\pm$ are two constants that is determined by the following jump conditions
\beq\label{jump}
&\tilde\Psi^\delta \bigg|_{\s_\Omega}=\mathcal{J}^\delta\bigg{|}_{\s_\Omega}=\frac{\tilde H}{\alpha_2}\bigg|_{\s_\Omega}~,\\
&\tilde\Psi^{+}-\tilde\Psi^{-}\bigg|_{\s_\Omega}=\mathcal{J}^0\bigg{|}_{\s_\Omega}\\
&=\left[\frac{\tilde F}{\alpha_2}-\frac{\alpha_1}{\alpha_2}\mathcal{J}^\delta-2\mathcal{J}^\delta_{,\s}\right]_{\s_\Omega}~,\\
&\tilde\Psi^{+}_{,\s}-\tilde\Psi^{-}_{,\s}\bigg|_{\s_\Omega}=\mathcal{J}^1\bigg{|}_{\s_\Omega}\\
&=\left[\frac{\tilde G}{\alpha_2}-\frac{\alpha_1}{\alpha_2}\mathcal{J}^0-\mathcal{J}^0_{,\s}-\mathcal{J}^\delta_{,\s\s}-\frac{\alpha_1}{\alpha_2}\mathcal{J}^\delta-\frac{\alpha_0}{\alpha_2}\mathcal{J}^\delta\right]_{\s_\Omega}~.
\eeq
For perturbation equation with unbounded source term, it is more convenient to consider following rescaling of the master function $\tilde\Psi^{U,(1,1)}=\tilde\psi^{U,(1,1)}/\s$ \cite{Leather:2024mls}. With this equation for $\tilde\psi\tpd$ becomes
\beq\label{unbounded_sources}
\mathcal{\tilde D}\tilde\psi^{U,(1,1)}&=\left[\gamma_2\frac{d^2}{d\sigma^2}+\gamma_1\frac{d}{d\sigma}+\gamma_0\right]\tilde\psi^{U,(1,1)}=\mathcal{\tilde{S}}^{U,(1,1)}\\
\mathcal{\tilde{S}}^{U,(1,1)} &=\s^2\tilde{S}^{U,(1,1)}
\eeq
The coefficients of the operator $\mathcal{\tilde D}$ is given by 
\beq
\gamma_2&=\sigma ^3-\sigma ^4\,,\quad
\gamma_1 =\zeta  \left(\sigma -2 \sigma ^3\right)-\sigma ^3,\\
\gamma_0&=\zeta ^2 (-\sigma ) (\sigma +1)-\zeta +\sigma  (\sigma -V_l(\sigma ))
\eeq
Note that, at infinity $\s=0$, the coefficients $\gamma_2$ and $\gamma_1$ vanishes. Thus, at the infinity $\tilde\psi^{U,(1,1)}=\mathcal{\tilde{S}}^{U,(1,1)}/\gamma_0$.
However, for the gravitational perturbation case, the source term $\mathcal{\tilde{S}}^{U,(1,1)}$ also vanishes at infinity. Thus, the rescaled master function can be obtained through L'Hopital rule
\beq
\lim_{\s\to 0}\tilde\Psi^{U,(1,1)}=\lim_{\s\to 0}\frac{\tilde\psi^{U,(1,1)}}{\s}=\tilde\psi^{U,(1,1)}_{,\s}
\eeq
\begin{figure*}[t!]
	\centering
	\minipage{0.48\textwidth}
	\includegraphics[width=\linewidth]{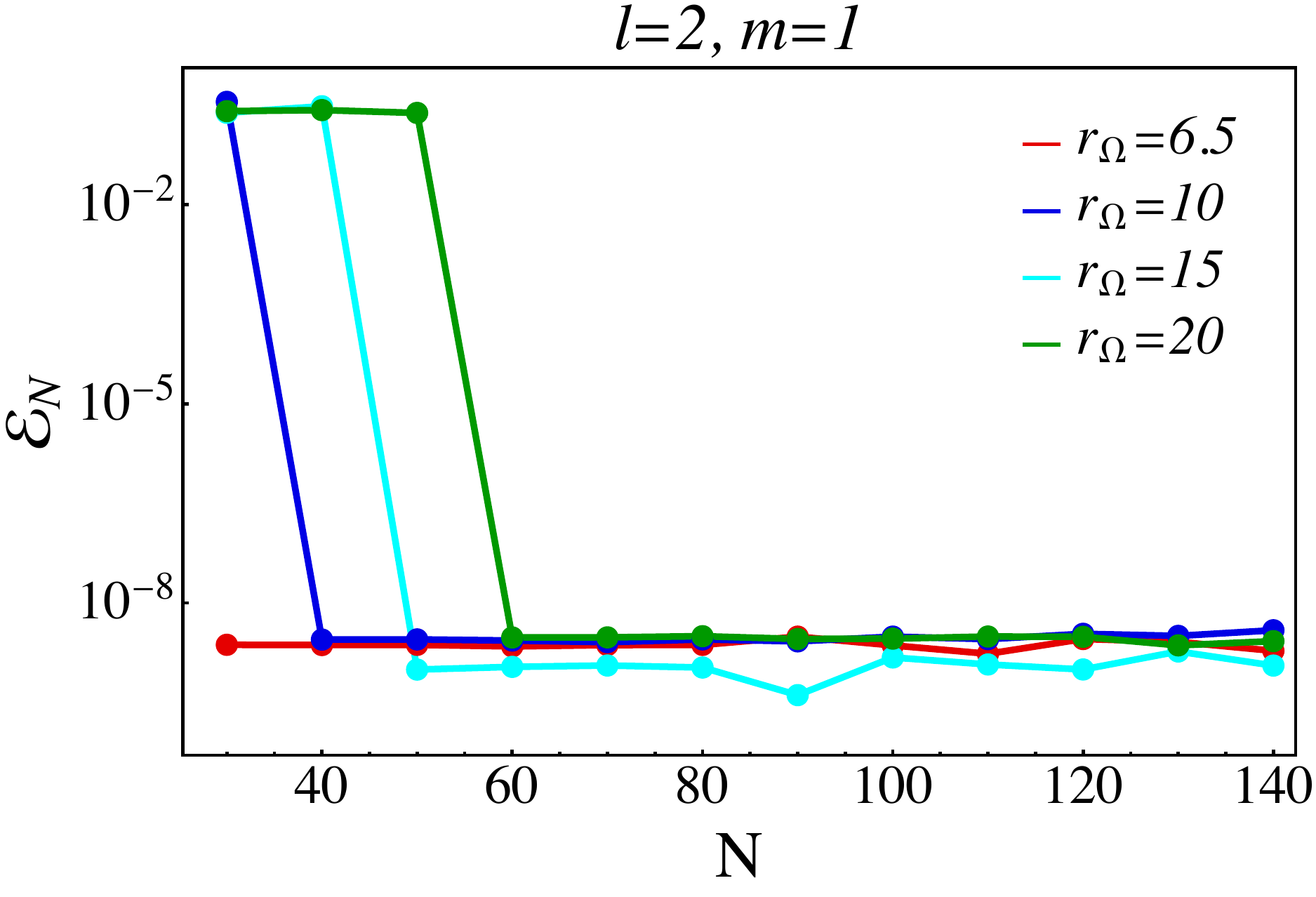}
	\endminipage\hfill
	\minipage{0.48\textwidth}
	\includegraphics[width=\linewidth]{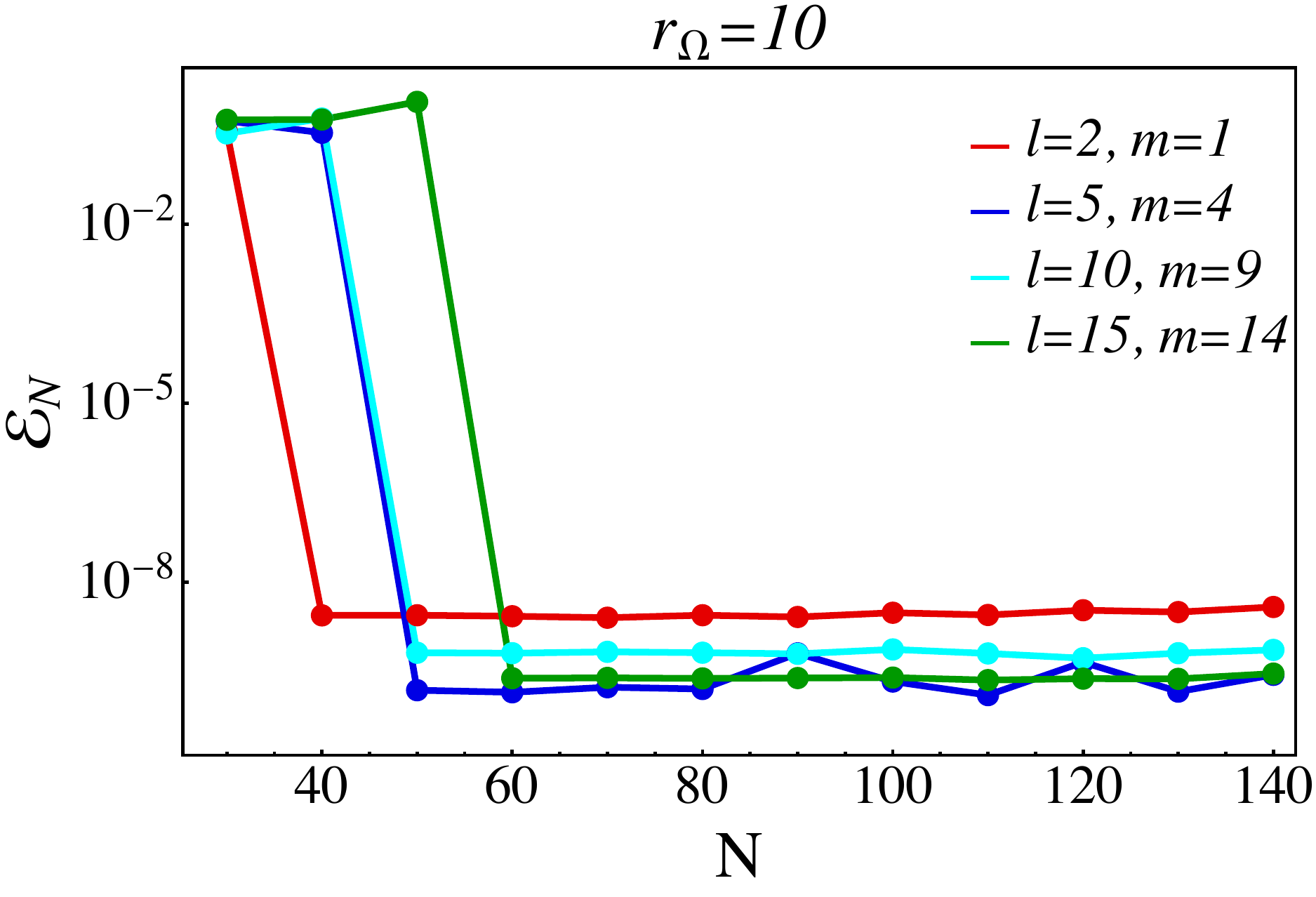}
	\endminipage
	\caption{The numerical convergence of the axial master function $\Psi_R\tpp$ is shown as a function of numerical resolution $N$. A high-resolution reference solution $\Psi_R^{(1,0),\textrm{ref}}$ computed with $N^{\textrm{ref}}=150$ is used to assess convergence. The left panel displays the convergence behavior for various values of $r_\Omega$ at fixed angular mode $l=2,~m=1$, while the right panel shows the convergence at fixed $r_\Omega = 10\Mbh$ across different angular modes. In both cases, we observe rapid numerical convergence. However, the convergence rate slows for larger $r_\Omega$ and higher $l$ modes.  This decline is due to the steepening gradient of the master function in these regimes, requiring finer resolution (i.e., more grid points) to maintain accuracy.
 }\label{fig: error}
\end{figure*}
\begin{widetext}
For the perturbation in matter sector, it is more convenient to consider the following transformation 
\beq
\Psi_F\tpd &=\frac{\mathscr{Z}_F(\sigma)}{\s} \tilde \Psi_F\tpd(\sigma)~,\\
\mathscr{Z}_F(\sigma)&=\mathscr{Z}(\sigma)e^{\eta_1/\s} (\sigma -1)^{\eta_2}~,
\eeq
where
\beq
\eta_1 &=\frac{(c_r-1) \zeta }{2 c_r }~,\\
\eta_2 &={-\frac{\sqrt{c_r^8+c_r^6 \left(4 \zeta ^2-2\right)+c_r^4}}{4 c_r^4}-\frac{\zeta }{2}}~.
\eeq
With this transformation, the equation for $\tilde\psi\tpd$ becomes
\beq\label{unbounded_sources_1}
\mathcal{\tilde D}_F\tilde \Psi_F\tpd&=\left[\beta_2\frac{d^2}{d\sigma^2}+\beta_1\frac{d}{d\sigma}+\beta_0\right]\tilde \Psi_F\tpd={\tilde{S}}_F^{(1,1)}\\
{\tilde{S}}_F^{(1,1)} &=\frac{\s^2}{\mathscr{Z}_F}S_F^{(1,1)}
\eeq
The coeffectient of the operator $\mathcal{\tilde D}_F$ is given by 
\beq
\beta_2&=\sigma ^3-\sigma ^4\,,\quad
\beta_1 =\sigma ^3 \left(\frac{\sqrt{c_r^8+c_r^6 \left(4 \zeta ^2-2\right)+c_r^4}}{2 c_r^4}-1\right)-\frac{\zeta  (\sigma -1) \sigma  (c_r \sigma +1)}{c_r},\\
\beta_0&=\frac{\zeta ^2 \left(c_r^2 (\sigma -1) \sigma +2 c_r (\sigma -1)+\sigma ^2+\sigma +2\right)}{4 c_r^2}+\frac{\sigma  \left(3 c_r^4 \sigma -c_r^2 \left(c_t^2 \left(2 l^2+2 l+5 \sigma -2\right)+\sigma \right)+c_t^2 \left(2 c_t^2 (\sigma -1)+\sigma \right)\right)}{2 c_r^4}\\&+\frac{\zeta  \left(-2 c_r^5 \sigma +2 c_r^4 (\sigma -2)+c_r \sigma ^2 \sqrt{c_r^8+c_r^6 \left(4 \zeta ^2-2\right)+c_r^4}+\sigma  \sqrt{c_r^8+c_r^6 \left(4 \zeta ^2-2\right)+c_r^4}\right)}{4 c_r^5}
\eeq
\end{widetext}
\subsection{Multi-domain spectral method}
\begin{table*}[ht!]
\centering
\def\arraystretch{1.0}      	
	\setlength{\tabcolsep}{1.3em}
\begin{tabular}{|c|c|c|c|c|}
\hline \hline
$r_\Omega$ & $\mathcal{F}\tpp_{\textrm{tot}}$ & $\mathcal{F}\tpd_{\textrm{Hern}}$ & $\mathcal{F}\tpd_{\textrm{NFW}}$ & $\Delta \mathcal{F}\tpp=\left|\frac{\mathcal{F}\tpp_{\textrm{tot}}-\mathcal{F}\tpp_{\textrm{BHPT}}}{\mathcal{F}\tpp_{\textrm{tot}}}\right|$ \\
\hline\hline
6  & 0.000470166     & $-2.73152\times10^{-5}$  & $-2.51363\times10^{-5}$  & $4.97836\times10^{-8}$ \\
7  & 0.000200081     & $-1.15485\times10^{-5}$  & $-1.06072\times10^{-5}$  & $4.08734\times10^{-10}$ \\
8  & 9.80522$\times10^{-5}$ & $-5.2709\times10^{-6}$   & $-4.83578\times10^{-6}$  & $5.70747\times10^{-10}$ \\
9  & 5.29666$\times10^{-5}$ & $-2.63661\times10^{-6}$  & $-2.41632\times10^{-6}$  & $6.82132\times10^{-11}$ \\
10 & 3.07582$\times10^{-5}$ & $-1.38422\times10^{-6}$  & $-1.26712\times10^{-6}$  & $7.3182\times10^{-10}$ \\
12 & 1.21459$\times10^{-5}$ & $-4.44748\times10^{-7}$  & $-4.06689\times10^{-7}$  & $3.69465\times10^{-10}$ \\
15 & 3.94451$\times10^{-6}$ & $-1.07252\times10^{-7}$  & $-9.77412\times10^{-8}$  & $7.44564\times10^{-10}$ \\
20 & 9.35735$\times10^{-7}$ & $-1.7566\times10^{-8}$   & $-1.59797\times10^{-8}$  & $1.58\times10^{-9}$ \\
\hline\hline
\end{tabular}
\caption{The total gravitational wave flux for different values of orbital radius $r_{\Omega}$ is presented. Here, $\mathcal{F}\tpp_{\textrm{tot}}$ represents the leading-order contribution to the flux, corresponding to the flux from a point particle orbiting a Schwarzschild black hole in a circular orbit. $\mathcal{F}\tpd_{\textrm{Hern}}$ and $\mathcal{F}\tpd_{\textrm{NFW}}$ represent the modifications to the gravitational wave flux (to linear order in $\xi$) due to the presence of dark matter for the Hernquist and NFW profiles, respectively. The parameters for the dark matter profiles are provided in \autoref{tab:Fitting_Parameters}. We compare our results with those obtained from the \texttt{Black Hole Perturbation Toolkit} \cite{BHPT}, where $\Delta \mathcal{F}\tpp=\left|\frac{\mathcal{F}\tpp_{\textrm{tot}}-\mathcal{F}\tpp_{\textrm{BHPT}}}{\mathcal{F}\tpp_{\textrm{tot}}}\right|$ denotes the relative error between the two results.}
\label{tab:flux0_com}
\end{table*}
In the previous section, we presented the perturbation equations introduced in \autoref{Sec:RWZ_formalism} within the hyperboloidal framework. In this section, we outline the details of the numerical approach, so-called the \textit{multi-domain spectral method}, used to solve these differential equations. Our analysis closely  follows that of \cite{PanossoMacedo:2022fdi, Leather:2024mls}, who implemented this method to calculate scalar and gravitational self-force. \par
We begin by dividing the compact interval $[0,1]$ into two subdomains, $\sigma \in \mathbf{D}_1 \cup \mathbf{D}_2$, where $\mathbf{D}_1 \equiv [\sigma_0, \sigma_\Omega]$ and $\mathbf{D}_2 \equiv [\sigma_\Omega, \sigma_1]$. The values of the endpoints $\sigma_0$ and $\sigma_1$ depend on the type of perturbation equation under consideration: $\{\sigma_0, \sigma_1\} = \{0,1\}$ for gravitational perturbations, and $\{\sigma_0,   \sigma_1\} = \{1/2, 2M/r_c\}$ for perturbations in the matter sector. Within each subdomain, the radial coordinate $\s\in[\s_i,\s_f]$ is mapped to a new coordinate $x\in[-1,1]$ through the following relation
\beq
\s &=\frac{1}{2}\left[\s_f(1+x)+\s_i(1-x)\right]\\
x &=\frac{2\s-(\s_i+\s_f)}{\s_f-\s_i}
\eeq
where $\s_i$ and $\s_f$ are the end points of an arbitrary subdomain $[\s_i,\s_f]$. With this remapping of the radial coordinate, we can discretize the linear differential operators introduced in the previous section through Chebyshev spectral method \cite{trefethenMATLAB10.5555/357801}. In each subdomain, we can approximate an arbitrary function $\varphi(x)$ through a truncated Chebyshev series $\varphi_N(x)=\sum_{i=0}^{N} c_i T_i(x)$, where $T_i$ is the Chebyshev polynomial of first kind, $c_i$ are the Chebyshev coefficients and $N$ is the order of expansion. To fix the Chebyshev coefficients, we discretize the compact interval $x\in[-1,1]$ in terms of Chebyshev-Lobatto grid points $\{x_j\}$, where $x_j$'s are given by
\beq
x_j=\cos{\left(\frac{j\pi}{N}\right)}\,,\qquad{j}\in \{0,1,...,N\}
\eeq
and demand that the approximate function $\varphi_N(x)$ coincide with the exact function $\varphi(x)$ at the grid points, i.e., $\varphi_N(x_j)=\varphi(x_j)$. However, it is important to note that we are considering a situation where the explicit form of $\varphi(x)$ is unknown; instead, we know only the differential equation $\mathscr{L}\varphi=\mathcal{S}$ that $\varphi(x)$ satisfies. Thus, we need to discretize the aforementioned differential equation. This can be done through the introduction of Chebyshev-Lobatto differential matrix $\mathbb{D}$ which gives the functional derivative of the function on the discrete grid points $\varphi'(x_i)=\sum_{j=0}^N \mathbb{D}_{i,j} \varphi_j$,
where $\varphi_j=\varphi(x_j)$. Second order derivatives can be obtained by applying this matrix again on $\varphi'(x_i)$. Higher order derivatives can be obtained using the same method. With the aid of Chebyshev-Lobatto differential matrix, we can discretize the differential operator $\mathscr{L}$ as $\mathbb{L}$ which satisfies the algebraic relation 
\beq
\mathbb{L}\cdot \vec{\varphi}=\vec{\mathbb{S}}~,
\eeq
where $\vec{\varphi}=\{\varphi(x_0),...,\varphi(x_N)\}$ and $\vec{\mathbb{S}}=\{\mathcal{S}(x_0),...,\mathcal{S}(x_N)\}$ are $N+1$ dimensional vectors. We can find the functional value of $\varphi(x)$ at each grid points by solving this algebraic equation.\par

Note that, in our context, the term $\vec{\mathbb{S}}$ is non-vanishing only for the unbounded sources (see \autoref{unbounded_sources} and \autoref{unbounded_sources_1}) within a specific subdomain. This equation can be solved directly using LU decomposition, which is implemented in \texttt{Mathematica} via the \texttt{LinearSolve} function \cite{Mathematica}.
\par
\begin{figure*}[htb!]
	\centering
	\minipage{0.48\textwidth}
	\includegraphics[width=\linewidth]{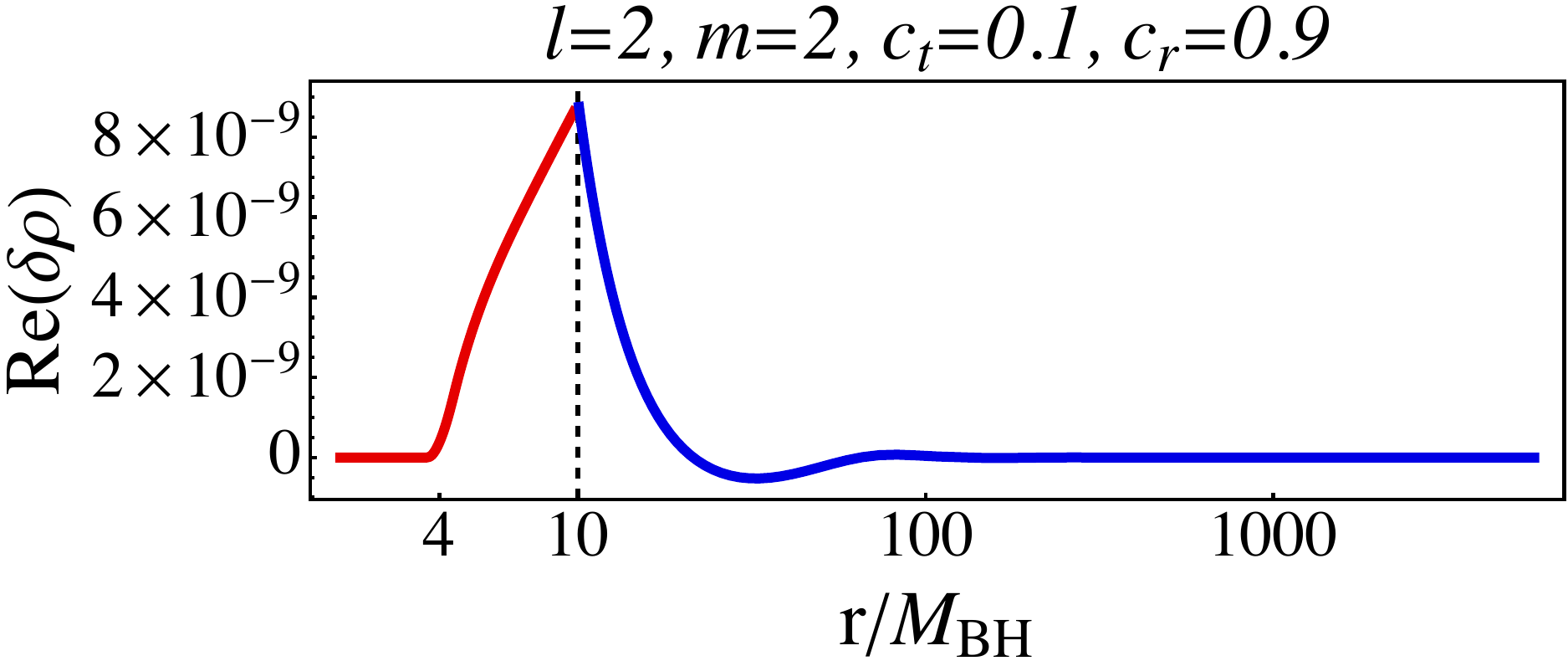}
	\endminipage\hfill
	\minipage{0.48\textwidth}
	\includegraphics[width=\linewidth]{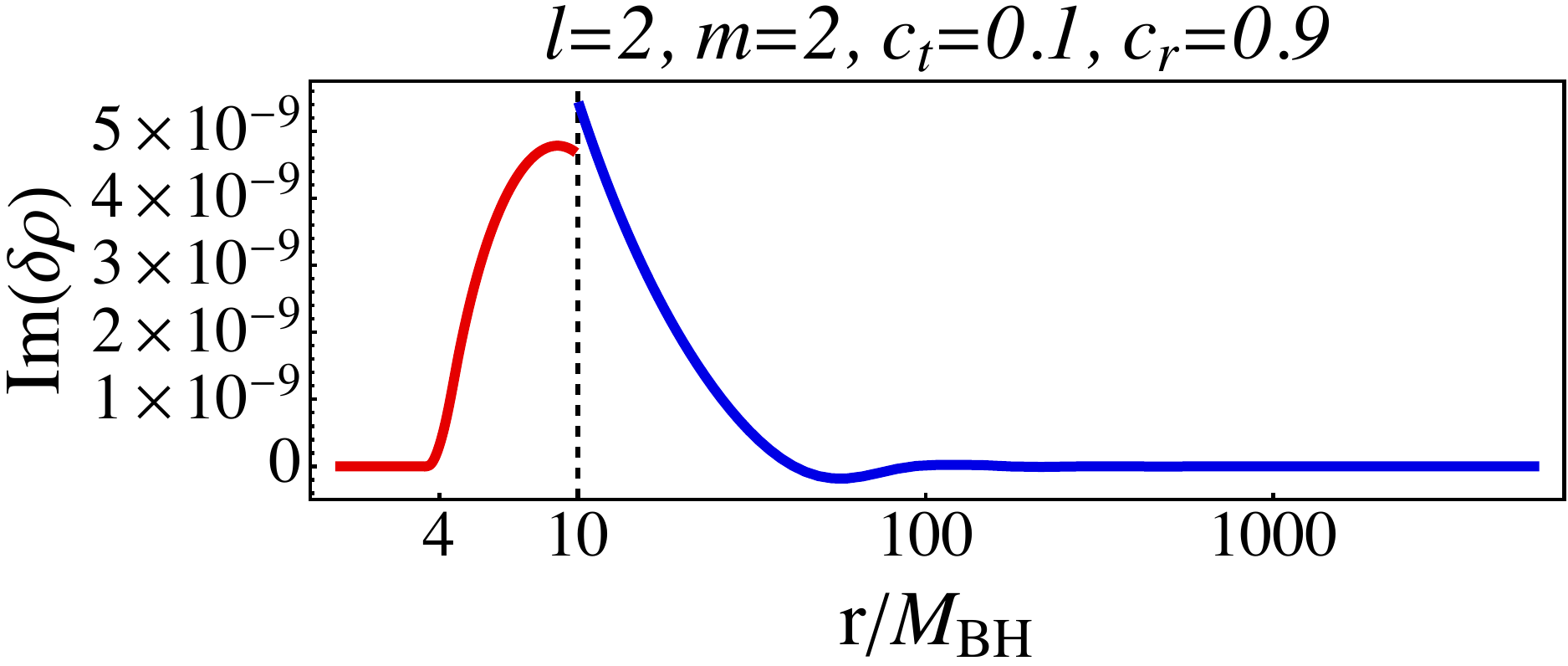}
	\endminipage
	\caption{The plot of real (left panel)  and the imaginary part (right panel) of density fluctuation corresponding to $(l=2,~m=2) $ mode as a function of $r$ is presented for Hernquist density profile. The parameters values are given in \autoref{tab:Fitting_Parameters}. The radial and tangential sound speed is taken to be $c_r=0.9$ and $c_t=0.1$, respectively. As before, we consider the secondary is located at $r_{\Omega}=10\Mbh$, represented by the black dashed line.  
 }\label{fig_density}
\end{figure*}
\begin{figure*}[htb!]
	\centering
	\minipage{0.48\textwidth}
	\includegraphics[width=\linewidth]{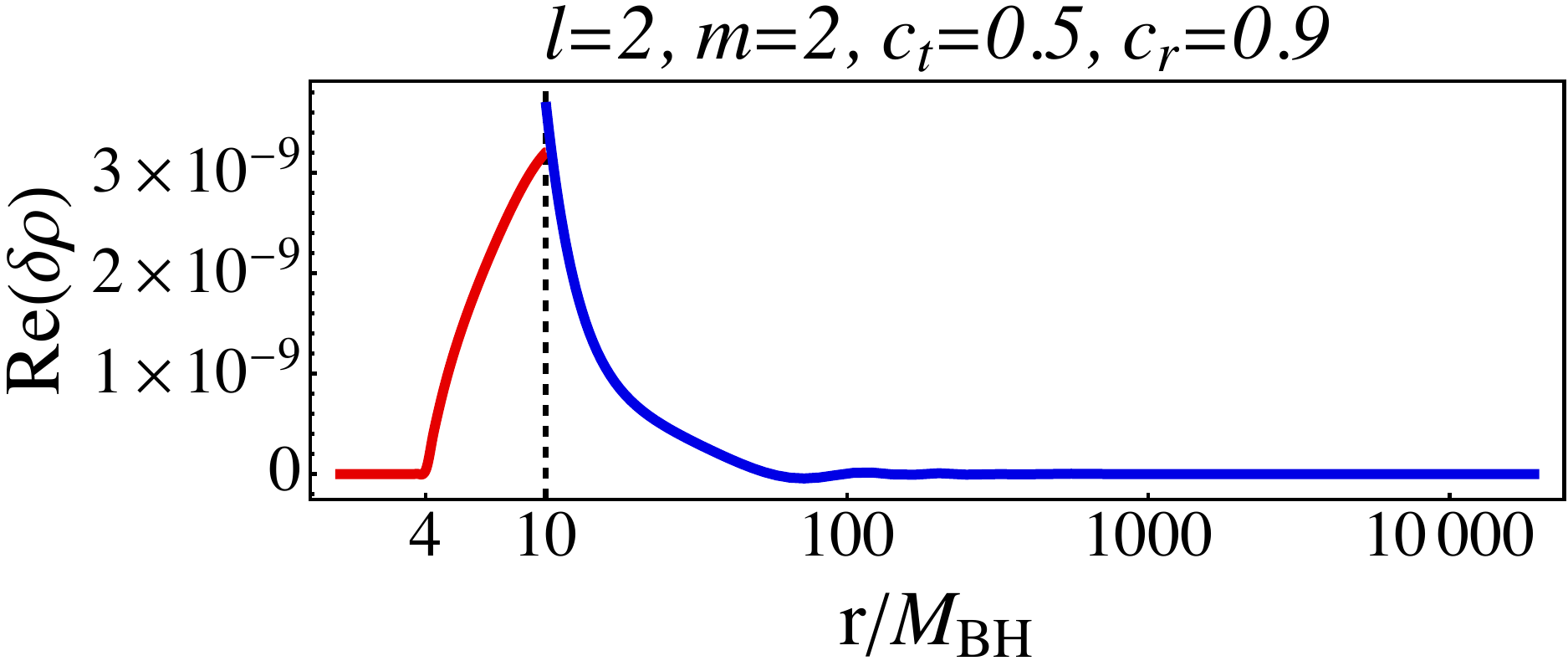}
	\endminipage\hfill
	\minipage{0.48\textwidth}
	\includegraphics[width=\linewidth]{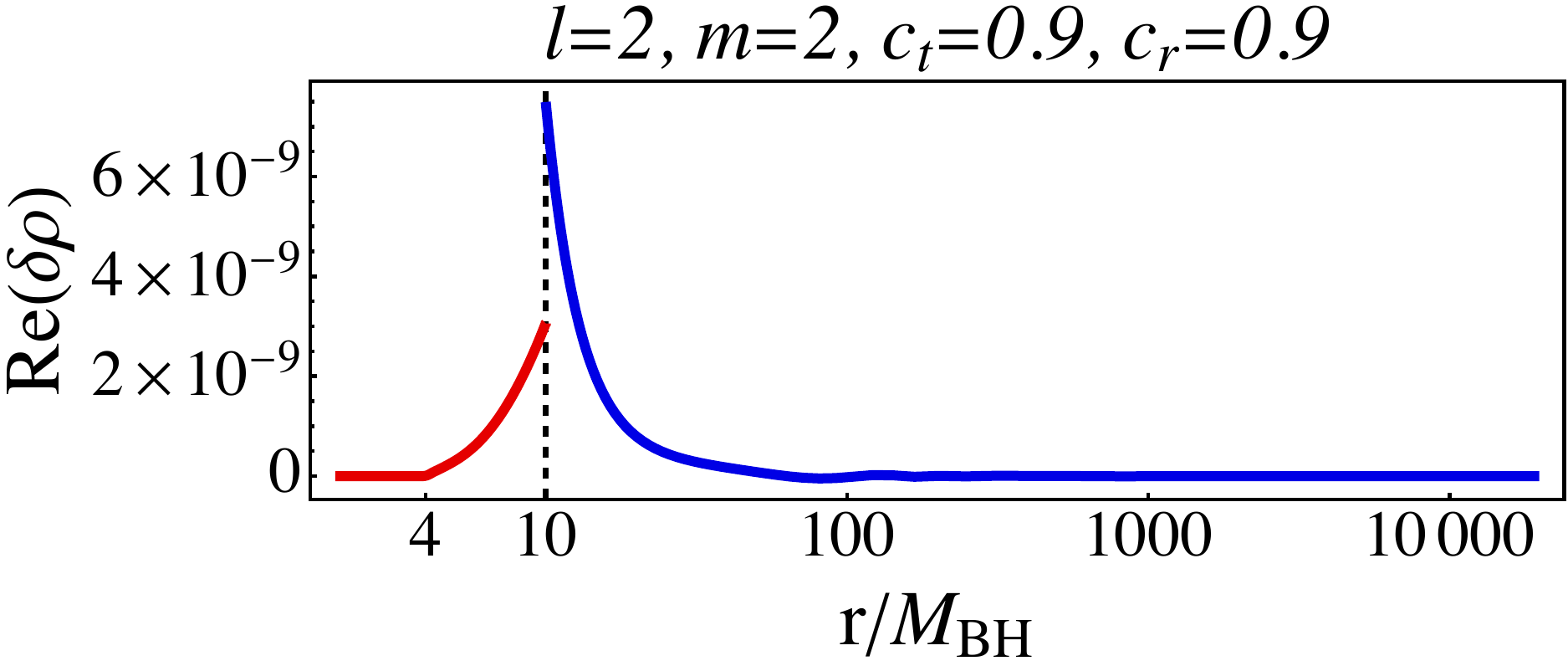}
	\endminipage
	\caption{The plot of real part of density fluctuation corresponding to $(l=2,~m=2) $ mode as a function of $r$ is presented for different sound speed $(c_r,~c_t)=(0.9,~0.5)$ (left panel) and $(c_r,~c_t)=(0.9,~0.9)$ (right panel). The density profile is taken to be 
    Hernquist density profile with  the parameters values are given in \autoref{tab:Fitting_Parameters}. As before, we consider the secondary is located at $r_{\Omega}=10\Mbh$, represented by the black dashed line.  
 }\label{fig_density_ct}
\end{figure*}
However, to solve equations with distributional sources (where the source term vanishes everywhere except at isolated points), we employ the singular value decomposition (SVD) method and impose the jump conditions given in \autoref{jump} at $\s_\Omega$ to obtain the solution of the inhomogeneous differential equation.\par
We have cross-checked our results using other numerical techniques, namely QR decomposition and LU decomposition. Within the machine precision of \texttt{Mathematica}, we find that the SVD method yields more accurate results, although it is marginally slower compared to LU and QR decomposition methods.\par

The main advantage of spectral method is that the approximate function $\varphi_N(x)$ converges very rapidly to the function $\varphi(x)$ as $N$ increases, given that $\varphi(x)$ is a smooth enough function. In particular, if the function is analytic, i.e., it locally admits a convergent power series, then the convergence is exponential. To get an estimate on the numerical error, we follow \cite{PanossoMacedo:2022fdi} and consider a reference solution $\varphi^{\textrm{ref}}(x)$ computed with high numerical resolution $N^{\textrm{ref}}$. With this, we can compute the numerical error using the relation 
\beq\label{error_numerical}
\mathcal{E}_N=\left|1-\frac{\varphi_N(x)}{\varphi^{\textrm{ref}} (x)}\right|\,,\quad N<N^{\textrm{ref}}
\eeq
In this paper, we set $N^{\textrm{ref}}=150$ following \cite{PanossoMacedo:2022fdi}. We expect that, for an analytical function $\varphi(x)$, the numerical error varies as $\mathcal{E}_N\sim \mathcal{C}^{-N}$, where $\mathcal{C}$ is a constant. \par
However, it is important to note that exponential convergence does not automatically implies highly accurate solution. Thus, in order confirm the accuracy of our result, we compare quantities like gravitational wave flux obtained using spectral method with other methods. Particularly, we compare our results with those obtained using \texttt{Black Hole Perturbation Toolkit} package \cite{BHPT} and calculate the relative error in the flux results.
\section{Results}\label{sec: results}
Having presented the governing equations necessary to study the influence of dark matter on gravitational waveforms, we now outline the main findings of our analysis. We begin by examining the modification in the gravitational wave flux due to the presence of dark matter and then discuss how this modification affects the gravitational waveform itself.
\begin{figure*}[t!]
	\centering
	\minipage{0.48\textwidth}
	\includegraphics[width=\linewidth]{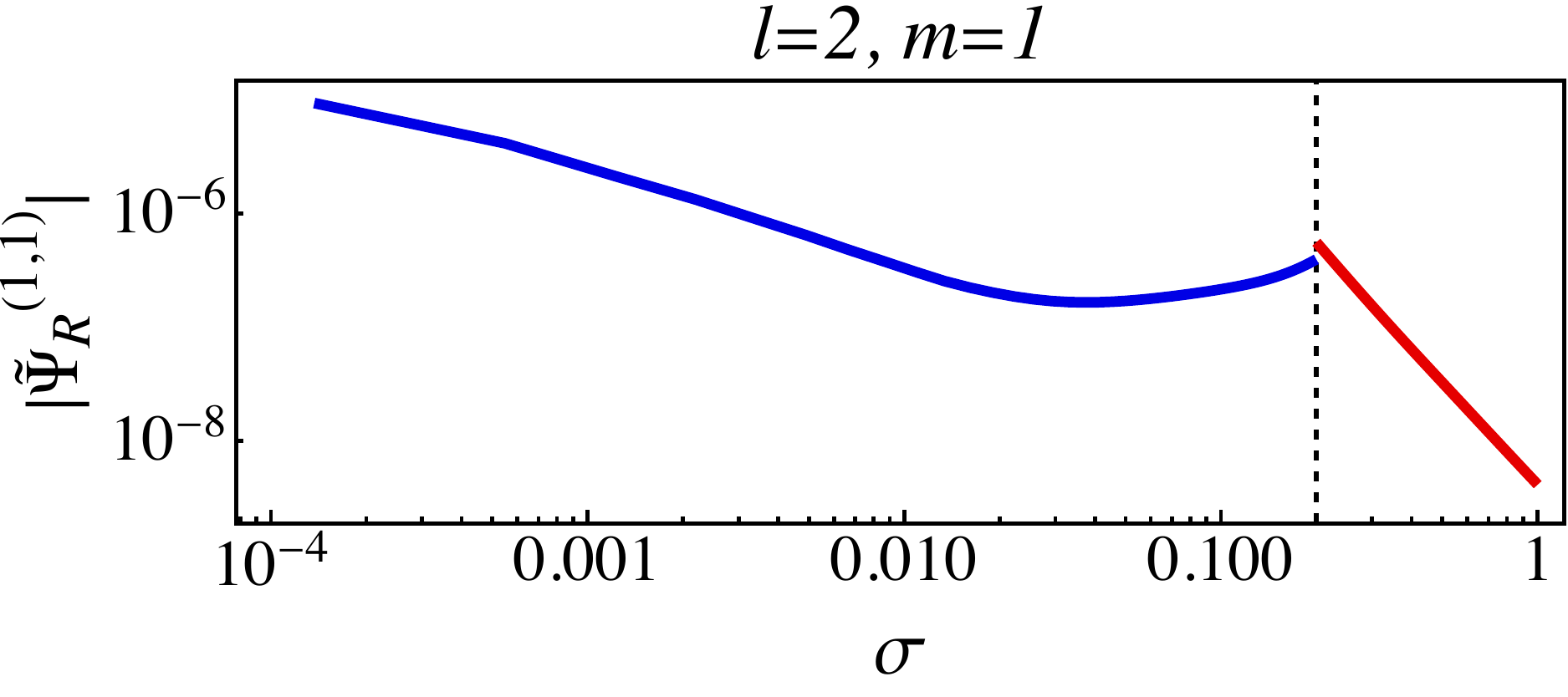}
	\endminipage\hfill
	\minipage{0.48\textwidth}
	\includegraphics[width=\linewidth]{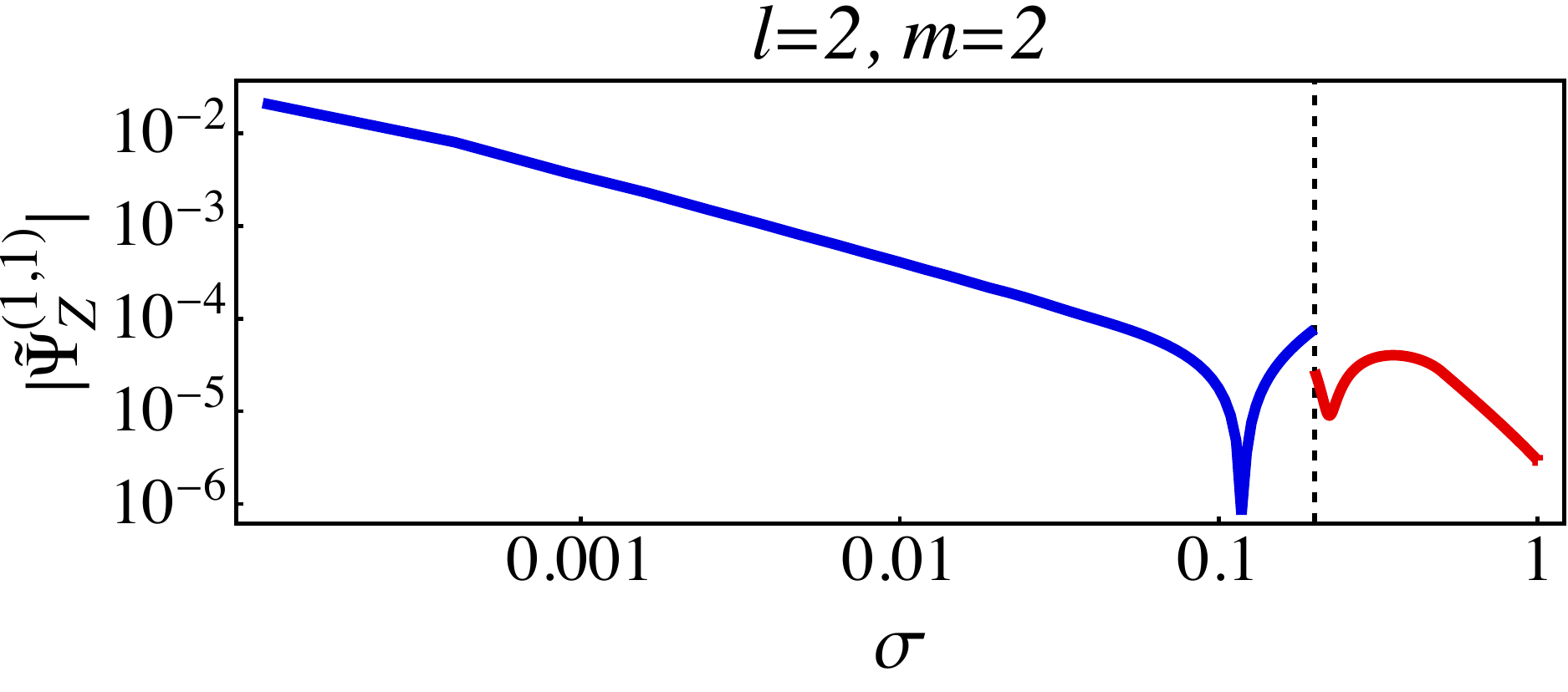}
	\endminipage
	\caption{The plots for the rescaled axial $\tilde{\Psi}_R\tpd$ (for $l=2,~m=1$ mode) and polar $\tilde{\Psi}_Z\tpd$ (for $l=2,~m=2$ mode) master functions at $\Od$ as functions of the compactified radial coordinate. The dark matter density profile is assumed to be Hernquist density profile with sound speed $(c_r,c_t)=(0.9,~0.1)$. 
 }\label{fig_grav_pert_ez}
\end{figure*}

\begin{figure*}[t!]
	\centering
	\minipage{0.48\textwidth}
	\includegraphics[width=\linewidth]{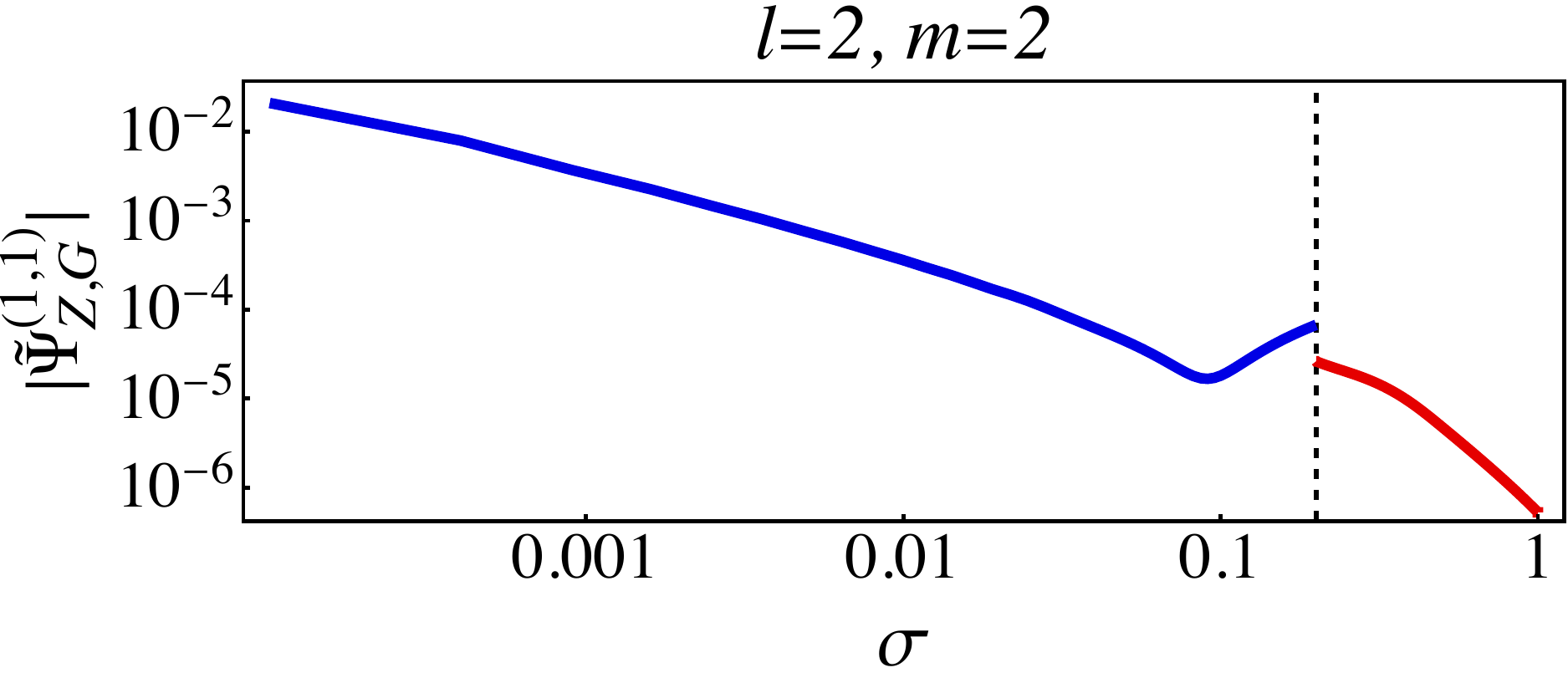}
	\endminipage\hfill
	\minipage{0.48\textwidth}
	\includegraphics[width=\linewidth]{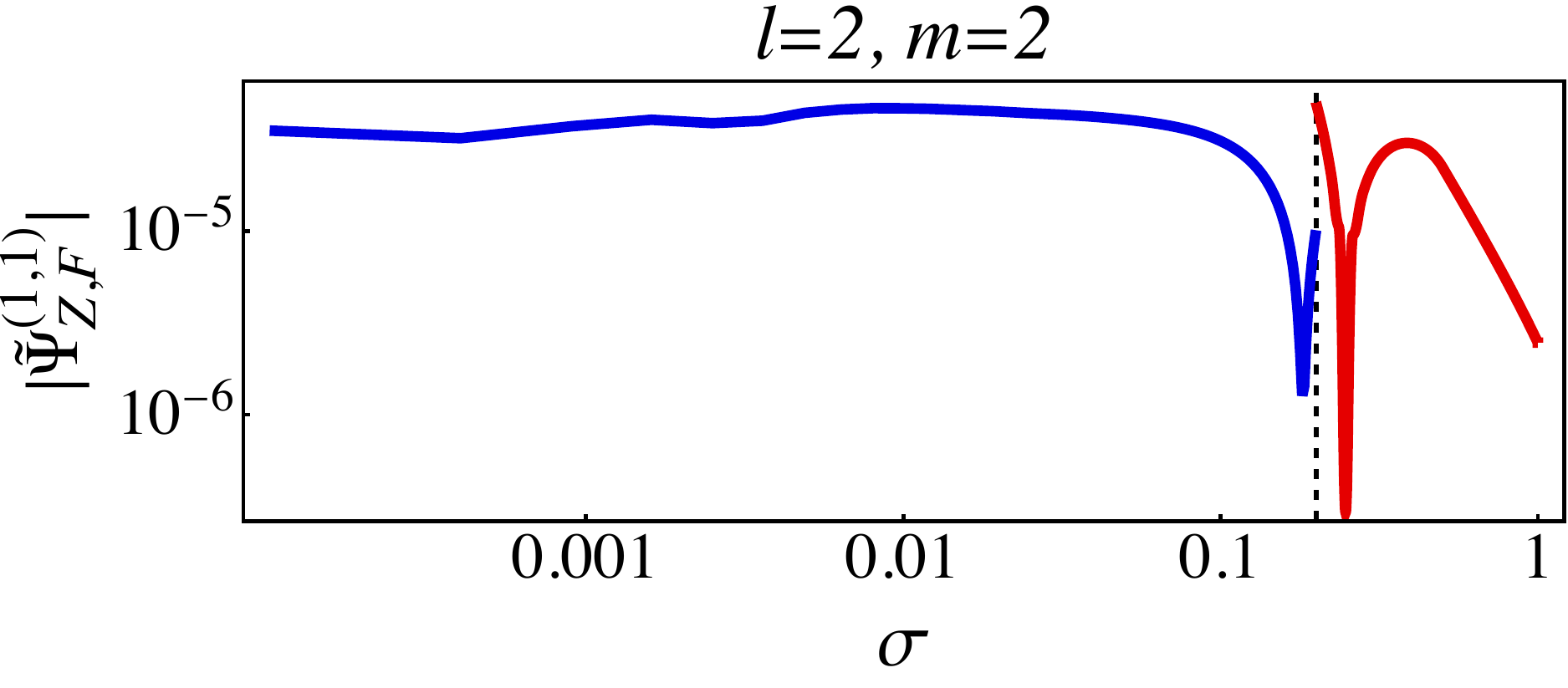}
	\endminipage
	\caption{
    The plot of $\tilde{\Psi}_{Z,G}\tpd$ (left panel) and $\tilde{\Psi}_{Z,F}\tpd$ (right panel) as functions of the compactified radial coordinate is presented for the $l=2,~m=2$ mode, considering the Hernquist density profile with sound speed $(c_r,c_t)=(0.9,~0.1)$. As described in the main text, at $\Od$, the gravitational perturbation arises from two sources: one due to the modification of the background spacetime by the presence of dark matter, represented by $\tilde{\Psi}_{Z,G}\tpd$, and the other due to perturbations from fluid fluctuations, represented by $\tilde{\Psi}_{Z,F}\tpd$.
 }\label{fig_grav_fluid_ez}
\end{figure*}
Before proceeding further, it is important to recall that we have assumed the dark matter perturbation parameter to scale as $\xi \sim \mathcal{O}(\epsilon)$. At order $\mathcal{O}(\xi)$, the force term $F^{\mu}$ induces a shift in the orbital parameters. At orders $\mathcal{O}(\epsilon)$, this force contains both conservative and dissipative components. The calculation of these components is well-established in the literature. Furthermore, for waveform modeling at first post-adiabatic (1PA) order, only the dissipative part of the self-force at $\mathcal{O}(\epsilon^2)$ is required. Similarly, for our purposes, only the dissipative contributions are relevant at $\Od$ for 1PA waveform constructions. In this work, we focus exclusively on the adiabatic contributions; that is, the dissipative components at $\mathcal{O}(\epsilon)$ and include the corresponding post-adiabatic effects due to dark matter.
\subsection{Gravitational wave flux}
First, we consider the perturbation at $\Op$. At this order, the perturbation equations are governed by \autoref{HS_10} with distributional source term given in \autoref{HyED10}. To solve the equation, we split the interval $[0,1]$ into two domain $\mathbf{D}_1\equiv [0,\s_\Omega]$ and $\mathbf{D}_2\equiv [\s_\Omega, 1]$.  In our implementation of multi-domain spectral method, each subdomain is discretized using $60+1$ grid points, and the precision is set to the machine precision of the \texttt{Mathematica} software.\par
In \autoref{fig:master_G_Oe}, we show the absolute value of the axial   $\tilde{\Psi}_R\tpp$ and polar $\tilde{\Psi}_Z\tpp$ as a function of the compactified radial coordinate $\s$. The plot shows the result for the angular mode $(l,m)=(2,1)$ and $(l,m)=(2,2)$, respectively. 
We consider the particle is at $r_\Omega=10$ ($\s_\Omega=0.2$), represented by the black dashed line in the figure. As can be seen, the master function (along with its derivative; not shown in the plot) is discontinuous at the location of the particle due to the distributional nature of the source term. \par
To see the convergence of the code, we calculate the relative error $\mathcal{E}_N$ (see \autoref{error_numerical}) for $\tilde\Psi_R(\sigma)$ at the location $\s=\s_\Omega-\varepsilon$ considering a reference function computed with $N=150$, where $\varepsilon\approx 10^{-10}$. In  \autoref{fig: error}, we present the result. As can be seen, the convergence is quite rapid even with moderate values of numerical resolution $N$. However, we notice that, to obtain solution within a given precision for larger values of angular number and $r_\Omega$, we need to have more numerical resolution. This is due to fact that, for larger values $l$ and $r_\Omega$, the gradient of the function is steep around the particle. Thus, larger number of grid points is required to achieve the numerical convergence.  In \cite{PanossoMacedo:2022fdi}, the authors solved this problem using analytical mesh refinement method. However, in this paper, we consider the evolution of the system in the domain $r_\Omega\in [20\Mbh, \textrm{ISCO}]$, where we compute the flux value by summing over all modes up to $l_{\textrm{max}}=10$. Here, ISCO denotes the position of innermost stable circular orbit, which is located at $r=6\Mbh$ for Schwarzschild black hole. Notice that, in this range, we obtain very accurate solutions even with moderate values of $N\sim 60$.
\begin{figure*}[t!]
	\centering
	\minipage{0.48\textwidth}
	\includegraphics[width=\linewidth]{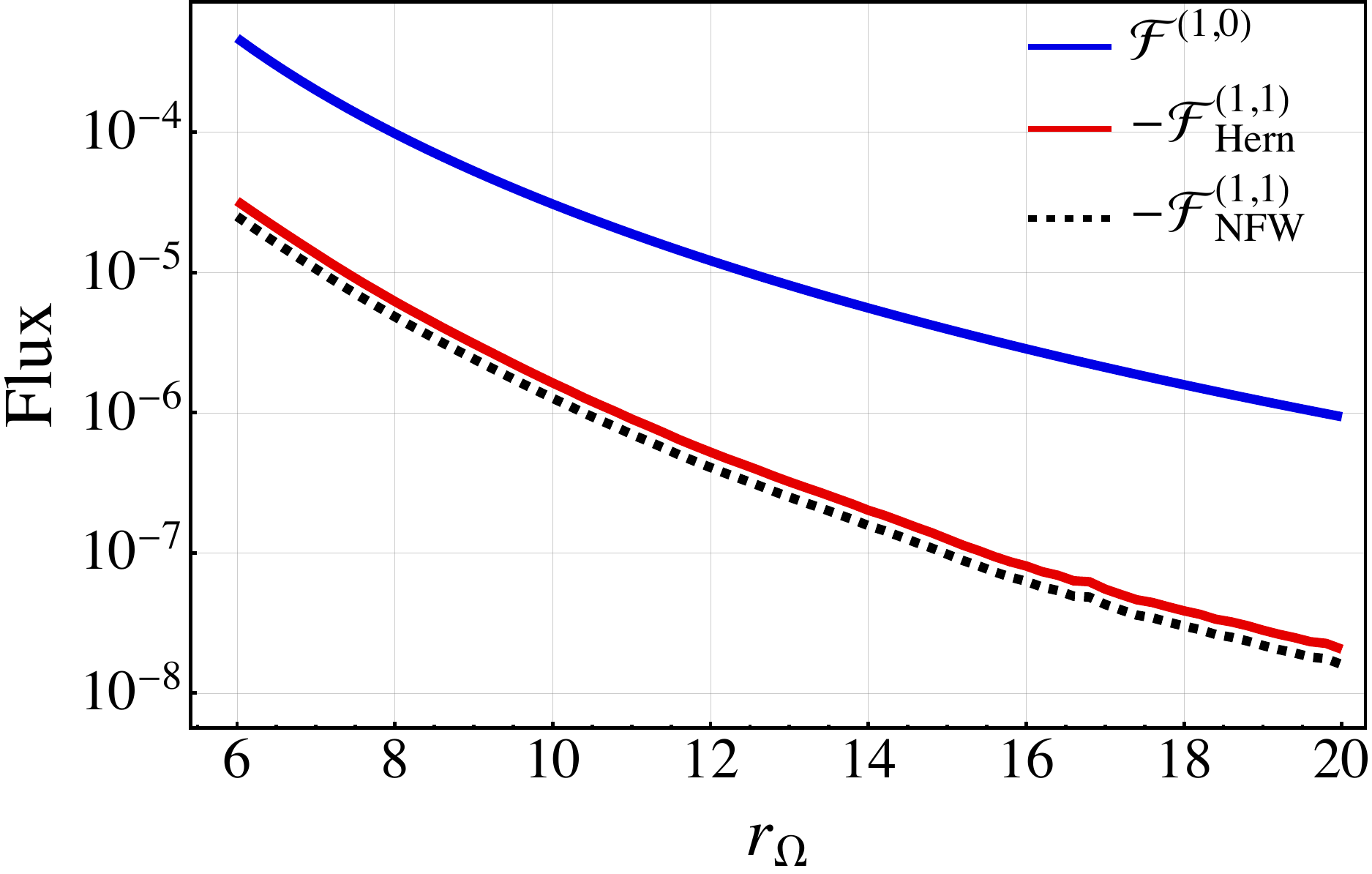}
	\endminipage\hfill
    \minipage{0.48\textwidth}
	\includegraphics[width=\linewidth]{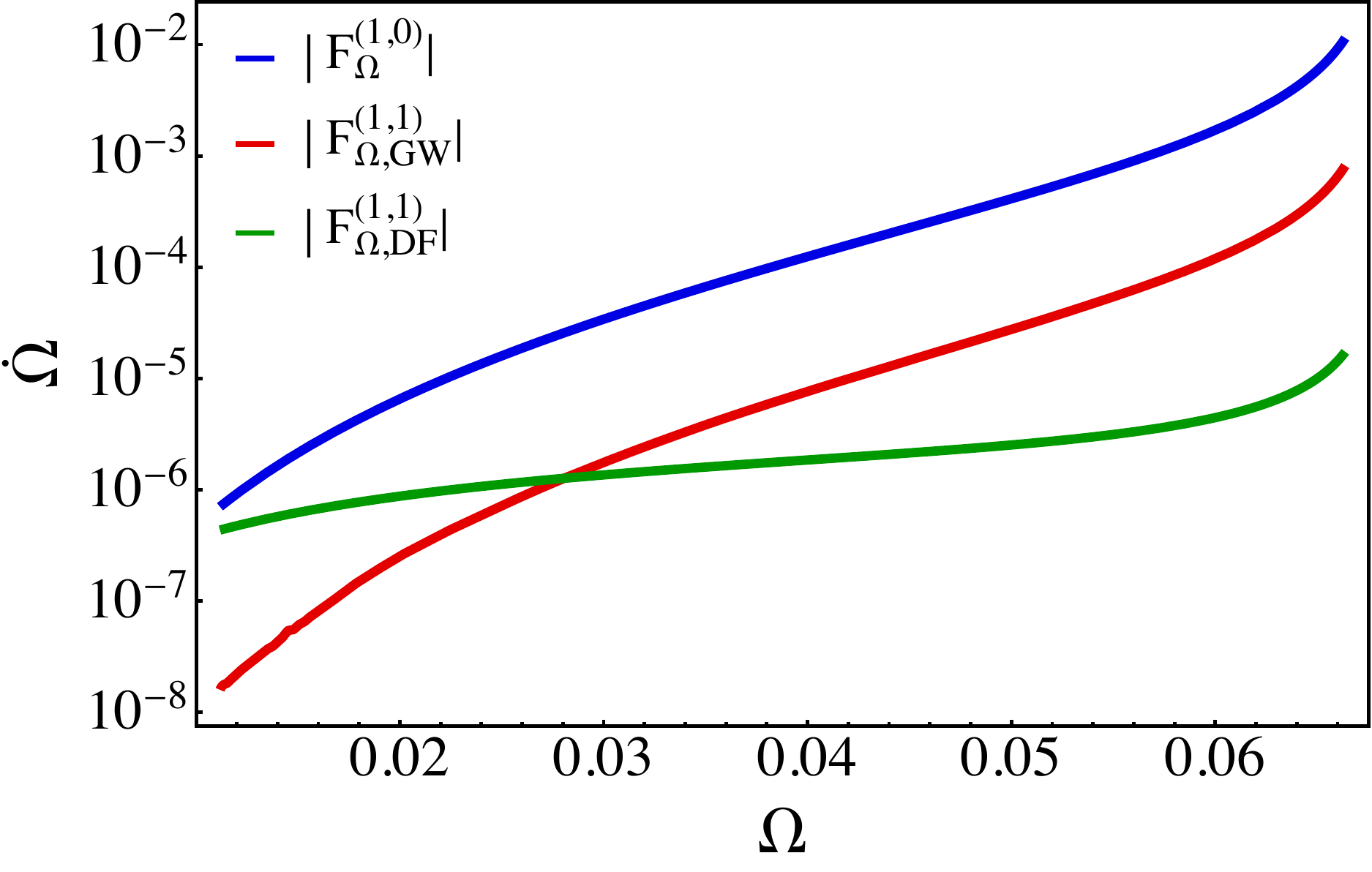}
	\endminipage
	\caption{Left panel: The plot of gravitational wave flux as a function of orbital radius $r_\Omega$. Here, the leading-order gravitational wave flux $\mathcal{F}\tpp_{\textrm{tot}}$ corresponds to the inspiral of a point particle around a Schwarzschild black hole, shown as a solid blue line. The modification in gravitational wave flux due to the presence of dark matter is plotted for two different dark matter density profiles: the Hernquist profile $\mathcal{F}\tpd_{\textrm{Hern}}$, represented by a solid red line, and the NFW profile $\mathcal{F}\tpd_{\textrm{NFW}}$, represented by a black dashed line. Right panel:
The contributions of different dissipative effects in shifting the orbital frequency are plotted as a function of orbital frequency. In the strong gravity regime, the leading contribution to the inspiral rate is attributed to $\F\tpp$, due to the leading-order gravitational wave flux $\mathcal{F}\tpp_{\textrm{tot}}$, shown in the blue curve. The modification in the inspiral rate, $\mathsf{F}_{\Omega,\textrm{GW}}\tpd$, arising from the dark matter-induced modification of gravitational wave flux, is shown as a solid red curve. The green curve shows the contribution from dynamical friction, $\mathsf{F}_{\Omega,\textrm{DF}}\tpd$. Here, we consider the Hernquist density profile. As can be seen, near the ISCO ($\Omega = 0.068$), the contribution from the modification of gravitational wave flux outweighs the contribution from dynamical friction.
 }\label{fig_flux}
\end{figure*}
Thus, we have not utilized the analytical mesh refinement method to obtain the results.\par
The gravitational wave flux to both future null infinity and the black hole horizon is computed using the expression in \autoref{flux_formula}. As the flux involves a sum over angular modes, we truncate the series at $l_{\textrm{max}} = 10$ for each orbital frequency $\Omega$ (equivalently, each $r_\Omega$), since the contribution from higher $l$ modes is negligible. For each $l$, the azimuthal index $m$ runs from $1$ to $l$.
\par
At leading order $\mathcal{O}(\epsilon)$, the flux corresponds to that of a point particle inspiraling into a Schwarzschild black hole. The computed results are shown in \autoref{tab:flux0_com}, where $\mathcal{F}\tpp_{\textrm{tot}}$ denotes the total gravitational wave flux. To validate the accuracy and robustness of our numerical implementation, we compare our results with those obtained using the \texttt{Black Hole Perturbation Toolkit}\cite{BHPT}. The relative errors, also presented in \autoref{tab:flux0_com}, are found to be of the order $\mathcal{O}(10^{-10})$, demonstrating the high accuracy of our method. Nevertheless, we observe a slight degradation in accuracy near the ISCO and for larger values of $r_\Omega$. As shown in \autoref{fig: error}, the reduced accuracy at large $r_\Omega$ is due to the limitations of the multi-domain spectral method implemented in this work, which becomes less effective as the function becomes more sharply peaked. Conversely, near the ISCO, the decreased precision arises from limitations of the \texttt{Black Hole Perturbation Toolkit} \cite{BHPT}, which employs the Mano–Suzuki–Takasugi (MST) method to compute the gravitational wave flux. This method is known to lose accuracy at high gravitational wave frequencies, $\omega = m\Omega$. This explain the relative discrepancies between the results obtained via the multi-domain spectral method and those from the \texttt{Black Hole Perturbation Toolkit} near the ISCO.
\par
The corrections due to the presence of dark matter appears at $\Od$. We first compute the matter perturbation equation \autoref{eq_o_ez_f}, as the density perturbation acts a source for gravitational perturbation for polar modes at $\Od$. As discussed in \autoref{Sec:RWZ_formalism}, we set Dirichlet boundary conditions $\delta \rho=0$ at $\s_0=2\Mbh/r_c$ and  $\s_1=1/2$. Thus, we split the interval $[\s_0,\s_1]$ into two domain $\mathbf{D}_1\equiv [\s_0,\s_\Omega]$ and $\mathbf{D}_2\equiv [\s_\Omega, \s_1]$ and solve the perturbation equation using multi-domain spectral method.

In \autoref{fig_density}, we plot real and imaginary part of density fluctuation as function of $r$ for $l=2,~m=2$. Here, we consider a Hernquiest type dark matter spike profile (see \autoref{tab:Fitting_Parameters}) with total halo mass $\Mh=10^4\Mbh$, $\Mh/a_0=0.001$, and  $r_c=100\Mh a_0/\Mbh$. Following \cite{PhysRevD.67.104017}, we set the radial and tangential sound speed as $c_r=0.9$ and $c_t=0.1$. 
The effect of sound speed on the density perturbation is plotted in \autoref{fig_density_ct}. As can be seen from the figure, the shape of the density fluctuation and the discontinuity at the location of the secondary object greatly depends on the sound speed.\par
\begin{figure*}[t!]
	\centering
	\minipage{0.48\textwidth}
	\includegraphics[width=\linewidth]{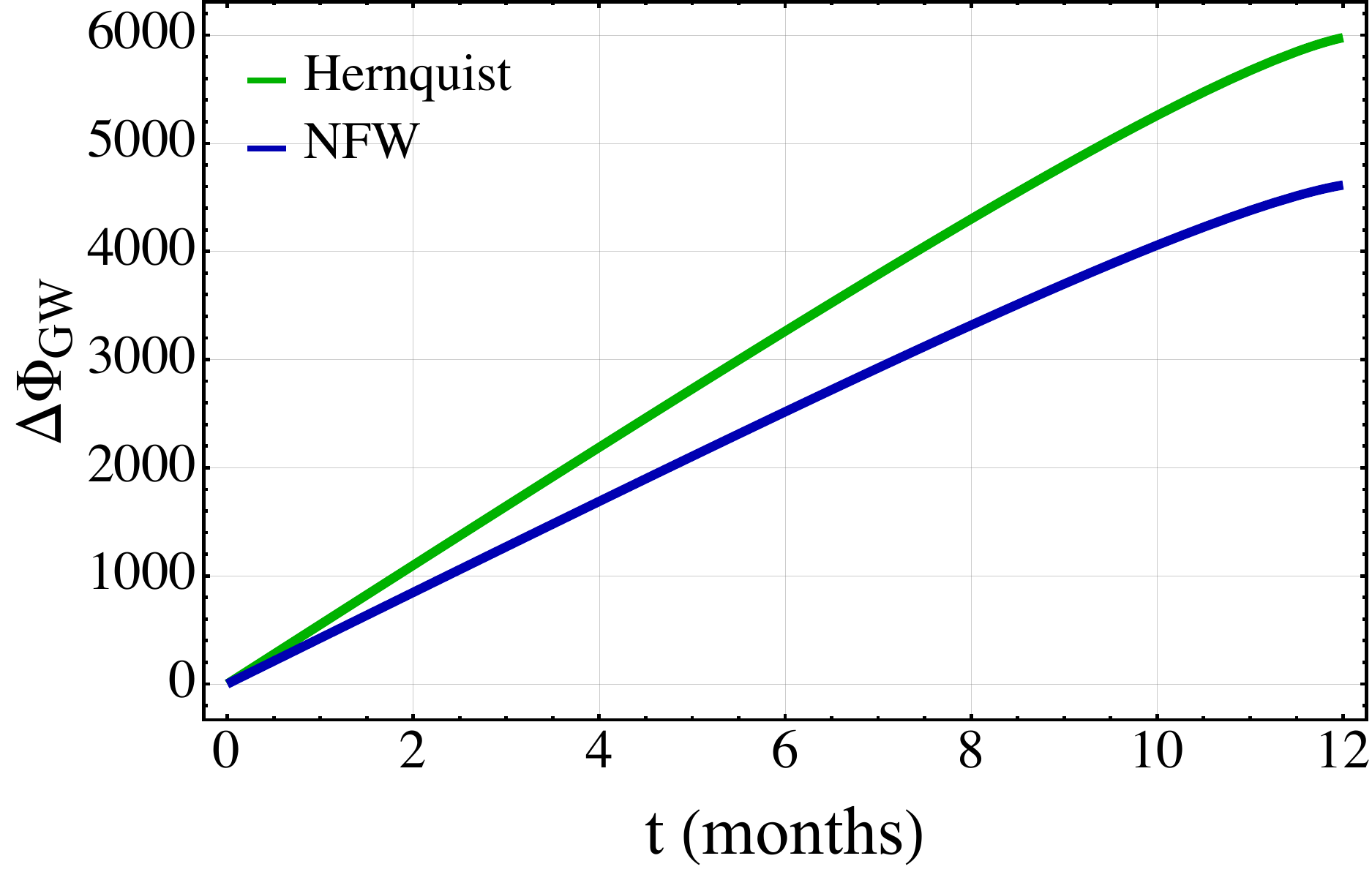}
	\endminipage\hfill
    \minipage{0.48\textwidth}
	\includegraphics[width=\linewidth]{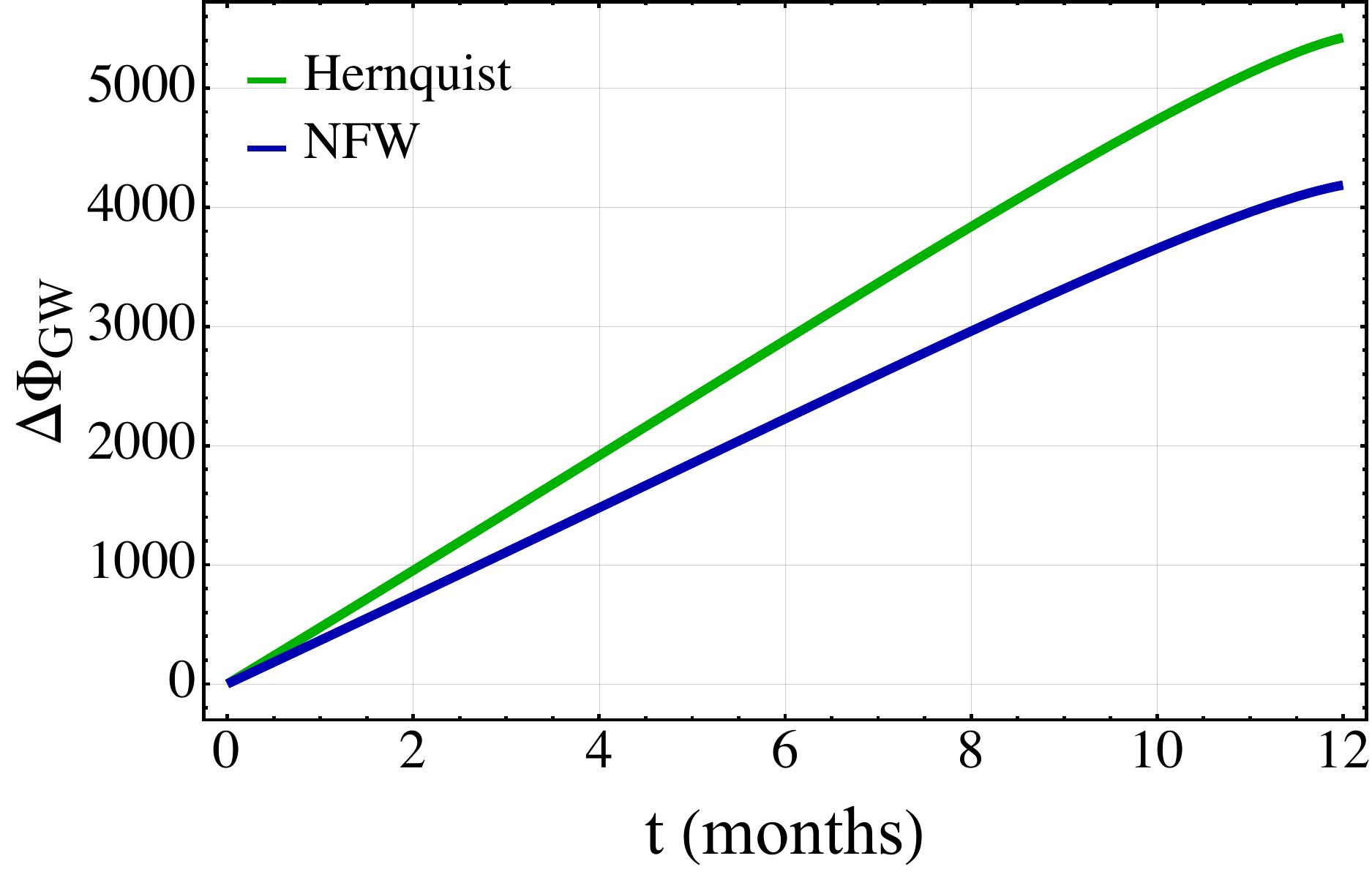}
	\endminipage
	\caption{Gravitational wave dephasing due to the presence of dark matter is plotted for a one-year observation period, considering an EMRI system with component masses $10^6 M_{\odot}$ and $10M_{\odot}$. Here, we consider both the Hernquist and NFW dark matter density profiles. The left panel shows the dephasing considering only the effect of gravitational wave emission. The right panel shows the combined effect of gravitational wave emission and dynamical friction.
 }\label{fig_phase}
\end{figure*}
At order $\Od$, the term $\tilde{\Psi}_R^{(1,0)}$ acts as the source for the gravitational perturbation $\tilde{\Psi}_R^{(1,1)}$ in the axial sector. In the polar sector, both $\tilde{\Psi}_Z^{(1,0)}$ and $\tilde{\Psi}_F^{(1,1)}$ serve as sources for the gravitational perturbation $\tilde{\Psi}_Z^{(1,1)}$. The gravitational perturbation equations are solved using the same multi-domain setup as before: the interval $[0, 1]$ is divided into two subdomains, $\mathbf{D}_1 \equiv [0, \sigma_\Omega]$ and $\mathbf{D}_2 \equiv [\sigma_\Omega, 1]$, each discretized with $60 + 1$ grid points. The results are shown in \autoref{fig_grav_pert_ez} . Here, the plot shows the gravitational perturbations for $(l, m) = (2, 1)$ and $(2, 2)$. As before, The particle's location is set at $r_\Omega = 10$ and we have used the same dark matter profile and sound speed parameters as in \autoref{fig_density}. As discussed earlier, $\Psi^{(1,1)}_{Z}$ can be decomposed accordingly as $\tilde\Psi^{(1,1)}_{Z}= \tilde\Psi^{(1,1)}_{Z,F}+\tilde\Psi^{(1,1)}_{Z,G}$ where $\Psi^{(1,1)}_{Z,F}$ represents the contribution to the metric perturbation induced by the fluid (i.e., density) fluctuation, while $\Psi^{(1,1)}_{Z,G}$ corresponds to the perturbation arising from the modification of the background spacetime due to the presence of dark matter. The contribution $\Psi^{(1,1)}_{Z,G}$ gives rise to gravitational self-force effects, whereas $\Psi^{(1,1)}_{Z,F}$ captures phenomena like dynamical friction \cite{Barausse:2007ph}. In \autoref{fig_grav_fluid_ez}, we show the $\Psi^{(1,1)}_{Z,G}$ and $\Psi^{(1,1)}_{Z,F}$ as function of compactified coordinate $\s$ for $l=2,~m=2$. \par
We compute the correction to the total gravitational wave flux due to the presence of dark matter using \autoref{flux_formula}. The results are presented in \autoref{tab:flux0_com} and in the left panel of \autoref{fig_flux}. In this analysis, we consider both Hernquist and NFW dark matter density profiles, with parameter values listed in \autoref{tab:Fitting_Parameters}. As evident from the results, the modification to the gravitational wave flux induced by dark matter is several orders of magnitude smaller than the vacuum general relativity contribution.
\par
Dissipative effects, such as gravitational wave emission and dynamical friction, cause the secondary object to lose orbital energy, initiating the inspiral phase. As a result, the orbital frequency evolves in accordance to \autoref{freq_change}. At leading order $\mathcal{O}(\epsilon)$, the change in orbital frequency $\dot{\Omega}$ is governed by $\mathsf{F}^{(1,0)}_{\Omega}$, which describes the flux-driven adiabatic evolution and is identical to that in vacuum Schwarzschild spacetime. At next order $\mathcal{O}(\xi \epsilon)$, both the gravitational wave emission correction $\mathsf{F}_{\Omega,\textrm{GW}}\tpd$ and the dynamical friction contribution $\mathsf{F}_{\Omega,\textrm{DF}}\tpd$ must be taken into account. In the right panel of \autoref{fig_flux}, we plot the leading-order contribution $\mathsf{F}^{(1,0)}_{\Omega}$ together with the dark matter correction $\mathsf{F}_{\Omega,\textrm{GW}}\tpd$ for a Hernquist density profile over the orbital frequency range $\Omega \in [\Omega_{\textrm{ini}}, \Omega_{\textrm{end}}]$, where $\Omega_{\textrm{ini}}$ and $\Omega_{\textrm{end}}$ correspond to the orbital frequencies associated with radii $r = 20\Mbh$ and the ISCO, respectively. For comparison, we also include the contribution from dynamical friction $\mathsf{F}_{\Omega,\textrm{DF}}\tpd$, computed using \autoref{dynamic_friction}. As the plot demonstrates, 
in the strong-gravity regime, the term $\mathsf{F}^{(1,0)}_{\Omega}$ remains the dominant driver of the inspiral. For  larger orbital separations (i.e., for small $\Omega$), the dynamical friction term $\mathsf{F}_{\Omega,\textrm{DF}}\tpd$ contributes more to orbital evolution than dark matter induced modification to gravitational wave flux. 
 Nevertheless,  the contribution from $\mathsf{F}_{\Omega,\textrm{GW}}\tpd$ outweighs that of $\mathsf{F}_{\Omega,\textrm{DF}}\tpd$ as the object moves closer to the ISCO.
\par
The evolution of the orbital phase $\phi(t)$ is obtained by numerically integrating \autoref{freq_change}. We assume that the secondary object begins its inspiral toward the supermassive black hole from an initial radius $r_\Omega = 10\Mbh$, and the integration proceeds until the object reaches the ISCO. The initial phase is set to $\phi(0) = 0$. The gravitational wave phase $\Phi_{\textrm{GW}}(t)$ for the dominant $(l = 2, m = 2)$ mode is related to the orbital phase by $\Phi_{\textrm{GW}}(t) = 2\phi(t)$.
To quantify the impact of dark matter on the gravitational waveform, we compute the gravitational wave phase shift relative to a reference waveform using
\begin{equation}\label{phase_shift}
\Delta\Phi_{\textrm{GW}}(t_{\textrm{obs}}) = \Phi_{\textrm{GW}}(t_{\textrm{obs}}) - \Phi_{\textrm{GW}}^{\textrm{ref}}(t_{\textrm{obs}}),
\end{equation}
where $\Phi_{\textrm{GW}}(t_{\textrm{obs}})$ denotes the accumulated gravitational wave phase over the observation time $t_{\textrm{obs}}$, which we take to be 1 year. The reference waveform corresponds to a point particle inspiraling into a vacuum Schwarzschild black hole.

The results are shown in \autoref{fig_phase}, where we plot the dephasing $\Delta\Phi_{\textrm{GW}}(t)$ as a function of time for both Hernquist and NFW dark matter density profiles. Here, we consider that a steller mass black hole of mass $10M_{\odot}$ inspiralling onto a supermassive black hole of of mass $10^6M_{\odot}$. Two scenarios are considered: the left panel shows the dephasing due to gravitational wave emission alone, while the right panel includes the combined effects of gravitational wave emission and dynamical friction. In both cases, we observe that the accumulated dephasing over a one-year observation period can reach $\sim \mathcal{O}(4000)$. As evident from the plots, the dominant contribution to the dephasing arises from gravitational wave emission. This is consistent with expectations, as the calculation is performed in the strong-field regime where, as shown earlier in \autoref{fig_flux}, the contribution from gravitational wave emission outweighs that from dynamical friction. 
\par
This dephasing provides a rough estimate of the potential of space-based gravitational-wave detectors such as LISA to detect the presence of dark matter through EMRI observations.  
Given that the average signal-to-noise ratio (SNR) for LISA is expected to be $\sim 30$, and that LISA can detect waveform dephasing as small as $\Delta\Phi_{\textrm{GW}} \geq 0.1$, we expect that LISA will be capable of detecting the presence of dark matter environments through their imprint on gravitational wave signals. However, it should be noted that the presence of dark matter effectively adds additional mass to the EMRI system. From an observational point-of-view, the resulting dephasing could be degenerate with that produced by a more massive primary black hole, rather than uniquely signaling the presence of dark matter, if not treated carefully. Therefore, a more comprehensive analysis accounting for such degeneracies is required to robustly confirm the presence of dark matter. In particular, a more rigorous assessment than the order-of-magnitude dephasing estimates presented here, such as a Fisher matrix or Bayesian analysis, is necessary to fully quantify the ability of space-based gravitational wave detectors to detect the presence of dark matter. 
\section{Conclusion}\label{conclusion}
No astrophysical system evolve in isolation. This is particularly true for EMRIs which often find refuge in the center of the galaxies, surrounded by other astrophysical objects, accretion disks and dark matter. Thus, consideration of the influence of such astrophysical environments on the gravitational wave emitted by such system is of paramount importance. In the paper, we have studied the effect of dark matter on the EMRI dynamics and  laid out a framework to model 1PA EMRI waveforms in the presence of dark matter. \par
Following Sadeghian et al.\cite{Sadeghian:2013laa}, we model the adiabatic growth of a black hole inside a cold and collisionless dark matter environment and numerically compute the resulting enhancement in the dark matter density profile due to the presence of the black hole. In this work, we consider two distinct dark matter density profiles: the Hernquist and NFW profiles. By describing the dark matter surrounding the black hole as an anisotropic fluid with vanishing radial pressure, we construct a self-gravitating system representing a black hole immersed in dark matter. Specifically, we show that the mass function for such a system can be expressed in terms of the Appell hypergeometric function. However, since our primary motivation is to model 1PA gravitational waveforms from such systems, we find that a perturbative approach is more suitable for capturing the effects of dark matter in this context, with the parameter $\xi$ introduced in \autoref{mass_func} characterizes the modification of Schwarzschild geometry induced by the dark matter.\par
When we consider an EMRI system in this dark matter environment, the resultant spacetime can be characterized by two parameters: $\xi$ which capture the modification in the Schwarzschild spacetime due to dark matter, and the parameter $\e$ which describes the modification of the spacetime due to the presence of the secondary (see \autoref{metric_emri}). Furthermore, the presence of the secondary also introduces perturbation to the dark matter. Due to the coupling between the secondary’s trajectory $z^\mu$, the metric, and the fluid perturbations, a complete description of the orbital dynamics requires solving the covariant acceleration equation \autoref{sf_eqn} together with the Einstein equations (\autoref{vac}–\autoref{matpp}) and the covariant conservation laws \autoref{cov_cons_law}, all within the two-timescale framework outlined in \autoref{sec: two_timescale}. We find that the presence of dark matter introduces additional force terms into the self-force equation \autoref{sf_eqn}. At leading order $\mathcal{O}(\xi)$, this force is purely conservative and results in shifts in conserved quantities such as energy and angular momentum. At next order $\mathcal{O}(\epsilon\xi)$, the correction includes both conservative and dissipative components. However, for constructing 1PA waveform models, only the orbit-averaged dissipative contributions are required.  Furthermore, to single out the contributions of the dark matter, we adopt the fixed-frequency framework described in \autoref{sec: two_timescale}.\par
The perturbations are analyzed within the Regge–Wheeler–Zerilli formalism and can be described by five master functions: four associated with metric perturbations, $(\Psi_{R}^{(1,0)}, \Psi_{R}^{(1,1)})$ and $(\Psi_{Z}^{(1,0)}, \Psi_{Z}^{(1,1)})$, and one associated with fluid perturbations, $\Psi_{F}^{(1,1)}$. All satisfy Regge–Wheeler–Zerilli-type equations (see \autoref{Sec:RWZ_formalism}). Notably, the second-order metric perturbations at $\mathcal{O}(\epsilon\xi)$ are sourced by both the dark matter–induced modification of the background spacetime and the density fluctuations in the dark matter, the latter being associated with phenomena such as dynamical friction. \par
However, one of main disadvantage of this formalism is the appearance of unbounded source terms for the perturbation equation at $\Od$. We circumvented the problem by adopting the multi-domain spectral method. Within the machine precision of \texttt{Mathematica} software \cite{Mathematica}, this method is quite fast and produces highly accurate result. We cross checked our results by comparing it with results from the \texttt{Black Hole Perturbation Toolkit} (see \autoref{tab:flux0_com}). \par
We computed the gravitational wave flux for both the Hernquist and NFW dark matter profiles and compare it with the energy loss due to dynamical friction. Our results indicate that in the strong-field regime, gravitational wave emission dominates the inspiral dynamics. Notably, for small orbital separations, the modification to the gravitational wave flux due to dark matter exceeds the contribution from dynamical friction. Using the computed fluxes, we then determined the orbital evolution of an EMRI system immersed in a dark matter environment. 
\par
Finally, we provided a rough order-of-magnitude estimate of the detectability of dark matter by space-based gravitational wave detectors using dephasing arguments.
Considering an EMRI system with component masses $10^6\mathrm{M}_{\odot}$ and $10\mathrm{M}_{\odot}$ over a one-year observation period, we computed the gravitational wave phase evolution in the presence of dark matter and compared it with a reference waveform from a vacuum EMRI system with the same component masses. We find that the presence of dark matter induces a phase shift with respect to the reference waveform of $\Delta\Phi_{\textrm{GW}} \sim \mathcal{O}(4000)$ for both the Hernquist and NFW profile. Given that the detection threshold for LISA is $\Delta\Phi_{\textrm{GW}} \sim 0.1$ at a signal-to-noise ratio of $\sim 30$, we expect that LISA will be capable of detecting dark matter environments surrounding supermassive black holes.
\par
However, it should be noted that the presence of dark matter effectively adds mass to the EMRI system, the induced waveform dephasing may be observationally indistinguishable from that arising from a heavier primary black hole in vacuum. Disentangling these effects therefore requires a more detailed analysis that explicitly incorporates parameter correlations and degeneracies. In particular, going beyond the order-of-magnitude dephasing estimates presented here, a systematic Fisher-matrix or fully Bayesian parameter-estimation study will be essential to reliably assess the sensitivity of space-based gravitational wave detectors to dark matter signatures in EMRI signals. However, we leave it for future work.\par
This paper provides a starting point for modeling first 1PA waveforms in the presence of a dark matter environment, with several potential directions for future extensions. In this study, we made a number of simplifying assumptions. In particular, to construct a self-gravitating system where the black hole is embedded in a dark matter environment, we assumed that all dark matter particles move in circular orbits around the supermassive black hole, resulting in vanishing radial pressure. A more realistic and detailed configuration would require solving the full Einstein-Vlasov system \cite{Rendall:1996gx, Rendall}. Since 1PA waveform modeling only requires perturbative information about the effect of dark matter, it would be particularly interesting to explore perturbative methods for solving the Einstein–Vlasov equation in this context.\par 
Furthermore, we have assumed the spacetime to be static and spherically symmetric. However, since astrophysical black holes are generally rotating, extending the model to include a rotating black hole is essential. This is especially relevant given that the dark matter density profile is expected to increase with black hole spin. A recent study \cite{Mitra:2025tag} has investigated this scenario and computed the gravitational wave dephasing, but it neglected the effect of dark matter on the background metric. As we have shown, in the strong gravity regime, particularly near the ISCO, the modification of the background geometry due to dark matter can outweigh the contribution from dynamic friction. Therefore, it would be important and interesting to incorporate such effects into our framework in future work.\par
Furthermore, we have assumed that the changes in the primary object's mass and angular momentum due to incident radiation and dark matter accretion are negligible. However, as noted in \cite{PhysRevD.103.064048, Mathews:2021rod}, over the long inspiral timescale, such effects can have a significant impact on the gravitational waveform. We therefore plan to incorporate these effects in future work.
\par
Lastly, this study does not fully account for the influence of dynamic friction in the strong gravity regime. Within the framework developed here, we have computed the density fluctuations and the resulting gravitational perturbations they induce. A natural next step is to isolate and quantify the specific contribution of dynamic friction in this regime, which we leave for future investigation.
\section*{Acknowledgements} 
We like to thank Takahiro Tanaka, Adam Pound, Vitor Cardoso, Hidetoshi Omiya, Rodrigo Panosso Macedo, Sumanta Chakraborty and Subhodeep Sarkar for helpful discussions.  
M.~R. supported by the JSPS KAKENHI Grant No.~ JP23KF0233.
T.~T is supported by JSPS KAKENHI Grant Nos.~ JP25KJ0067 and JP25K17397.

\appendix
	\labelformat{section}{Appendix #1}
	\labelformat{subsection}{Appendix #1}
	\labelformat{subsubsection}{Appendix #1}
\section{Perturbation equation: Regge-Wheeler-Zerilli Formalism}\label{app:1}
\subsection{Metric perturbation}
In this appendix, we present the equations governing the metric perturbations induced by the presence of the secondary object within the Regge–Wheeler formalism. As described in the main text, the metric perturbation in this framework is decomposed into axial (odd parity) and polar (even parity) sectors, i.e., $h_{\mu\nu} = h_{\mu\nu}^{ \textrm{odd}}+h_{\mu\nu}^{ \textrm{even}}$, the expression of which are given by \cite{PhysRevD.67.104017, Cardoso:2022whc}
\begin{equation} \label{prt1}
    \begin{aligned}
    h_{\mu\nu}^{ \textrm{odd}}(t,r,\theta,\phi) =& \sum_{lm} \frac{2\sqrt{(n+1)}}{r}\Big(ih_{1}^{lm}\mathbf c_{lm}-h_{0}^{lm}\mathbf c^{0}_{lm}\Big)  \\
    h_{\mu\nu}^{ \textrm{even}}(t,r,\theta,\phi) =& \sum_{lm}\Big( a H_{0}^{lm}\mathbf a^{0}_{lm}-i\sqrt{2}H_{1}^{lm}\mathbf a^{1}_{lm}\\&+\frac{H_{2}^{lm}}{b}\mathbf a_{lm}+\sqrt{2}K^{lm}\mathbf g_{lm}\Big).
\end{aligned}
\end{equation}
where, $n=l(l+1)/2-1$. Note that, the odd parity perturbation components are described in terms of functions $h\todd^{lm}=(h_{0}^{lm}, h_{1}^{lm})$ while the even parity components are are defined in terms of $h\teven^{lm}=(K^{lm}, H_{0}^{lm}, H_{1}^{lm}, H_{2}^{lm})$, which are functions of are functions of ($t, r$). For the sake of notational simplicity, we drop the superscript $lm$ in the terms describing the even and odd parity perturbations $h\todd^{lm}$ and $h\teven^{lm}$. Furthermore, we can write these terms as $h\todd=h\todd\tpp+\xi h\todd\tpd$ ($h\teven=h\teven\tpp+\xi h\teven\tpd$), where $h\todd\tpp$ $(h\teven\tpp)$ describe the perturbation at $\Op$, whereas $h\todd\tpd$ $(h\teven\tpd)$ describe the perturbation at $\Od$. In the similar manner, we can write the energy momentum tensor of the point particle in terms of  angular basis \cite{PhysRevD.67.104017}
\beq\label{harmonicexp}
\Tpp_{\mu\nu}(t, r, \theta, \phi)&=\sum_{lm}
\bigg[{A}^{0}_{l m }\mathbf{a}^{0}_{l m}+{A}^{1}_{l m}\mathbf{a}^{1}_{l m} +{A}_{l m }\mathbf{a}_{l m}\\&+{B}^{0}_{l m }\mathbf{b}^0_{l m}
+{B}_{l m }\mathbf{b}_{l m}+{Q}^{0}_{l m }\mathbf{c}^0_{l m}+{Q}_{l m }\mathbf{c}_{l m}\\
&+{D}_{l m }\mathbf{d}_{l m}+{G}_{l m}\mathbf{g}_{l m}+{F}_{l m }\mathbf{f}_{l m}\bigg]. 
\eeq
In \autoref{prt1} and \autoref{harmonicexp}, $\mathcal{Y}_{lm}=\{\mathbf a^{0}_{lm}, \mathbf a^{1}_{lm},...,\mathbf f_{lm}\}$ are the ten tensor harmonics, the expression of which can be found in \cite{PhysRevD.67.104017}. In the above equation, the expansion coefficients $\mathbf{\mathfrak{Y}}_{lm}=\{{A}^{0}_{l m }, {A}^{1}_{l m },... , {F}_{l m }\}$ can be obtained by projecting each tensor harmonics on the point particle energy-momentum tensor, e.g. \cite{PhysRevD.67.104017, Cardoso:2022whc}
\beq\label{scalar_product}
\mathfrak{Y}_{lm}=\int d\mathbf{\Omega}~\eta^{\mu\alpha}\eta^{\nu\beta}~\mathcal{Y}^{*}_{lm}~\Tpp_{\alpha\beta}
\eeq
Note that, we can write the expansion coefficients as $\mathfrak{Y}_{lm}=\mathfrak{Y}_{p,lm}+\xi \mathfrak{Y}_{d,lm}$, where $\mathfrak{Y}_{p,lm}$ and $\mathfrak{Y}_{d,lm}$ denotes the expansion coefficients at $\Op$ and $\Od$, respectively. The non vanishing components of $\mathfrak{Y}_{lm}$ for a point particle in a  circular orbit are given by \cite{ Cardoso:2022whc}
\beq\label{expansion_coef}
A^{0}_{lm}&=\frac{\mu u_0 a \sqrt{ab}}{r^2}Y^*_{lm}\delta_r\,,\\ B^{0}_{lm} &=\frac{i\mu u_3 \sqrt{ab}}{\sqrt{n+1}r}\partial_\phi Y^*_{lm}\delta_r\\Q^{0}_{lm}&=-\frac{i\mu u_3 \sqrt{ab}}{\sqrt{n+1}r}\partial_\theta Y^*_{lm}\delta_r\,,\\ D_{lm}&=\frac{i\mu u_3^2 \sqrt{ba}}{\sqrt{2n(n+1)}u_0 }\partial_{\phi\theta} Y^*_{lm}\delta_r\\
G_{lm}&=\frac{\mu u_3^2 \sqrt{b/a}}{\sqrt{2}u_0}Y^*_{lm}\delta_r\,,\\~ F_{lm}&=\frac{\mu u_3^2 \sqrt{b/a}}{2\sqrt{2n(n+1)}u_0 }(\partial_{\phi\phi}-\partial_{\theta\theta}) Y^*_{lm}\delta_r
\eeq
 where, $Y^*_{lm}$ is the conjugate of spherical harmonics $Y_{lm}(\theta,\phi)$, $\delta_r=\delta(r-r_p)$. We can obtain the values of $\mathfrak{Y}_{p,lm}$ and $\mathfrak{Y}_{d,lm}$ by substituting \autoref{Gtt1p}, \autoref{r_expand} and \autoref{u_expand} in the above equation and expanding up to $\Od$.
\subsection{Perturbation equations for the fluid}\label{app: fluid_pert}
In order to study the fluid perturbation equation , we adopt to the Lagrangian perturbation theory \cite{1975ApJ...200..204F}. We consider the Lagrangian displacement vector $\chi^{\mu}$, which is the generator of diffeomorphism that would take the would lines of the unperturbed fluid to the perturbed worldlines. The change in any quantity $Q$ is defined by the (i) Eulerian variation, $\delta Q$ which defines the change in $Q$ at a fixed point in the manifold, (ii) Lagrangian variation, $\Delta Q$, which accounts for the variation of the field with respect to a frame that is itself dragged along with the fluid. the relationship between the two is given by $\Delta=\delta+\mathcal{L}_{\chi}$, where $\mathcal{L}_{\chi}$ is the Lie derivative along the Lagrangian displacement vector $\chi$. The change in fluid is attributed with $\Delta$, which the change is field is attributed with $\delta$. Note that, the Lagrangian displacement is invariant is invariant under gauge transformation. There is further freedom, $\chi^{\mu}+\mathcal{L}_{\chi} \mathfrak{u}^{\mu}$ leads to the same perturbed wouldlines as $\chi^{\mu}$. thus, one is free to arbitrarily restrict the timelike part of $\chi^{\mu}$, which we set such that the fluid four velocity satisfies the relation  $\chi_{\mu}\mathfrak{u}^{\mu}=0$. The expressions for the Lagrangian displacement vector can be written in terms of spherical harmonics, and using the relations \cite{1975ApJ...200..204F}
\beq
\delta \mathfrak{u}^{\alpha}=\left(g^\alpha_\beta+\mathfrak{u}^\alpha\mathfrak
 u_\beta\right)&\left(\chi^\beta_{;\gamma}\mathfrak u^\gamma-\mathfrak u^\beta_{;\gamma}\chi^\gamma\right)+\frac{1}{2}\mathfrak u^\alpha\mathfrak u^\gamma \mathfrak u ^{\beta}h_{\gamma\beta}\\
(\mathfrak{u}^{\mu}+\delta \mathfrak{u}^{\mu})&(\mathfrak{u}_{\mu}+\delta \mathfrak{u}_{\mu})=-1\\
(\mathfrak{u}^{\mu}+\delta \mathfrak{u}^{\mu})&({k}_{\mu}+\delta {k}_{\mu})=0\\
({k}^{\mu}+\delta {k}^{\mu})&({k}_{\mu}+\delta {k}_{\mu})=1~,
\eeq
we obtain the following expression for $\mathfrak u^\mu$ and $k^\mu$ as 
\beq\label{fluid_vel}
\delta\mathfrak{u}^0&=\frac{1}{2\sqrt{a}}\sum_{lm}H_{0} Y_{lm}\,,\quad \delta\mathfrak u^1=\frac{\sqrt{a}}{16\pi r^2\rho_{\textrm{BH}}b}\sum_{lm}W_{lm} Y_{lm}\\
\delta\mathfrak u^2&=\frac{\sqrt{a}}{4\pi r^2\kappa}\sum_{lm}\left(V_{lm}\partial_\theta Y_{lm}-\frac{U_{lm}}{\sin{\theta}}\partial_\phi Y_{lm}\right)
\\
\delta\mathfrak u^3&=\frac{\sqrt{a}}{4\pi r^2\kappa\sin^2{\theta}}\sum_{lm}\left(V_{lm}\partial_\phi Y_{lm}+\frac{U_{lm}}{\sin{\theta}}\partial_\theta Y_{lm}\right)\\
\delta k_0&=\frac{    a Y_{lm} W_{lm}}  {16 \pi  r^2 b(r)^{3/2} \rho (r)}\,,\quad 
\delta k_1 =\frac{    Y_{lm} H_2(r)}{2 \sqrt{b(r)}}
\eeq
where $\kappa=\rho_{\textrm{BH}}+p_t$ and where we retain the contribution to $\delta\mathfrak{u}^{\mu}$ up to $\Od$. We have used some functional redefinition to represent these equations in the form of \cite{Cardoso:2022whc}. Furthermore, the perturbation changes the density and pressure as 
\beq \label{pert_dense}
\delta \rho_{\textrm{BH}}&=\sum_{lm}\delta \rho_{lm}Y_{lm}\,,\\ \delta p_r &=\sum_{lm}\delta p_{r,lm}Y_{lm}\,,\\ \delta p_t &=\sum_{lm}\delta p_{t,lm}Y_{lm}
\eeq
Further relation between the perturbed density and pressure is obtained through barotropic equation of state which is given as follows
\beq \label{pert_press}
\delta p_{r,t}=c_{r,t}^2 (r) \delta \rho
\eeq
where $c_{r}$ and $c_t$ are radial and tangential sound speeds. Following \cite{ Cardoso:2022whc}, we consider these as constants.
\begin{widetext}
\subsection{Perturbation equations}
\subsubsection{Perturbation equation in the gravity sector
}
Replacing \autoref{prt1} and \autoref{harmonicexp} in \autoref{pp}, we find the perturbation equation at $\Op$ for the axial sector can be written in terms of master function $\Psi_R\tpp$ \cite{PhysRevD.67.104017}
\beq\label{odd_eq1}
\mathcal{L}_R\Psi_R\tpp\equiv \left[\partial^2_r{*}+\omega^2-V_{R}(r)\right]\Psi_R\tpp=S_R\tpp~,
 \eeq
 where 
 \beq\label{odd_pot}
V_{R}(r)&=\frac{f}{r^2}\left[2(n+1)-\frac{6M}{r}\right]\\
S_R\tpp &=\frac{4 i \pi  \sqrt{2} r f^2 D_{lm}^{'(1,0)}}{\sqrt{n (n+1)}}+\frac{4 i \pi  \sqrt{2} r D_{lm}\tpp f f'}{\sqrt{n (n+1)}}+\frac{8 i \pi  f^2 Q_{lm}\tpp}{\sqrt{n+1}}.
\eeq
The metric perturbation functions $h\todd\tpp=(h_{0}\tpp, h_{1}\tpp)$ in terms of $\Psi\tpp_R$ is given by Eq.~(A33) and Eq.~(A34) of \cite{PhysRevD.67.104017}.\par 

Similarly, the perturbation in polar sector can be described in terms of the master function $\Psi_Z\tpp$, which the governing perturbation equation us as follows \cite{PhysRevD.67.104017}
\beq\label{even_eq}
\mathcal{L}_Z\Psi_Z\tpp\equiv \left[\partial^2_r{*}+\omega^2-V_{R}(r)\right]\Psi_Z\tpp=S_Z\tpp~,
 \eeq
  \beq\label{even_pot}
V_{Z}(r)&=\frac{f \left(18 \Mbh^3+18 \Mbh^2 n r+6 \Mbh n^2 r^2+2 (n+1) n^2 r^3\right)}{r^3 (3 \Mbh+n r)^2}\\
S_Z\tpp &=\frac{f^2 r^2 \partial_r C_{1,lm}\tpp}{3 \Mbh+n r}-\frac{3 f \Mbh C_{1,lm}\tpp (\Mbh-(n+2) r)}{(3 \Mbh+n r)^2}\\&-\frac{i f^3 r \partial_r C_{2,lm}\tpp}{3 \Mbh+n r}+\frac{i f^2 C_{2,lm}\tpp (n (n+1) r-\Mbh (n+3))}{(3 \Mbh+n r)^2}.
\eeq
where expressions of the $C_{1,lm}\tpp$ and  $C_{2,lm}\tpp$ are given by Eq.~(A43) and Eq.~(A44) in \cite{PhysRevD.67.104017}. \par
Similarly, replacing \autoref{prt1} and \autoref{harmonicexp} in \autoref{matpp}, we find the perturbation equation at $\Od$ for the axial sector can be written in terms of master function $\Psi_R\tpd$
\beq\label{odd_eq_dm}
\mathcal{L}_R\Psi_R\tpd &=S_R\tpd~,\\
\mathcal{L}_F\Psi_F\tpd &=S_F\tpd~,\\
\mathcal{L}_Z\Psi_Z\tpd &=S_Z\tpd~,
 \eeq
 where the expression of $\mathcal{L}_F$ is given in \autoref{eq_o_ez_f}. The expressions for $S\tpd$ are quite lengthy. Thus, we have provided a \texttt{Mathematica} notebook where one can find the expressions for this term \cite{RaBHman}. Interestingly, 
 $S_Z\tpd$ satisfies a equation similar to $S_Z\tpp$ in \autoref{even_pot} with $C_{1,lm}\tpp$ and  $C_{2,lm}\tpp$ replaced by $C_{1,lm}\tpd$ and  $C_{2,lm}\tpd$. The expressions of $C_{1,lm}\tpd$ and  $C_{2,lm}\tpd$ is also provided in that notebook.
%
 \subsubsection{Fluid perturbation equation }
 The fluid perturbation equation is given by \autoref{eq_o_ez_f}, where the fluid perturbation quantities are related to the master function 
 \beq
 \delta \rho &=\Psi^{(1,1)}_F r^{-2+\frac{c_t^2}{c_r^2}}f^{-\frac{3}{4}-\frac{1}{4c_r^2}}\\
 W_{lm}&=-\frac{2 f \delta m' \left(2 r \omega  H_1-i f r \partial_r B_{lm}\tpp\right)}{r \omega }-\frac{2 i f C_{1,lm}\tpp\left(\Mbh^2 (6 n+3)-4 \Mbh n r+n r^2\right) \delta m'}{\omega  (3 \Mbh+n r)^2}\\&-\frac{2 C_{2,lm}\tpp(r-2 \Mbh)^2 \left(3 \Mbh^2-\Mbh n (2 n+1) r+n (n+1) r^2\right) \delta m'}{r^3 \omega  (3 \Mbh+n r)^2}\\&+\frac{8 i \pi  f^{\frac{1}{4}-\frac{1}{4 c_r^2}} \Psi_F\tpp r^{\frac{c_t^2}{c_r^2}-2} \left(\Mbh \left(c_r^2-4 c_t^2-1\right)+2 c_t^2 r\right)}{\omega }-\frac{16 i \pi  c_r^2 f^{\frac{5}{4}-\frac{1}{4 c_r^2}} r^{\frac{c_t^2}{c_r^2}} \partial_r \Psi_F\tpp}{\omega }
 \\&+\frac{2 i f \partial_r\Psi_Z\tpp \delta m' \left(3 \Mbh^2 (n+2)-\Mbh r \left(n^2+3 r^2 \omega ^2\right)+n r^2 \left(n-r^2 \omega ^2+1\right)\right)}{r^2 \omega  (3 \Mbh+n r)}\\&+\frac{2 i \Psi_Z\tpp \delta m' \left(18 \Mbh^4 (n+2)+3 \Mbh^3 r \left(4 n^2+3 n+12 r^2 \omega ^2-6\right)+3 \Mbh^2 r^2 \left(2 n^3+n \left(10 r^2 \omega ^2-3\right)-3 r^2 \omega ^2\right)\right)}{r^4 \omega  (3 \Mbh+n r)^2}\\&+\frac{2 i \Psi_Z\tpp \delta m' \left(-\Mbh n r^3 \left(5 n^2+n \left(5-6 r^2 \omega ^2\right)+9 r^2 \omega ^2\right)+n^2 r^4 \left(n-2 r^2 \omega ^2+1\right)\right)}{r^4 \omega  (3 \Mbh+n r)^2}\\
 V_{lm} &=\frac{i B_{lm}\tpp \delta m'}{2 r^2 \omega }+\frac{i C_{1,lm}\tpp\left(3 \Mbh^2+3 \Mbh n r-n r^2\right) \delta m'}{2 r \omega  (3 \Mbh+n r)^2}+\frac{f C_{2,lm}\tpp\left(3 \Mbh^2+3 \Mbh n r-n r^2\right) \delta m'}{2 r^2 \omega  (3 \Mbh+n r)^2}\\&-\frac{4 i \pi  c_t^2 f^{-\frac{1}{4 c_r^2}-\frac{3}{4}} \Psi_F\tpd r^{\frac{c_t^2}{c_r^2}-2}}{\omega }+\frac{2 i \pi  \sqrt{2} \Mbh F_{lm}\tpp \delta m'}{f \sqrt{n (n+1)} r \omega }+\frac{i \Psi_Z\tpp \delta m' \left(\frac{9 \Mbh^3+9 \Mbh^2 n r+3 \Mbh n^2 r^2+n^2 (n+1) r^3}{r^4 (3 \Mbh+n r)^2}-\frac{\omega ^2}{f r}\right)}{2 \omega }\\&+\frac{i \partial_r\Psi_Z\tpp \left(1-\frac{3 \Mbh (\Mbh+n r+r)}{r (3 \Mbh+n r)}\right) \delta m'}{2 r^2 \omega }
 \eeq

\subsubsection{Metric reconstruction }
The metric $h\todd\tpp$ is related to the master function $\Psi_R\tpp$ through the relation
\beq\label{met_odd}
h_1\tpp &=\frac{r \Psi_R\tpp}{f}\\
h_0\tpp &=\frac{4 \pi  \sqrt{2} f r^2 D_{lm}(r)}{\sqrt{n (n+1)} \omega }+\frac{i f r \partial_r\Psi_R\tpp}{\omega }+\frac{i f \Psi_R\tpp}{\omega }
\\
h_1\tpd &=\frac{r \Psi_R\tpd}{f}\\
h_0\tpd &=\frac{4 \pi  \sqrt{2} f r^2 r_{0,1} \partial_r D_{lm}\tpp}{\sqrt{n (n+1)} \omega }+\frac{4 \pi  \sqrt{2} D_{lm}\tpp \left(f r^2 \delta f+2 f r r_{0,1}-2 r \delta m\right)}{\sqrt{n (n+1)} \omega }\\&+\frac{i \Psi_R\tpp\left(2 f^2 r^2 \delta f+f r^2 \left(f r \delta f'(r)-4 \delta m'(r)\right)+4 (4 \Mbh-r) \delta m\right)}{2 f r^2 \omega }\\&+\frac{4 \pi  \sqrt{2} f r^2 D_{lm}\tpd}{\sqrt{n (n+1)} \omega }+\frac{i f r \partial_r \Psi_R\tpd}{\omega }+\frac{i f \Psi_R\tpp}{\omega }-\frac{i\partial_r \Psi_R\tpp (4 \delta m-f r \delta f)}{\omega }
\eeq
 The metric $h\teven^{(A)}$ is related to the master function $\Psi_z^{(A)}$ through the relation
\beq\label{met_even}
K^{(A)} &=-\frac{2 f r C_{1,lm}^{(A)}}{\frac{6 \Mbh}{r}+2 n}+\frac{2 i f^2 C_{2,lm}^{(A)}}{\frac{6 \Mbh}{r}+2 n}+f \partial_r \Psi_z^{(A)}+\frac{\Psi_z^{(A)} \left(2 f n+\frac{12 \Mbh^2}{r^2}+\frac{10 \Mbh n}{r}+2 n^2\right)}{6 \Mbh+2 n r}\\
H_0^{(A)} &=B_{lm}^{(A)}+\frac{r C_{1,lm}^{(A)} \left(3 \Mbh^2+3 \Mbh n r-n r^2\right)}{(3 \Mbh+n r)^2}-\frac{i f C_{2,lm}^{(A)} \left(3 \Mbh^2+3 \Mbh n r-n r^2\right)}{(3 \Mbh+n r)^2}\\&+\Psi_z^{(A)} \left(\frac{9 \Mbh^3+9 \Mbh^2 n r+3 \Mbh n^2 r^2+n^2 (n+1) r^3}{r^2 (3 \Mbh+n r)^2}-\frac{r \omega ^2}{f}\right)+\partial_r \Psi_z^{(A)} \left(1-\frac{3 \Mbh (\Mbh+n r+r)}{r (3 \Mbh+n r)}\right)\\
H_2^{(A)} &=B_{lm}^{(A)}+\frac{r C_{1,lm}^{(A)} \left(3 \Mbh^2+3 \Mbh n r-n r^2\right)}{(3 \Mbh+n r)^2}-\frac{i f C_{2,lm}^{(A)} \left(3 \Mbh^2+3 \Mbh n r-n r^2\right)}{(3 \Mbh+n r)^2}-\frac{8 \sqrt{2} \pi  r^2 F_{lm}}{\sqrt{n (n+1)}}\\&+\Psi_z^{(A)} \left(\frac{9 \Mbh^3+9 \Mbh^2 n r+3 \Mbh n^2 r^2+n^2 (n+1) r^3}{r^2 (3 \Mbh+n r)^2}-\frac{r \omega ^2}{f}\right)+\partial_r \Psi_z^{(A)} \left(1-\frac{3 \Mbh (\Mbh+n r+r)}{r (3 \Mbh+n r)}\right)\\
&-\varepsilon \frac{8 \sqrt{2} \pi  r r_{(0,1)} \left(r F_{lm}'(r)+2 F_{lm}\right)}{\sqrt{n (n+1)}}
\eeq
where if $(A)=(1,0)$, then $\varepsilon=0$ and if $(A)=(1,1)$, then $\varepsilon=1$
\section{Distributional properties of Dirac delta function}\label{app: Dirac delta}
In this section, we outline the distributional properties of the Dirac delta function.
For any generic contentiously differentiable function $f(r)$, the Dirac delta function satisfies the following relations
\beq\label{Dirac_delta_1}
f(r)\delta(r-r_0)&=f\left(r_0\right) \delta \left(r-r_0\right)\\
f(r)\delta'(r-r_0)&=f\left(r_0\right) \delta '\left(r-r_0\right)-\delta \left(r-r_0\right) f'\left(r_0\right)\\
f(r)\delta'(r-r_0)&=\delta \left(r-r_0\right) f''\left(r_0\right)-2 f'\left(r_0\right) \delta '\left(r-r_0\right)+f\left(r_0\right) \delta ''\left(r-r_0\right)
\eeq
Furthermore, for any generic smooth and contentiously differentiable function $g(r)$, the Dirac delta functions satisfies the following relations
\beq\label{Dirac_delta_2}
\delta(g(r))&=\frac
{\delta \left(r-r_0\right)}{\left| g'\left(r_0\right)\right| }\\
\delta'(g(r))&=\frac{\delta '\left(r-r_0\right)}{g'\left(r_0\right) \left| g'\left(r_0\right)\right| }+\frac{\delta \left(r-r_0\right) g''\left(r_0\right)}{g'\left(r_0\right){}^2 \left| g'\left(r_0\right)\right| }\\
\delta''(g(r))&=\frac{\delta ''\left(r-r_0\right)}{g'\left(r_0\right){}^2 \left| g'\left(r_0\right)\right| }+\frac{3 g''\left(r_0\right) \delta '\left(r-r_0\right)}{g'\left(r_0\right){}^3 \left| g'\left(r_0\right)\right| }+\frac{\delta \left(r-r_0\right) \left(\frac{3 g''\left(r_0\right){}^2}{g'\left(r_0\right){}^3}-\frac{g^{(3)}\left(r_0\right)}{g'\left(r_0\right){}^2}\right)}{g'\left(r_0\right) \left| g'\left(r_0\right)\right| }
\eeq
 \end{widetext}


	\bibliography{0_ref}

@article{Kawamura:2011zz,
    author = "Kawamura, Seiji and others",
    editor = "Buchman, Sasha and Sun, Ke-Xun",
    title = "{The Japanese space gravitational wave antenna: DECIGO}",
    doi = "10.1088/0264-9381/28/9/094011",
    journal = "Class. Quant. Grav.",
    volume = "28",
    pages = "094011",
    year = "2011"
}

@article{Lacroix:2018qqh,
    author = "Lacroix, Thomas and Stref, Martin and Lavalle, Julien",
    title = "{Anatomy of Eddington-like inversion methods in the context of dark matter searches}",
    eprint = "1805.02403",
    archivePrefix = "arXiv",
    primaryClass = "astro-ph.GA",
    reportNumber = "LUPM:18-022",
    doi = "10.1088/1475-7516/2018/09/040",
    journal = "JCAP",
    volume = "09",
    pages = "040",
    year = "2018"
}

@article{1990ApJ...356..359H,
       author = {{Hernquist}, Lars},
        title = "{An Analytical Model for Spherical Galaxies and Bulges}",
      journal = {Astrophys. J.},
         year = 1990,
        month = jun,
       volume = {356},
        pages = {359},
          doi = {10.1086/168845},
       adsurl = {https://ui.adsabs.harvard.edu/abs/1990ApJ...356..359H}
}

@article{PhysRevD.105.L061501,
  title = {Black holes in galaxies: Environmental impact on gravitational-wave generation and propagation},
  author = {Cardoso, Vitor and Destounis, Kyriakos and Duque, Francisco and Macedo, Rodrigo Panosso and Maselli, Andrea},
  journal = {Phys. Rev. D},
  volume = {105},
  issue = {6},
  pages = {L061501},
  numpages = {7},
  year = {2022},
  month = {Mar},
  publisher = {American Physical Society},
  doi = {10.1103/PhysRevD.105.L061501},
  url = {https://link.aps.org/doi/10.1103/PhysRevD.105.L061501}
}

@article{LIGOScientific:2017vox,
    author = "Abbott, B. . P. . and others",
    collaboration = "LIGO Scientific, Virgo",
    title = "{GW170608: Observation of a 19-solar-mass Binary Black Hole Coalescence}",
    eprint = "1711.05578",
    archivePrefix = "arXiv",
    primaryClass = "astro-ph.HE",
    reportNumber = "LIGO-DOCUMENT-P170608-V8",
    doi = "10.3847/2041-8213/aa9f0c",
    journal = "Astrophys. J. Lett.",
    volume = "851",
    pages = "L35",
    year = "2017"
}

@article{Martel:2005ir,
    author = "Martel, Karl and Poisson, Eric",
    title = "{Gravitational perturbations of the Schwarzschild spacetime: A Practical covariant and gauge-invariant formalism}",
    eprint = "gr-qc/0502028",
    archivePrefix = "arXiv",
    doi = "10.1103/PhysRevD.71.104003",
    journal = "Phys. Rev. D",
    volume = "71",
    pages = "104003",
    year = "2005"
}

@article{PhysRev.108.1063,
  title = {Stability of a Schwarzschild Singularity},
  author = {Regge, Tullio and Wheeler, John A.},
  journal = {Phys. Rev.},
  volume = {108},
  issue = {4},
  pages = {1063--1069},
  numpages = {0},
  year = {1957},
  month = {Nov},
  publisher = {American Physical Society},
  doi = {10.1103/PhysRev.108.1063},
  url = {https://link.aps.org/doi/10.1103/PhysRev.108.1063}
}

@article{PhysRevLett.24.737,
  title = {Effective Potential for Even-Parity Regge-Wheeler Gravitational Perturbation Equations},
  author = {Zerilli, Frank J.},
  journal = {Phys. Rev. Lett.},
  volume = {24},
  issue = {13},
  pages = {737--738},
  numpages = {0},
  year = {1970},
  month = {Mar},
  publisher = {American Physical Society},
  doi = {10.1103/PhysRevLett.24.737},
  url = {https://link.aps.org/doi/10.1103/PhysRevLett.24.737}
}

@article{PhysRevD.2.2141,
  title = {Gravitational Field of a Particle Falling in a Schwarzschild Geometry Analyzed in Tensor Harmonics},
  author = {Zerilli, Frank J.},
  journal = {Phys. Rev. D},
  volume = {2},
  issue = {10},
  pages = {2141--2160},
  numpages = {0},
  year = {1970},
  month = {Nov},
  publisher = {American Physical Society},
  doi = {10.1103/PhysRevD.2.2141},
  url = {https://link.aps.org/doi/10.1103/PhysRevD.2.2141}
}

@article{Amaro_Seoane_2012,
	doi = {10.1088/0264-9381/29/12/124016},
	url = {https://doi.org/10.1088/0264-9381/29/12/124016},
	year = 2012,
	month = {jun},
	publisher = {{IOP} Publishing},
	volume = {29},
	number = {12},
	pages = {124016},
	author = {Pau Amaro-Seoane and Sofiane Aoudia and Stanislav Babak and Pierre Bin{\'{e}}truy and Emanuele Berti and Alejandro Boh{\'{e}} and Chiara Caprini and Monica Colpi and Neil J Cornish and Karsten Danzmann and Jean-Fran{\c{c}}ois Dufaux and Jonathan Gair and Oliver Jennrich and Philippe Jetzer and Antoine Klein and Ryan N Lang and Alberto Lobo and Tyson Littenberg and Sean T McWilliams and Gijs Nelemans and Antoine Petiteau and Edward K Porter and Bernard F Schutz and Alberto Sesana and Robin Stebbins and Tim Sumner and Michele Vallisneri and Stefano Vitale and Marta Volonteri and Henry Ward},
	title = {Low-frequency gravitational-wave science with {eLISA}/{NGO}},
	journal = {Classical and Quantum Gravity}
}

@article{Luo_2016,
	doi = {10.1088/0264-9381/33/3/035010},
	url = {https://doi.org/10.1088/0264-9381/33/3/035010},
	year = 2016,
	month = {jan},
	publisher = {{IOP} Publishing},
	volume = {33},
	number = {3},
	pages = {035010},
	author = {Jun Luo and Li-Sheng Chen and Hui-Zong Duan and Yun-Gui Gong and Shoucun Hu and Jianghui Ji and Qi Liu and Jianwei Mei and Vadim Milyukov and Mikhail Sazhin and Cheng-Gang Shao and Viktor T Toth and Hai-Bo Tu and Yamin Wang and Yan Wang and Hsien-Chi Yeh and Ming-Sheng Zhan and Yonghe Zhang and Vladimir Zharov and Ze-Bing Zhou},
	title = {{TianQin}: a space-borne gravitational wave detector},
	journal = {Classical and Quantum Gravity}
}

@article{PhysRevD.95.103012,
  title = {Science with the space-based interferometer LISA. V. Extreme mass-ratio inspirals},
  author = {Babak, Stanislav and Gair, Jonathan and Sesana, et al.},
  journal = {Phys. Rev. D},
  volume = {95},
  issue = {10},
  pages = {103012},
  numpages = {21},
  year = {2017},
  month = {May},
  publisher = {American Physical Society},
  doi = {10.1103/PhysRevD.95.103012},
  url = {https://link.aps.org/doi/10.1103/PhysRevD.95.103012}
}

@article{PhysRevD.102.024041,
  title = {Extreme mass ratio inspirals with spinning secondary: A detailed study of equatorial circular motion},
  author = {Piovano, Gabriel Andres and Maselli, Andrea and Pani, Paolo},
  journal = {Phys. Rev. D},
  volume = {102},
  issue = {2},
  pages = {024041},
  numpages = {28},
  year = {2020},
  month = {Jul},
  publisher = {American Physical Society},
  doi = {10.1103/PhysRevD.102.024041},
  url = {https://link.aps.org/doi/10.1103/PhysRevD.102.024041}
}

@article{Cardoso:2022whc,
    author = "Cardoso, Vitor and Destounis, Kyriakos and Duque, Francisco and Panosso Macedo, Rodrigo and Maselli, Andrea",
    title = "{Gravitational Waves from Extreme-Mass-Ratio Systems in Astrophysical Environments}",
    eprint = "2210.01133",
    archivePrefix = "arXiv",
    primaryClass = "gr-qc",
    doi = "10.1103/PhysRevLett.129.241103",
    journal = "Phys. Rev. Lett.",
    volume = "129",
    number = "24",
    pages = "241103",
    year = "2022"
}

@article{PhysRevD.67.104017,
  title = {Gauge problem in the gravitational self-force: Harmonic gauge approach in the Schwarzschild background},
  author = {Sago, Norichika and Nakano, Hiroyuki and Sasaki, Misao},
  journal = {Phys. Rev. D},
  volume = {67},
  issue = {10},
  pages = {104017},
  numpages = {14},
  year = {2003},
  month = {May},
  publisher = {American Physical Society},
  doi = {10.1103/PhysRevD.67.104017},
  url = {https://link.aps.org/doi/10.1103/PhysRevD.67.104017}
}

@article{Amaro-Seoane:2007osp,
    author = "Amaro-Seoane, Pau and Gair, Jonathan R. and Freitag, Marc and Coleman Miller, M. and Mandel, Ilya and Cutler, Curt J. and Babak, Stanislav",
    title = "{Astrophysics, detection and science applications of intermediate- and extreme mass-ratio inspirals}",
    eprint = "astro-ph/0703495",
    archivePrefix = "arXiv",
    doi = "10.1088/0264-9381/24/17/R01",
    journal = "Class. Quant. Grav.",
    volume = "24",
    pages = "R113--R169",
    year = "2007"
}

@article{Gair:2017ynp,
    author = "Gair, Jonathan R. and Babak, Stanislav and Sesana, Alberto and Amaro-Seoane, Pau and Barausse, Enrico and Berry, Christopher P. L. and Berti, Emanuele and Sopuerta, Carlos",
    editor = "Giardini, Domencio and Jetzer, Philippe",
    title = "{Prospects for observing extreme-mass-ratio inspirals with LISA}",
    eprint = "1704.00009",
    archivePrefix = "arXiv",
    primaryClass = "astro-ph.GA",
    doi = "10.1088/1742-6596/840/1/012021",
    journal = "J. Phys. Conf. Ser.",
    volume = "840",
    number = "1",
    pages = "012021",
    year = "2017"
}

@article{Babak:2017tow,
    author = "Babak, Stanislav and Gair, Jonathan and Sesana, Alberto and Barausse, Enrico and Sopuerta, Carlos F. and Berry, Christopher P. L. and Berti, Emanuele and Amaro-Seoane, Pau and Petiteau, Antoine and Klein, Antoine",
    title = "{Science with the space-based interferometer LISA. V: Extreme mass-ratio inspirals}",
    eprint = "1703.09722",
    archivePrefix = "arXiv",
    primaryClass = "gr-qc",
    doi = "10.1103/PhysRevD.95.103012",
    journal = "Phys. Rev. D",
    volume = "95",
    number = "10",
    pages = "103012",
    year = "2017"
}

@article{Gair:2010yu,
    author = "Gair, Jonathan R. and Tang, Christopher and Volonteri, Marta",
    title = "{LISA extreme-mass-ratio inspiral events as probes of the black hole mass function}",
    eprint = "1004.1921",
    archivePrefix = "arXiv",
    primaryClass = "astro-ph.GA",
    doi = "10.1103/PhysRevD.81.104014",
    journal = "Phys. Rev. D",
    volume = "81",
    pages = "104014",
    year = "2010"
}

@article{Barack:2009ux,
    author = "Barack, Leor",
    title = "{Gravitational self force in extreme mass-ratio inspirals}",
    eprint = "0908.1664",
    archivePrefix = "arXiv",
    primaryClass = "gr-qc",
    doi = "10.1088/0264-9381/26/21/213001",
    journal = "Class. Quant. Grav.",
    volume = "26",
    pages = "213001",
    year = "2009"
}

@article{Hinderer:2008dm,
    author = "Hinderer, Tanja and Flanagan, Eanna E.",
    title = "{Two timescale analysis of extreme mass ratio inspirals in Kerr. I. Orbital Motion}",
    eprint = "0805.3337",
    archivePrefix = "arXiv",
    primaryClass = "gr-qc",
    doi = "10.1103/PhysRevD.78.064028",
    journal = "Phys. Rev. D",
    volume = "78",
    pages = "064028",
    year = "2008"
}

@article{Drasco:2005kz,
    author = "Drasco, Steve and Hughes, Scott A.",
    title = "{Gravitational wave snapshots of generic extreme mass ratio inspirals}",
    eprint = "gr-qc/0509101",
    archivePrefix = "arXiv",
    doi = "10.1103/PhysRevD.73.024027",
    journal = "Phys. Rev. D",
    volume = "73",
    number = "2",
    pages = "024027",
    year = "2006",
    note = "[Erratum: Phys.Rev.D 88, 109905 (2013), Erratum: Phys.Rev.D 90, 109905 (2014)]"
}

@article{Gair:2004iv,
    author = "Gair, Jonathan R. and Barack, Leor and Creighton, Teviet and Cutler, Curt and Larson, Shane L. and Phinney, E. Sterl and Vallisneri, Michele",
    title = "{Event rate estimates for LISA extreme mass ratio capture sources}",
    eprint = "gr-qc/0405137",
    archivePrefix = "arXiv",
    doi = "10.1088/0264-9381/21/20/003",
    journal = "Class. Quant. Grav.",
    volume = "21",
    pages = "S1595--S1606",
    year = "2004"
}

@article{Sopuerta:2009iy,
    author = "Sopuerta, Carlos F. and Yunes, Nicolas",
    title = "{Extreme and Intermediate-Mass Ratio Inspirals in Dynamical Chern-Simons Modified Gravity}",
    eprint = "0904.4501",
    archivePrefix = "arXiv",
    primaryClass = "gr-qc",
    doi = "10.1103/PhysRevD.80.064006",
    journal = "Phys. Rev. D",
    volume = "80",
    pages = "064006",
    year = "2009"
}

@article{Rahman:2022fay,
    author = "Rahman, Mostafizur and Kumar, Shailesh and Bhattacharyya, Arpan",
    title = "{Gravitational wave from extreme mass-ratio inspirals as a probe of extra dimensions}",
    eprint = "2212.01404",
    archivePrefix = "arXiv",
    primaryClass = "gr-qc",
    doi = "10.1088/1475-7516/2023/01/046",
    journal = "JCAP",
    volume = "01",
    pages = "046",
    year = "2023"
}

@misc{BHPT,
	author = {{\relax Black Hole Perturbation Toolkit,}},
	howpublished = {\href{http://bhptoolkit.org/}{(http://bhptoolkit.org/})}
}

@article{PhysRevD.69.044025,
  title = {Gravitational waveforms from a point particle orbiting a Schwarzschild black hole},
  author = {Martel, Karl},
  journal = {Phys. Rev. D},
  volume = {69},
  issue = {4},
  pages = {044025},
  numpages = {20},
  year = {2004},
  month = {Feb},
  publisher = {American Physical Society},
  doi = {10.1103/PhysRevD.69.044025},
  url = {https://link.aps.org/doi/10.1103/PhysRevD.69.044025}
}

@article{Coogan:2021uqv,
    author = "Coogan, Adam and Bertone, Gianfranco and Gaggero, Daniele and Kavanagh, Bradley J. and Nichols, David A.",
    title = "{Measuring the dark matter environments of black hole binaries with gravitational waves}",
    eprint = "2108.04154",
    archivePrefix = "arXiv",
    primaryClass = "gr-qc",
    doi = "10.1103/PhysRevD.105.043009",
    journal = "Phys. Rev. D",
    volume = "105",
    number = "4",
    pages = "043009",
    year = "2022"
}

@article{Rahman:2021eay,
    author = "Rahman, Mostafizur and Bhattacharyya, Arpan",
    title = "{Prospects for determining the nature of the secondaries of extreme mass-ratio inspirals using the spin-induced quadrupole deformation}",
    eprint = "2112.13869",
    archivePrefix = "arXiv",
    primaryClass = "gr-qc",
    doi = "10.1103/PhysRevD.107.024006",
    journal = "Phys. Rev. D",
    volume = "107",
    number = "2",
    pages = "024006",
    year = "2023"
}

@article{Drummond:2023loz,
    author = "Drummond, Lisa V. and Hanselman, Alexandra G. and Becker, Devin R. and Hughes, Scott A.",
    title = "{Extreme mass-ratio inspiral of a spinning body into a Kerr black hole I: Evolution along generic trajectories}",
    eprint = "2305.08919",
    archivePrefix = "arXiv",
    primaryClass = "gr-qc",
    month = "5",
    year = "2023"
}

@article{Fan:2022wio,
    author = "Fan, Hui-Min and Zhong, Shiyan and Liang, Zheng-Cheng and Wu, Zheng and Zhang, Jian-dong and Hu, Yi-Ming",
    title = "{Extreme-mass-ratio burst detection with TianQin}",
    eprint = "2209.13387",
    archivePrefix = "arXiv",
    primaryClass = "gr-qc",
    doi = "10.1103/PhysRevD.106.124028",
    journal = "Phys. Rev. D",
    volume = "106",
    number = "12",
    pages = "124028",
    year = "2022"
}

@article{Liang:2022gdk,
    author = "Liang, Dicong and Xu, Rui and Mai, Zhan-Feng and Shao, Lijing",
    title = "{Probing vector hair of black holes with extreme-mass-ratio inspirals}",
    eprint = "2212.09346",
    archivePrefix = "arXiv",
    primaryClass = "gr-qc",
    doi = "10.1103/PhysRevD.107.044053",
    journal = "Phys. Rev. D",
    volume = "107",
    number = "4",
    pages = "044053",
    year = "2023"
}

@article{DeLuca:2023laa,
    author = "De Luca, Valerio and Khoury, Justin",
    title = "{Superfluid dark matter around black holes}",
    eprint = "2302.10286",
    archivePrefix = "arXiv",
    primaryClass = "astro-ph.CO",
    doi = "10.1088/1475-7516/2023/04/048",
    journal = "JCAP",
    volume = "04",
    pages = "048",
    year = "2023"
}

@misc{Mathematica,
  author = {Wolfram Research{,} Inc.},
  title = {Mathematica, {V}ersion 12.0},
  url = {https://www.wolfram.com/mathematica},
  note = {Champaign, IL, 2022}
}

@article{Maselli:2021men,
    author = "Maselli, Andrea and Franchini, Nicola and Gualtieri, Leonardo and Sotiriou, Thomas P. and Barsanti, Susanna and Pani, Paolo",
    title = "{Detecting fundamental fields with LISA observations of gravitational waves from extreme mass-ratio inspirals}",
    eprint = "2106.11325",
    archivePrefix = "arXiv",
    primaryClass = "gr-qc",
    doi = "10.1038/s41550-021-01589-5",
    journal = "Nature Astron.",
    volume = "6",
    number = "4",
    pages = "464--470",
    year = "2022"
}

@article{Apostolatos:2009vu,
    author = "Apostolatos, Theocharis A. and Lukes-Gerakopoulos, Georgios and Contopoulos, George",
    title = "{How to Observe a Non-Kerr Spacetime Using Gravitational Waves}",
    eprint = "0906.0093",
    archivePrefix = "arXiv",
    primaryClass = "gr-qc",
    doi = "10.1103/PhysRevLett.103.111101",
    journal = "Phys. Rev. Lett.",
    volume = "103",
    pages = "111101",
    year = "2009"
}

@article{Lukes-Gerakopoulos:2010ipp,
    author = "Lukes-Gerakopoulos, Georgios and Apostolatos, Theocharis A. and Contopoulos, George",
    title = "{Observable signature of a background deviating from the Kerr metric}",
    eprint = "1003.3120",
    archivePrefix = "arXiv",
    primaryClass = "gr-qc",
    doi = "10.1103/PhysRevD.81.124005",
    journal = "Phys. Rev. D",
    volume = "81",
    pages = "124005",
    year = "2010"
}

@article{Destounis:2020kss,
    author = "Destounis, Kyriakos and Suvorov, Arthur G. and Kokkotas, Kostas D.",
    title = "{Testing spacetime symmetry through gravitational waves from extreme-mass-ratio inspirals}",
    eprint = "2009.00028",
    archivePrefix = "arXiv",
    primaryClass = "gr-qc",
    doi = "10.1103/PhysRevD.102.064041",
    journal = "Phys. Rev. D",
    volume = "102",
    number = "6",
    pages = "064041",
    year = "2020"
}

@article{Destounis:2021mqv,
    author = "Destounis, Kyriakos and Suvorov, Arthur G. and Kokkotas, Kostas D.",
    title = "{Gravitational-wave glitches in chaotic extreme-mass-ratio inspirals}",
    eprint = "2103.05643",
    archivePrefix = "arXiv",
    primaryClass = "gr-qc",
    doi = "10.1103/PhysRevLett.126.141102",
    journal = "Phys. Rev. Lett.",
    volume = "126",
    number = "14",
    pages = "141102",
    year = "2021"
}

@article{Destounis:2021rko,
    author = "Destounis, Kyriakos and Kokkotas, Kostas D.",
    title = "{Gravitational-wave glitches: Resonant islands and frequency jumps in nonintegrable extreme-mass-ratio inspirals}",
    eprint = "2108.02782",
    archivePrefix = "arXiv",
    primaryClass = "gr-qc",
    doi = "10.1103/PhysRevD.104.064023",
    journal = "Phys. Rev. D",
    volume = "104",
    number = "6",
    pages = "064023",
    year = "2021"
}

@article{Destounis:2023gpw,
    author = "Destounis, Kyriakos and Huez, Giulia and Kokkotas, Kostas D.",
    title = "{Geodesics and gravitational waves in chaotic extreme-mass-ratio inspirals: the curious case of Zipoy-Voorhees black-hole mimickers}",
    eprint = "2301.11483",
    archivePrefix = "arXiv",
    primaryClass = "gr-qc",
    doi = "10.1007/s10714-023-03119-2",
    journal = "Gen. Rel. Grav.",
    volume = "55",
    number = "6",
    pages = "71",
    year = "2023"
}

@article{Destounis:2023khj,
    author = "Destounis, Kyriakos and Angeloni, Federico and Vaglio, Massimo and Pani, Paolo",
    title = "{Extreme-mass-ratio inspirals into rotating boson stars: Nonintegrability, chaos, and transient resonances}",
    eprint = "2305.05691",
    archivePrefix = "arXiv",
    primaryClass = "gr-qc",
    doi = "10.1103/PhysRevD.108.084062",
    journal = "Phys. Rev. D",
    volume = "108",
    number = "8",
    pages = "084062",
    year = "2023"
}

@article{Barausse:2014tra,
    author = "Barausse, Enrico and Cardoso, Vitor and Pani, Paolo",
    title = "{Can environmental effects spoil precision gravitational-wave astrophysics?}",
    eprint = "1404.7149",
    archivePrefix = "arXiv",
    primaryClass = "gr-qc",
    doi = "10.1103/PhysRevD.89.104059",
    journal = "Phys. Rev. D",
    volume = "89",
    number = "10",
    pages = "104059",
    year = "2014"
}

@article{Courty:2023rxk,
    author = "Courty, Aubin and Destounis, Kyriakos and Pani, Paolo",
    title = "{Spectral instability of quasinormal modes and strong cosmic censorship}",
    eprint = "2307.11155",
    archivePrefix = "arXiv",
    primaryClass = "gr-qc",
    doi = "10.1103/PhysRevD.108.104027",
    journal = "Phys. Rev. D",
    volume = "108",
    number = "10",
    pages = "104027",
    year = "2023"
}

@article{Sarkar:2023rhp,
    author = "Sarkar, Subhodeep and Rahman, Mostafizur and Chakraborty, Sumanta",
    title = "{Perturbing the perturbed: Stability of quasinormal modes in presence of a positive cosmological constant}",
    eprint = "2304.06829",
    archivePrefix = "arXiv",
    primaryClass = "gr-qc",
    doi = "10.1103/PhysRevD.108.104002",
    journal = "Phys. Rev. D",
    volume = "108",
    number = "10",
    pages = "104002",
    year = "2023"
}

@article{Vicente:2025gsg,
    author = "Vicente, Rodrigo and Karydas, Theophanes K. and Bertone, Gianfranco",
    title = "{A fully relativistic treatment of EMRIs in collisionless environments}",
    eprint = "2505.09715",
    archivePrefix = "arXiv",
    primaryClass = "gr-qc",
    month = "5",
    year = "2025"
}

@article{PhysRevLett.109.051101,
  title = {Second-Order Gravitational Self-Force},
  author = {Pound, Adam},
  journal = {Phys. Rev. Lett.},
  volume = {109},
  issue = {5},
  pages = {051101},
  numpages = {5},
  year = {2012},
  month = {Jul},
  publisher = {American Physical Society},
  doi = {10.1103/PhysRevLett.109.051101},
  url = {https://link.aps.org/doi/10.1103/PhysRevLett.109.051101}
}

@article{PhysRevD.95.104056,
  title = {Nonlinear gravitational self-force: Second-order equation of motion},
  author = {Pound, Adam},
  journal = {Phys. Rev. D},
  volume = {95},
  issue = {10},
  pages = {104056},
  numpages = {33},
  year = {2017},
  month = {May},
  publisher = {American Physical Society},
  doi = {10.1103/PhysRevD.95.104056},
  url = {https://link.aps.org/doi/10.1103/PhysRevD.95.104056}
}

@article{Dyson:2025dlj,
    author = "Dyson, Conor and Spieksma, Thomas F. M. and Brito, Richard and van de Meent, Maarten and Dolan, Sam",
    title = "{Environmental Effects in Extreme-Mass-Ratio Inspirals: Perturbations to the Environment in Kerr Spacetimes}",
    eprint = "2501.09806",
    archivePrefix = "arXiv",
    primaryClass = "gr-qc",
    doi = "10.1103/PhysRevLett.134.211403",
    journal = "Phys. Rev. Lett.",
    volume = "134",
    number = "21",
    pages = "211403",
    year = "2025"
}

@article{Mino:1996nk,
    author = "Mino, Yasushi and Sasaki, Misao and Tanaka, Takahiro",
    title = "{Gravitational radiation reaction to a particle motion}",
    eprint = "gr-qc/9606018",
    archivePrefix = "arXiv",
    reportNumber = "OU-TAP-38, KUNS-1394",
    doi = "10.1103/PhysRevD.55.3457",
    journal = "Phys. Rev. D",
    volume = "55",
    pages = "3457--3476",
    year = "1997"
}

@article{Quinn:1996am,
    author = "Quinn, Theodore C. and Wald, Robert M.",
    title = "{An Axiomatic approach to electromagnetic and gravitational radiation reaction of particles in curved space-time}",
    eprint = "gr-qc/9610053",
    archivePrefix = "arXiv",
    doi = "10.1103/PhysRevD.56.3381",
    journal = "Phys. Rev. D",
    volume = "56",
    pages = "3381--3394",
    year = "1997"
}

@article{Mathews:2021rod,
    author = "Mathews, Josh and Pound, Adam and Wardell, Barry",
    title = "{Self-force calculations with a spinning secondary}",
    eprint = "2112.13069",
    archivePrefix = "arXiv",
    primaryClass = "gr-qc",
    doi = "10.1103/PhysRevD.105.084031",
    journal = "Phys. Rev. D",
    volume = "105",
    number = "8",
    pages = "084031",
    year = "2022"
}

@article{Mathews:2025nyb,
    author = "Mathews, Josh and Pound, Adam",
    title = "{Post-adiabatic waveform-generation framework for asymmetric precessing binaries}",
    eprint = "2501.01413",
    archivePrefix = "arXiv",
    primaryClass = "gr-qc",
    month = "1",
    year = "2025"
}

@article{Hopper:2012ty,
    author = "Hopper, Seth and Evans, Charles R.",
    title = "{Metric perturbations from eccentric orbits on a Schwarzschild black hole: I. Odd-parity Regge-Wheeler to Lorenz gauge transformation and two new methods to circumvent the Gibbs phenomenon}",
    eprint = "1210.7969",
    archivePrefix = "arXiv",
    primaryClass = "gr-qc",
    reportNumber = "AEI-2012-190",
    doi = "10.1103/PhysRevD.87.064008",
    journal = "Phys. Rev. D",
    volume = "87",
    number = "6",
    pages = "064008",
    year = "2013"
}

@article{Warburton:2013lea,
    author = "Warburton, Niels and Wardell, Barry",
    title = "{Applying the effective-source approach to frequency-domain self-force calculations}",
    eprint = "1311.3104",
    archivePrefix = "arXiv",
    primaryClass = "gr-qc",
    doi = "10.1103/PhysRevD.89.044046",
    journal = "Phys. Rev. D",
    volume = "89",
    number = "4",
    pages = "044046",
    year = "2014"
}

@article{Durkan:2022fvm,
    author = "Durkan, Leanne and Warburton, Niels",
    title = "{Slow evolution of the metric perturbation due to a quasicircular inspiral into a Schwarzschild black hole}",
    eprint = "2206.08179",
    archivePrefix = "arXiv",
    primaryClass = "gr-qc",
    doi = "10.1103/PhysRevD.106.084023",
    journal = "Phys. Rev. D",
    volume = "106",
    number = "8",
    pages = "084023",
    year = "2022"
}

@article{Ansorg:2003br,
    author = "Ansorg, Marcus and Kleinwachter, A. and Meinel, R.",
    title = "{Highly accurate calculation of rotating neutron stars: detailed description of the numerical methods}",
    eprint = "astro-ph/0301173",
    archivePrefix = "arXiv",
    doi = "10.1051/0004-6361:20030618",
    journal = "Astron. Astrophys.",
    volume = "405",
    pages = "711",
    year = "2003"
}

@article{Ansorg:2006gd,
    author = "Ansorg, Marcus",
    editor = "Campanelli, Manuela and Rezzolla, Luciano",
    title = "{Multi-Domain Spectral Method for Initial Data of Arbitrary Binaries in General Relativity}",
    eprint = "gr-qc/0612081",
    archivePrefix = "arXiv",
    doi = "10.1088/0264-9381/24/12/S01",
    journal = "Class. Quant. Grav.",
    volume = "24",
    number = "12",
    pages = "S1--S14",
    year = "2007"
}

@article{PanossoMacedo:2022fdi,
    author = "Panosso Macedo, Rodrigo and Leather, Benjamin and Warburton, Niels and Wardell, Barry and Zengino\u{g}lu, An\i{}l",
    title = "{Hyperboloidal method for frequency-domain self-force calculations}",
    eprint = "2202.01794",
    archivePrefix = "arXiv",
    primaryClass = "gr-qc",
    doi = "10.1103/PhysRevD.105.104033",
    journal = "Phys. Rev. D",
    volume = "105",
    number = "10",
    pages = "104033",
    year = "2022"
}

@article{Leather:2024mls,
    author = "Leather, Benjamin",
    title = "{Gravitational self-force with hyperboloidal slicing and spectral methods}",
    eprint = "2411.14976",
    archivePrefix = "arXiv",
    primaryClass = "gr-qc",
    month = "11",
    year = "2024"
}

@article{PanossoMacedo:2024pox,
    author = "Panosso Macedo, Rodrigo and Bourg, Patrick and Pound, Adam and Upton, Samuel D.",
    title = "{Multidomain spectral method for self-force calculations}",
    eprint = "2404.10083",
    archivePrefix = "arXiv",
    primaryClass = "gr-qc",
    doi = "10.1103/PhysRevD.110.084008",
    journal = "Phys. Rev. D",
    volume = "110",
    number = "8",
    pages = "084008",
    year = "2024"
}

@article{PanossoMacedo:2018hab,
	author = "Panosso Macedo, Rodrigo and Jaramillo, Jos\'e Luis and Ansorg, Marcus",
	title = {{Hyperboloidal slicing approach to quasi-normal mode expansions: the Reissner-Nordstr\"om case}},
	eprint = "1809.02837",
	archivePrefix = "arXiv",
	primaryClass = "gr-qc",
	doi = "10.1103/PhysRevD.98.124005",
	journal = "Phys. Rev. D",
	volume = "98",
	number = "12",
	pages = "124005",
	year = "2018"
}

@article{Jaramillo:2020tuu,
	author = "Jaramillo, Jos\'e Luis and Panosso Macedo, Rodrigo and Al Sheikh, Lamis",
	title = "{Pseudospectrum and Black Hole Quasinormal Mode Instability}",
	eprint = "2004.06434",
	archivePrefix = "arXiv",
	primaryClass = "gr-qc",
	doi = "10.1103/PhysRevX.11.031003",
	journal = "Phys. Rev. X",
	volume = "11",
	number = "3",
	pages = "031003",
	year = "2021"
}

@article{Zenginoglu:2011jz,
	author = "Zenginoglu, Anil",
	title = "{A Geometric framework for black hole perturbations}",
	eprint = "1102.2451",
	archivePrefix = "arXiv",
	primaryClass = "gr-qc",
	doi = "10.1103/PhysRevD.83.127502",
	journal = "Phys. Rev. D",
	volume = "83",
	pages = "127502",
	year = "2011"
}

@book{trefethenMATLAB10.5555/357801,
author = {Trefethen, Lloyd N.},
title = {Spectral Methods in MATLAB},
year = {2000},
isbn = {0898714656},
publisher = {Society for Industrial and Applied Mathematics},
address = {USA}
}

@article{LIGOScientific:2018mvr,
    author = "Abbott, B. P. and others",
    collaboration = "LIGO Scientific, Virgo",
    title = "{GWTC-1: A Gravitational-Wave Transient Catalog of Compact Binary Mergers Observed by LIGO and Virgo during the First and Second Observing Runs}",
    eprint = "1811.12907",
    archivePrefix = "arXiv",
    primaryClass = "astro-ph.HE",
    reportNumber = "LIGO-P1800307",
    doi = "10.1103/PhysRevX.9.031040",
    journal = "Phys. Rev. X",
    volume = "9",
    number = "3",
    pages = "031040",
    year = "2019"
}

@article{LIGOScientific:2020ibl,
    author = "Abbott, R. and others",
    collaboration = "LIGO Scientific, Virgo",
    title = "{GWTC-2: Compact Binary Coalescences Observed by LIGO and Virgo During the First Half of the Third Observing Run}",
    eprint = "2010.14527",
    archivePrefix = "arXiv",
    primaryClass = "gr-qc",
    reportNumber = "P2000061",
    doi = "10.1103/PhysRevX.11.021053",
    journal = "Phys. Rev. X",
    volume = "11",
    pages = "021053",
    year = "2021"
}

@article{Cardoso:2016oxy,
    author = "Cardoso, Vitor and Hopper, Seth and Macedo, Caio F. B. and Palenzuela, Carlos and Pani, Paolo",
    title = "{Gravitational-wave signatures of exotic compact objects and of quantum corrections at the horizon scale}",
    eprint = "1608.08637",
    archivePrefix = "arXiv",
    primaryClass = "gr-qc",
    doi = "10.1103/PhysRevD.94.084031",
    journal = "Phys. Rev. D",
    volume = "94",
    number = "8",
    pages = "084031",
    year = "2016"
}

@article{Cheung:2021bol,
    author = "Cheung, Mark Ho-Yeuk and Destounis, Kyriakos and Macedo, Rodrigo Panosso and Berti, Emanuele and Cardoso, Vitor",
    title = "{Destabilizing the Fundamental Mode of Black Holes: The Elephant and the Flea}",
    eprint = "2111.05415",
    archivePrefix = "arXiv",
    primaryClass = "gr-qc",
    doi = "10.1103/PhysRevLett.128.111103",
    journal = "Phys. Rev. Lett.",
    volume = "128",
    number = "11",
    pages = "111103",
    year = "2022"
}

@article{Zenginoglu:2007jw,
    author = "Zenginoglu, Anil",
    title = "{Hyperboloidal foliations and scri-fixing}",
    eprint = "0712.4333",
    archivePrefix = "arXiv",
    primaryClass = "gr-qc",
    reportNumber = "AEI-2007-177",
    doi = "10.1088/0264-9381/25/14/145002",
    journal = "Class. Quant. Grav.",
    volume = "25",
    pages = "145002",
    year = "2008"
}

@article{Ansorg:2016ztf,
    author = "Ansorg, Marcus and Panosso Macedo, Rodrigo",
    title = "{Spectral decomposition of black-hole perturbations on hyperboloidal slices}",
    eprint = "1604.02261",
    archivePrefix = "arXiv",
    primaryClass = "gr-qc",
    doi = "10.1103/PhysRevD.93.124016",
    journal = "Phys. Rev. D",
    volume = "93",
    number = "12",
    pages = "124016",
    year = "2016"
}

@article{PanossoMacedo:2023qzp,
    author = "Panosso Macedo, Rodrigo",
    title = "{Hyperboloidal approach for static spherically symmetric spacetimes: a didactical introductionand applications in black-hole physics}",
    eprint = "2307.15735",
    archivePrefix = "arXiv",
    primaryClass = "gr-qc",
    doi = "10.1098/rsta.2023.0046",
    journal = "Phil. Trans. Roy. Soc. Lond. A",
    volume = "382",
    number = "2267",
    pages = "20230046",
    year = "2024"
}

@article{Barausse:2007ph,
    author = "Barausse, Enrico",
    title = "{Relativistic dynamical friction in a collisional fluid}",
    eprint = "0709.0211",
    archivePrefix = "arXiv",
    primaryClass = "astro-ph",
    doi = "10.1111/j.1365-2966.2007.12408.x",
    journal = "Mon. Not. Roy. Astron. Soc.",
    volume = "382",
    pages = "826--834",
    year = "2007"
}

@article{KAGRA:2021vkt,
    author = "Abbott, R. and others",
    collaboration = "KAGRA, VIRGO, LIGO Scientific",
    title = "{GWTC-3: Compact Binary Coalescences Observed by LIGO and Virgo during the Second Part of the Third Observing Run}",
    eprint = "2111.03606",
    archivePrefix = "arXiv",
    primaryClass = "gr-qc",
    reportNumber = "LIGO-P2000318",
    doi = "10.1103/PhysRevX.13.041039",
    journal = "Phys. Rev. X",
    volume = "13",
    number = "4",
    pages = "041039",
    year = "2023"
}

@article{Chakravarti:2024ncc,
    author = "Chakravarti, Kabir and Reza, Amit and Trombetta, Leonardo G.",
    title = "{Mergers of hairy black holes: Constraining topological couplings from entropy}",
    eprint = "2405.10127",
    archivePrefix = "arXiv",
    primaryClass = "gr-qc",
    doi = "10.1103/PhysRevD.110.064032",
    journal = "Phys. Rev. D",
    volume = "110",
    number = "6",
    pages = "064032",
    year = "2024"
}

@article{Pretorius:2005gq,
    author = "Pretorius, Frans",
    title = "{Evolution of binary black hole spacetimes}",
    eprint = "gr-qc/0507014",
    archivePrefix = "arXiv",
    doi = "10.1103/PhysRevLett.95.121101",
    journal = "Phys. Rev. Lett.",
    volume = "95",
    pages = "121101",
    year = "2005"
}

@article{Campanelli:2005dd,
  author    = "Campanelli, Manuela and Lousto, Carlos O. and Marronetti, Pedro and Zlochower, Yosef",
  title     = "Accurate evolutions of orbiting black-hole binaries without excision",
  journal   = "Phys. Rev. Lett.",
  volume    = "96",
  year      = "2006",
  pages     = "111101",
  doi       = "10.1103/PhysRevLett.96.111101",
  eprint    = "gr-qc/0511048"
}

@article{Buonanno:2006ui,
  author    = "Buonanno, Alessandra and Cook, Gregory B. and Pretorius, Frans",
  title     = "Inspiral, merger and ring-down of equal-mass black-hole binaries",
  journal   = "Phys. Rev. D",
  volume    = "75",
  year      = "2007",
  pages     = "124018",
  doi       = "10.1103/PhysRevD.75.124018",
  eprint    = "gr-qc/0610122"
}

@article{Centrella:2010mx,
  author    = "Centrella, Joan and Baker, John G. and Kelly, Bernard J. and van Meter, James R.",
  title     = "Black-hole binaries, gravitational waves, and numerical relativity",
  journal   = "Rev. Mod. Phys.",
  volume    = "82",
  year      = "2010",
  pages     = "3069--3119",
  doi       = "10.1103/RevModPhys.82.3069",
  eprint    = "1010.5260 [gr-qc]"
}

@article{TianQin:2015yph,
    author = "Luo, Jun and others",
    collaboration = "TianQin",
    title = "{TianQin: a space-borne gravitational wave detector}",
    eprint = "1512.02076",
    archivePrefix = "arXiv",
    primaryClass = "astro-ph.IM",
    doi = "10.1088/0264-9381/33/3/035010",
    journal = "Class. Quant. Grav.",
    volume = "33",
    number = "3",
    pages = "035010",
    year = "2016"
}

@article{Ruan:2018tsw,
    author = "Ruan, Wen-Hong and Guo, Zong-Kuan and Cai, Rong-Gen and Zhang, Yuan-Zhong",
    title = "{Taiji program: Gravitational-wave sources}",
    eprint = "1807.09495",
    archivePrefix = "arXiv",
    primaryClass = "gr-qc",
    doi = "10.1142/S0217751X2050075X",
    journal = "Int. J. Mod. Phys. A",
    volume = "35",
    number = "17",
    pages = "2050075",
    year = "2020"
}

@article{LIGOScientific:2016aoc,
    author = "Abbott, B. P. and others",
    collaboration = "LIGO Scientific, Virgo",
    title = "{Observation of Gravitational Waves from a Binary Black Hole Merger}",
    eprint = "1602.03837",
    archivePrefix = "arXiv",
    primaryClass = "gr-qc",
    reportNumber = "LIGO-P150914",
    doi = "10.1103/PhysRevLett.116.061102",
    journal = "Phys. Rev. Lett.",
    volume = "116",
    number = "6",
    pages = "061102",
    year = "2016"
}

@Inbook{Carson:2020rea,
author="Carson, Zack
and Yagi, Kent",
editor="Bambi, Cosimo
and Katsanevas, Stavros
and Kokkotas, Konstantinos D.",
title="Testing General Relativity with Gravitational Waves",
bookTitle="Handbook of Gravitational Wave Astronomy",
year="2020",
publisher="Springer Singapore",
address="Singapore",
pages="1--33",
isbn="978-981-15-4702-7",
doi="10.1007/978-981-15-4702-7_41-1",
url="https://doi.org/10.1007/978-981-15-4702-7_41-1"
}

@article{Berti:2015itd,
    author = "Berti, Emanuele and others",
    title = "{Testing General Relativity with Present and Future Astrophysical Observations}",
    eprint = "1501.07274",
    archivePrefix = "arXiv",
    primaryClass = "gr-qc",
    doi = "10.1088/0264-9381/32/24/243001",
    journal = "Class. Quant. Grav.",
    volume = "32",
    pages = "243001",
    year = "2015"
}

@article{Perkins:2020tra,
    author = "Perkins, Scott E. and Yunes, Nicol\'as and Berti, Emanuele",
    title = "{Probing Fundamental Physics with Gravitational Waves: The Next Generation}",
    eprint = "2010.09010",
    archivePrefix = "arXiv",
    primaryClass = "gr-qc",
    doi = "10.1103/PhysRevD.103.044024",
    journal = "Phys. Rev. D",
    volume = "103",
    number = "4",
    pages = "044024",
    year = "2021"
}

@article{Barack:2018yly,
    author = "Barack, Leor and others",
    title = "{Black holes, gravitational waves and fundamental physics: a roadmap}",
    eprint = "1806.05195",
    archivePrefix = "arXiv",
    primaryClass = "gr-qc",
    doi = "10.1088/1361-6382/ab0587",
    journal = "Class. Quant. Grav.",
    volume = "36",
    number = "14",
    pages = "143001",
    year = "2019"
}

@article{PhysRevD.78.064028,
  title = {Two-timescale analysis of extreme mass ratio inspirals in Kerr spacetime: Orbital motion},
  author = {Hinderer, Tanja and Flanagan, \'Eanna \'E.},
  journal = {Phys. Rev. D},
  volume = {78},
  issue = {6},
  pages = {064028},
  numpages = {39},
  year = {2008},
  month = {Sep},
  publisher = {American Physical Society},
  doi = {10.1103/PhysRevD.78.064028}
}

@article{Liebling:2012fv,
    author = "Liebling, Steven L. and Palenzuela, Carlos",
    title = "{Dynamical Boson Stars}",
    eprint = "1202.5809",
    archivePrefix = "arXiv",
    primaryClass = "gr-qc",
    doi = "10.12942/lrr-2012-6",
    journal = "Living Rev. Rel.",
    volume = "15",
    pages = "6",
    year = "2012"
}

@article{Cardoso:2019rvt,
    author = "Cardoso, Vitor and Pani, Paolo",
    title = "{Testing the nature of dark compact objects: a status report}",
    eprint = "1904.05363",
    archivePrefix = "arXiv",
    primaryClass = "gr-qc",
    doi = "10.1007/s41114-019-0020-4",
    journal = "Living Rev. Rel.",
    volume = "22",
    number = "1",
    pages = "4",
    year = "2019"
}

@article{PhysRevD.103.064048,
  title = {Two-timescale evolution of extreme-mass-ratio inspirals: Waveform generation scheme for quasicircular orbits in Schwarzschild spacetime},
  author = {Miller, Jeremy and Pound, Adam},
  journal = {Phys. Rev. D},
  volume = {103},
  issue = {6},
  pages = {064048},
  numpages = {43},
  year = {2021},
  month = {Mar},
  publisher = {American Physical Society},
  doi = {10.1103/PhysRevD.103.064048}
}

@article{LIGOScientific:2021sio,
    author = "Abbott, R. and others",
    collaboration = "LIGO Scientific, VIRGO, KAGRA",
    title = "{Tests of General Relativity with GWTC-3}",
    eprint = "2112.06861",
    archivePrefix = "arXiv",
    primaryClass = "gr-qc",
    reportNumber = "LIGO-P2100275",
    month = "12",
    year = "2021"
}

@article{Poisson:2011nh,
    author = "Poisson, Eric and Pound, Adam and Vega, Ian",
    title = "{The Motion of point particles in curved spacetime}",
    eprint = "1102.0529",
    archivePrefix = "arXiv",
    primaryClass = "gr-qc",
    doi = "10.12942/lrr-2011-7",
    journal = "Living Rev. Rel.",
    volume = "14",
    pages = "7",
    year = "2011"
}

@article{PhysRevD.81.084039,
  title = {Self-force on a scalar charge in Kerr spacetime: Circular equatorial orbits},
  author = {Warburton, Niels and Barack, Leor},
  journal = {Phys. Rev. D},
  volume = {81},
  issue = {8},
  pages = {084039},
  numpages = {17},
  year = {2010},
  month = {Apr},
  publisher = {American Physical Society},
  doi = {10.1103/PhysRevD.81.084039},
  url = {https://link.aps.org/doi/10.1103/PhysRevD.81.084039}
}

@article{Speeney:2022ryg,
    author = "Speeney, Nicholas and Antonelli, Andrea and Baibhav, Vishal and Berti, Emanuele",
    title = "{Impact of relativistic corrections on the detectability of dark-matter spikes with gravitational waves}",
    eprint = "2204.12508",
    archivePrefix = "arXiv",
    primaryClass = "gr-qc",
    doi = "10.1103/PhysRevD.106.044027",
    journal = "Phys. Rev. D",
    volume = "106",
    number = "4",
    pages = "044027",
    year = "2022"
}

@article{Gondolo:1999ef,
    author = "Gondolo, Paolo and Silk, Joseph",
    title = "{Dark matter annihilation at the galactic center}",
    eprint = "astro-ph/9906391",
    archivePrefix = "arXiv",
    reportNumber = "MPI-PHT-99-10, OUAST-99-9",
    doi = "10.1103/PhysRevLett.83.1719",
    journal = "Phys. Rev. Lett.",
    volume = "83",
    pages = "1719--1722",
    year = "1999"
}

@article{Speeney:2024mas,
    author = "Speeney, Nicholas and Berti, Emanuele and Cardoso, Vitor and Maselli, Andrea",
    title = "{Black holes surrounded by generic matter distributions: polar perturbations and energy flux}",
    eprint = "2401.00932",
    archivePrefix = "arXiv",
    primaryClass = "gr-qc",
    month = "1",
    year = "2024"
}

@misc{nakagawa2023appelllauricella,
      title={Appell-Lauricella hypergeometric functions over finite fields and algebraic varieties}, 
      author={Akio Nakagawa},
      year={2023},
      eprint={2210.13993},
      archivePrefix={arXiv},
      primaryClass={math.NT}
}

@article{Sadeghian:2013laa,
    author = "Sadeghian, Laleh and Ferrer, Francesc and Will, Clifford M.",
    title = "{Dark matter distributions around massive black holes: A general relativistic analysis}",
    eprint = "1305.2619",
    archivePrefix = "arXiv",
    primaryClass = "astro-ph.GA",
    doi = "10.1103/PhysRevD.88.063522",
    journal = "Phys. Rev. D",
    volume = "88",
    number = "6",
    pages = "063522",
    year = "2013"
}

@article{Barack:2018yvs,
    author = "Barack, Leor and Pound, Adam",
    title = "{Self-force and radiation reaction in general relativity}",
    eprint = "1805.10385",
    archivePrefix = "arXiv",
    primaryClass = "gr-qc",
    doi = "10.1088/1361-6633/aae552",
    journal = "Rept. Prog. Phys.",
    volume = "82",
    number = "1",
    pages = "016904",
    year = "2019"
}

@article{Sago:2005fn,
    author = "Sago, Norichika and Tanaka, Takahiro and Hikida, Wataru and Ganz, Katsuhiko and Nakano, Hiroyuki",
    title = "{The Adiabatic evolution of orbital parameters in the Kerr spacetime}",
    eprint = "gr-qc/0511151",
    archivePrefix = "arXiv",
    doi = "10.1143/PTP.115.873",
    journal = "Prog. Theor. Phys.",
    volume = "115",
    pages = "873--907",
    year = "2006"
}

@article{Isoyama:2018sib,
    author = "Isoyama, Soichiro and Fujita, Ryuichi and Nakano, Hiroyuki and Sago, Norichika and Tanaka, Takahiro",
    title = "{{\textquotedblleft}Flux-balance formulae{\textquotedblright} for extreme mass-ratio inspirals}",
    eprint = "1809.11118",
    archivePrefix = "arXiv",
    primaryClass = "gr-qc",
    reportNumber = "KUNS-2734, YITP-18-96",
    doi = "10.1093/ptep/pty136",
    journal = "PTEP",
    volume = "2019",
    number = "1",
    pages = "013E01",
    year = "2019"
}

@article{vandeMeent:2017bcc,
    author = "van de Meent, Maarten",
    title = "{Gravitational self-force on generic bound geodesics in Kerr spacetime}",
    eprint = "1711.09607",
    archivePrefix = "arXiv",
    primaryClass = "gr-qc",
    doi = "10.1103/PhysRevD.97.104033",
    journal = "Phys. Rev. D",
    volume = "97",
    number = "10",
    pages = "104033",
    year = "2018"
}

@article{PhysRevD.96.084057,
  title = {Evolution of small-mass-ratio binaries with a spinning secondary},
  author = {Warburton, Niels and Osburn, Thomas and Evans, Charles. R.},
  journal = {Phys. Rev. D},
  volume = {96},
  issue = {8},
  pages = {084057},
  numpages = {17},
  year = {2017},
  month = {Oct},
  publisher = {American Physical Society},
  doi = {10.1103/PhysRevD.96.084057},
  url = {https://link.aps.org/doi/10.1103/PhysRevD.96.084057}
}

@article{Witzany:2019dii,
    author = "Witzany, Vojt{\v{e}}ch",
    title = "{Spin-perturbed orbits near black holes}",
    eprint = "1903.03649",
    archivePrefix = "arXiv",
    primaryClass = "gr-qc",
    month = "3",
    year = "2019"
}

@article{PhysRevD.102.064013,
  title = {Dissipation in extreme mass-ratio binaries with a spinning secondary},
  author = {Akcay, Sarp and Dolan, Sam R. and Kavanagh, Chris and Moxon, Jordan and Warburton, Niels and Wardell, Barry},
  journal = {Phys. Rev. D},
  volume = {102},
  issue = {6},
  pages = {064013},
  numpages = {25},
  year = {2020},
  month = {Sep},
  publisher = {American Physical Society},
  doi = {10.1103/PhysRevD.102.064013},
  url = {https://link.aps.org/doi/10.1103/PhysRevD.102.064013}
}

@article{PhysRevD.89.104020,
  title = {Practical, covariant puncture for second-order self-force calculations},
  author = {Pound, Adam and Miller, Jeremy},
  journal = {Phys. Rev. D},
  volume = {89},
  issue = {10},
  pages = {104020},
  numpages = {25},
  year = {2014},
  month = {May},
  publisher = {American Physical Society},
  doi = {10.1103/PhysRevD.89.104020},
  url = {https://link.aps.org/doi/10.1103/PhysRevD.89.104020}
}

@article{PhysRevD.92.104047,
  title = {Second-order perturbation theory: Problems on large scales},
  author = {Pound, Adam},
  journal = {Phys. Rev. D},
  volume = {92},
  issue = {10},
  pages = {104047},
  numpages = {26},
  year = {2015},
  month = {Nov},
  publisher = {American Physical Society},
  doi = {10.1103/PhysRevD.92.104047},
  url = {https://link.aps.org/doi/10.1103/PhysRevD.92.104047}
}

@article{Tomaselli:2023ysb,
    author = "Tomaselli, Giovanni Maria and Spieksma, Thomas F. M. and Bertone, Gianfranco",
    title = "{Dynamical friction in gravitational atoms}",
    eprint = "2305.15460",
    archivePrefix = "arXiv",
    primaryClass = "gr-qc",
    doi = "10.1088/1475-7516/2023/07/070",
    journal = "JCAP",
    volume = "07",
    pages = "070",
    year = "2023"
}

@article{Mitra:2025tag,
    author = "Mitra, Soumodeep and Speeney, Nicholas and Chakraborty, Sumanta and Berti, Emanuele",
    title = "{Extreme mass ratio inspirals in rotating dark matter spikes}",
    eprint = "2505.04697",
    archivePrefix = "arXiv",
    primaryClass = "gr-qc",
    month = "5",
    year = "2025"
}

@article{Rahman:2023sof,
    author = "Rahman, Mostafizur and Kumar, Shailesh and Bhattacharyya, Arpan",
    title = "{Probing astrophysical environment with eccentric extreme mass-ratio inspirals}",
    eprint = "2306.14971",
    archivePrefix = "arXiv",
    primaryClass = "gr-qc",
    doi = "10.1088/1475-7516/2024/01/035",
    journal = "JCAP",
    volume = "01",
    pages = "035",
    year = "2024"
}

@article{Chakraborty:2024gcr,
    author = "Chakraborty, Sumanta and Comp{\`e}re, Geoffrey and Machet, Ludovico",
    title = "{Tidal Love numbers and quasi-normal modes of the Schwarzschild-Hernquist black hole}",
    eprint = "2412.14831",
    archivePrefix = "arXiv",
    primaryClass = "gr-qc",
    month = "12",
    year = "2024"
}

@inproceedings{Rendall:1996gx,
    author = "Rendall, Alan D.",
    title = "{An Introduction to the Einstein-Vlasov system}",
    booktitle = "{Mathematical Aspects of Theories of Gravitation}",
    eprint = "gr-qc/9604001",
    archivePrefix = "arXiv",
    reportNumber = "AEI-005",
    month = "2",
    year = "1996"
}

@InProceedings{Rendall,
author="Rendall, Alan D.",
editor="Chru{\'{s}}ciel, Piotr T.
and Friedrich, Helmut",
title="The Einstein-Vlasov System",
booktitle="The Einstein Equations and the Large Scale Behavior of Gravitational Fields",
year="2004",
publisher="Birkh{\"a}user Basel",
address="Basel",
pages="231--250",
isbn="978-3-0348-7953-8"
}

@ARTICLE{1975ApJ...200..204F,
       author = {{Friedman}, J.~L. and {Schutz}, B.~F.},
        title = "{On the stability of relativistic systems.}",
      journal = {\apj},
         year = 1975,
        month = aug,
       volume = {200},
        pages = {204-220},
          doi = {10.1086/153778},
       adsurl = {https://ui.adsabs.harvard.edu/abs/1975ApJ...200..204F}
}

@article{Navarro:1996gj,
    author = "Navarro, Julio F. and Frenk, Carlos S. and White, Simon D. M.",
    title = "{A Universal density profile from hierarchical clustering}",
    eprint = "astro-ph/9611107",
    archivePrefix = "arXiv",
    doi = "10.1086/304888",
    journal = "Astrophys. J.",
    volume = "490",
    pages = "493--508",
    year = "1997"
}

@article{Chandrasekhar:1943ys,
    author = "Chandrasekhar, Subrahmanyan",
    title = "{Dynamical Friction. I. General Considerations: the Coefficient of Dynamical Friction}",
    doi = "10.1086/144517",
    journal = "Astrophys. J.",
    volume = "97",
    pages = "255",
    year = "1943"
}

@misc{RaBHman,
  author = {Rahman, Mostafizur and Takahashi, Takuya},
  title = {Source Term Coefficients and Master Variables code},
howpublished = {GitHub repository},
  url = {https://github.com/RaBHman/Dark_Matter_Environment.git},
year         = {2025}
}

@article{Berezhiani:2023vlo,
    author = "Berezhiani, Lasha and Cintia, Giordano and De Luca, Valerio and Khoury, Justin",
    title = "{Dynamical friction in dark matter superfluids: The evolution of black hole binaries}",
    eprint = "2311.07672",
    archivePrefix = "arXiv",
    primaryClass = "astro-ph.CO",
    doi = "10.1088/1475-7516/2024/06/024",
    journal = "JCAP",
    volume = "06",
    pages = "024",
    year = "2024"
}

@article{Gliorio:2025cbh,
    author = "Gliorio, Sara and Berti, Emanuele and Maselli, Andrea and Speeney, Nicholas",
    title = "{Extreme mass ratio inspirals in dark matter halos: dynamics and distinguishability of halo models}",
    eprint = "2503.16649",
    archivePrefix = "arXiv",
    primaryClass = "gr-qc",
    month = "3",
    year = "2025"
}

@article{Datta:2025ruh,
    author = "Datta, Sayak and Maselli, Andrea",
    title = "{A multi-parameter expansion for the evolution of asymmetric binaries in astrophysical environments}",
    eprint = "2507.04471",
    archivePrefix = "arXiv",
    primaryClass = "gr-qc",
    month = "7",
    year = "2025"
}

@article{DeFalco:2024ojf,
    author = "De Falco, Vittorio and Battista, Emmanuele and Usseglio, Davide and Capozziello, Salvatore",
    title = "{Radiative losses and radiation-reaction effects at the first post-Newtonian order in Einstein{\textendash}Cartan theory}",
    eprint = "2401.13374",
    archivePrefix = "arXiv",
    primaryClass = "gr-qc",
    doi = "10.1140/epjc/s10052-024-12476-4",
    journal = "Eur. Phys. J. C",
    volume = "84",
    number = "2",
    pages = "137",
    year = "2024"
}

@article{Battista:2023znv,
    author = "Battista, Emmanuele and De Falco, Vittorio and Usseglio, Davide",
    title = "{First post-Newtonian N-body problem in Einstein{\textendash}Cartan theory with the Weyssenhoff fluid: Lagrangian and first integrals}",
    eprint = "2301.08954",
    archivePrefix = "arXiv",
    primaryClass = "gr-qc",
    doi = "10.1140/epjc/s10052-023-11249-9",
    journal = "Eur. Phys. J. C",
    volume = "83",
    number = "2",
    pages = "112",
    year = "2023"
}

@article{Battista:2021rlh,
    author = "Battista, Emmanuele and De Falco, Vittorio",
    title = "{First post-Newtonian generation of gravitational waves in Einstein-Cartan theory}",
    eprint = "2109.01384",
    archivePrefix = "arXiv",
    primaryClass = "gr-qc",
    doi = "10.1103/PhysRevD.104.084067",
    journal = "Phys. Rev. D",
    volume = "104",
    number = "8",
    pages = "084067",
    year = "2021"
}
	\bibliographystyle{utphys1}
\end{document}